\documentclass{aims}
\usepackage{amsmath}
  \usepackage{paralist}
  \usepackage{graphics} 
  \usepackage{epsfig} 
\usepackage{graphicx}  \usepackage{epstopdf}
 \usepackage[colorlinks=true]{hyperref}
\hypersetup{urlcolor=blue, citecolor=red}

\usepackage{comment}

\def\currentvolume{X}
 \def\currentissue{X}
  \def\currentyear{200X}
   \def\currentmonth{XX}
    \def\ppages{X--XX}
     \def\DOI{10.3934/xx.xx.xx.xx}

\newtheorem{theorem}{Theorem}[section]

\newtheorem{lemma}[theorem]{Lemma}
\newtheorem{proposition}{Proposition}

\theoremstyle{definition}

\newtheorem{assumption}{Assumption} 
\newtheorem{rem}{Remark} 

\newcommand\independent{\protect\mathpalette{\protect\independenT}{\perp}}
    \def\independenT#1#2{\mathrel{\rlap{$#1#2$}\mkern2mu{#1#2}}}

\usepackage{constants}
\usepackage{amsthm}
\usepackage{amsmath}
\usepackage{amssymb}
\usepackage{algorithm}
\usepackage{subfig}

\newcommand{\basic}[1]{\arabic{#1}}
\newconstantfamily{event}{
symbol=\mathcal{E},
format=\basic
}
\newconstantfamily{cte}{
symbol=\operatorname{C},
format=\basic
}
\newconstantfamily{gam}{
symbol=\gamma,
format=\basic
}

\def\bias{\operatorname{B}}
\def\R{\mathbb{R}}
\def\D{\mathbb{D}}
\def\E{\mathbb{E}}
\def\B{\mathbb{B}}
\def\i{\operatorname{i}} 
\def\tr{\operatorname{trace}} 
\def\K{\operatorname{K}} 
\def\L{\operatorname{L}} 
\def\H{\operatorname{H}} 
\def\Nave{{N}_{X,\min,\epsilon}} 
\def\NN{\overline{N}} 

\def\l{\left}
\def\r{\right}

\title[Learning of the Second-Moment Matrix] 
      {Randomized Learning of the \\Second-Moment Matrix of a Smooth Function}

\author[A.\ Eftekhari and M.\ B.\ Wakin and P.\ Li and P.\ G.\ Constantine]{}


 \email{armin.eftekhari@epfl.ch}
 \email{mwakin@mines.edu}
 \email{pingli@stat.rutgers.edu}
 \email{paul.constantine@colorado.edu}

\thanks{AE was partially supported by the Alan Turing Institute under the EPSRC grant EP/N510129/1 and also by the Turing Seed Funding grant SF019.   MBW was partially supported by NSF grant CCF-1409258 and NSF CAREER grant CCF-1149225. PGC was partially supported by the U.S.\ Department of Energy Office of Science, Office of Advanced Scientific Computing Research, Applied Mathematics program under award DE-SC0011077 and the Defense Advanced Research Projects Agency's Enabling Quantification of Uncertainty in Physical Systems.
}

\thanks{$^*$ Corresponding author: Armin Eftekhari}

\begin{document}
\maketitle

\centerline{\scshape Armin Eftekhari$^*$}
\medskip
{\footnotesize
 \centerline{Institute of Electrical Engineering}
\centerline{\'{E}cole Polytechnique F\'{e}d\'{e}rale de Lausanne}
   \centerline{ 1015 Lausanne, Switzerland}
} 

\medskip

\centerline{\scshape Michael B.\ Wakin}
\medskip
{\footnotesize
 \centerline{Department of Electrical Engineering}
  \centerline{Colorado School of Mines}
   \centerline{Denver, CO 80401, USA}
}

\bigskip

\centerline{\scshape Ping Li}
\medskip
{\footnotesize
 \centerline{Departments of Statistics and Computer Science}
  \centerline{Rutgers University}
   \centerline{Piscataway,NJ 08854, USA}
}

\bigskip

\centerline{\scshape Paul G.\ Constantine}
\medskip
{\footnotesize
 \centerline{Department of Computer Science}
  \centerline{University of Colorado Boulder}
   \centerline{Boulder, CO 80309, USA}
}

\bigskip

 \centerline{(Communicated by the associate editor name)}

\begin{abstract}
Consider an open set $\D\subseteq\mathbb{R}^n$, equipped with a probability measure $\mu$. An important characteristic of a smooth function $f:\D\rightarrow\R$ is its \emph{second-moment matrix} $\Sigma_{\mu}:=\int \nabla f(x) \nabla f(x)^* \mu(dx) \in\mathbb{R}^{n\times n}$, where $\nabla f(x)\in\R^n$ is the gradient of $f(\cdot)$ at $x\in\mathbb{D}$ and $*$ stands for transpose. For instance, the span of the leading $r$ eigenvectors of $\Sigma_{\mu}$ forms an \emph{active subspace} of $f(\cdot)$, which contains the directions along which $f(\cdot)$ changes the most and is of particular interest in \emph{ridge approximation}. In this work, we propose a simple algorithm for  estimating $\Sigma_{\mu}$ from random point evaluations of  $f(\cdot)$ \emph{without} imposing any structural assumptions on $\Sigma_{\mu}$. Theoretical guarantees for this algorithm are established with the aid of the same technical tools that have proved valuable in the context of covariance matrix estimation from partial measurements.
\end{abstract}


\section{Introduction}
\label{sec:intro}

Central to approximation theory, machine learning, and computational sciences in general is the task of learning a function given its finitely many point samples. More concretely, consider an open set $\D\subseteq\R^n$, equipped with probability measure $\mu$.  The objective is to \emph{learn} (approximate) a smooth function $f:\D\rightarrow\R$ from the query points
$$
\{x_i\}_{i=1}^N \subset \D,
$$
and evaluation of $f(\cdot)$ at these points~\cite{traub1980general,cohen2012capturing,fornasier2012learning,haupt2011distilled,wendland2004scattered,rasmussen2006gaussian}.

An important quantity in this context is the \emph{second-moment matrix} of $f(\cdot)$ with respect to the measure $\mu$, defined as
\begin{equation}
\Sigma_{\mu} := \E_x \l[ \nabla f(x) \cdot \l(\nabla f(x)\r)^*\r]  = \int_{\D} \nabla f(x) \cdot \l(\nabla f(x)\r)^* \, \mu(dx) \in \mathbb{R}^{n\times n},
\label{eq:def of Sigma intro}
\end{equation}
where $\nabla f(x)\in\R^n$ is the gradient of $f(\cdot)$ at $x\in\D$  and the superscript $*$ denotes vector and matrix transpose.\footnote{As suggested above, we will often suppress the dependence on $f(\cdot)$ in our notation for the sake of brevity.} The $[i,j]$th entry of this matrix, namely $\Sigma_\mu[i,j]$,
measures the expected product between the $i$th and $j$th partial derivatives of $f(\cdot)$. Note that $\Sigma_\mu$ captures key information about how $f(\cdot)$ changes along different directions. Indeed, for an arbitrary vector $v\in\R^n$ with $\|v\|_2=1$, the \emph{directional derivative} of $f(\cdot)$ at $x\in\D$ and along $v$ is $v^*\nabla f(x)$, and it is easy to check that the directional derivative of $f(\cdot)$  along $v$, itself a scalar function on $\mathbb{D}$, has the average energy of ${v^*\Sigma_\mu v}$ with respect to the measure $\mu$. The directions with the most energy, that is the directions along which $f(\cdot)$ changes the most on average, are particularly important in \emph{ridge approximation}, where we are interested in approximating (the possibly  complicated function) $f(\cdot)$ with a (simpler) ridge function. More specifically, the leading $r$ eigenvectors of $\Sigma_\mu$ span an $r$-dimensional \emph{active subspace} of $f(\cdot)$ with respect to the measure $\mu$ \cite{constantine2015active}, which contains the directions along which $f(\cdot)$ changes the most. If $U_{\mu,r}\in\mathbb{R}^{n\times r}$ denotes an orthonormal basis for this active subspace, then it  might be possible to reliably approximate $f(x)$ with $h(U_{\mu,r}^*x)$ for all $x\in\mathbb{D}$ and for some smooth function $h:\mathbb{R}^r\rightarrow\mathbb{R}$. In this sense, we might think of ridge approximation and active subspaces as the extensions of, respectively, dimensionality reduction and principal components to high-dimensional functions. Beyond approximation theory, the significance of second-moment matrices (and related concepts) across a number of other disciplines is discussed in Section~\ref{sec:related work}.

With this introduction, the main objective of this paper is the following, which will be made precise later in Section~\ref{sec:problem statement}.

\begin{itemize}
\item[] \textbf{Objective:} \emph{Design query points $\{x_i\}_{i=1}^N$ and  learn from  $\{x_i,f(x_i)\}_{i=1}^N$ the second-moment matrix of $f(\cdot)$ with respect to the measure $\mu$.}
\end{itemize}

We must emphasize that we impose \emph{no structural assumptions} on the second-moment matrix (such as being low rank or sparse), a point that we shall revisit later in Section~\ref{sec:related work}. Our approach to this problem, alongside the results, is summarized next with minimal details for better accessibility. A rigorous account of the problem and our approach is then presented in  Sections \ref{sec:problem statement} and \ref{sec:results}.

\subsection{Approach}
\label{sec:approach}

We assume in this paper that points in the domain $\D$ are observed randomly according to the probability measure $\mu$. In particular, consider $N$ random points drawn independently from $\mu$ and stored as the columns of a matrix $X\in\R^{n\times N}$. It is then easy to verify~\cite{constantine2014computing} that
\begin{equation}
\dot{\Sigma}_{X} := \frac{1}{N} \sum_{x\in X}
\nabla f\l( x\r)  \cdot \nabla f\l( x\r)^*
\label{eq:def of Sigma dot intro}
\end{equation}
is an unbiased estimator of $\Sigma_{\mu}$ in \eqref{eq:def of Sigma intro}.\footnote{As indicated above, we slightly abuse the standard notation by treating matrices and sets interchangeably. For example, the expression $x\in X$ can also be interpreted as $x$ being a column of matrix $X$. } In fact,  a standard large deviation analysis reveals that $\|\dot{\Sigma}_{X}-\Sigma_{\mu}\| \propto \frac{1}{\sqrt{N}}$,  with overwhelming probability and for any matrix norm $\|\cdot\|$.

\begin{figure}[t]
\begin{center}
\includegraphics[scale=0.8]{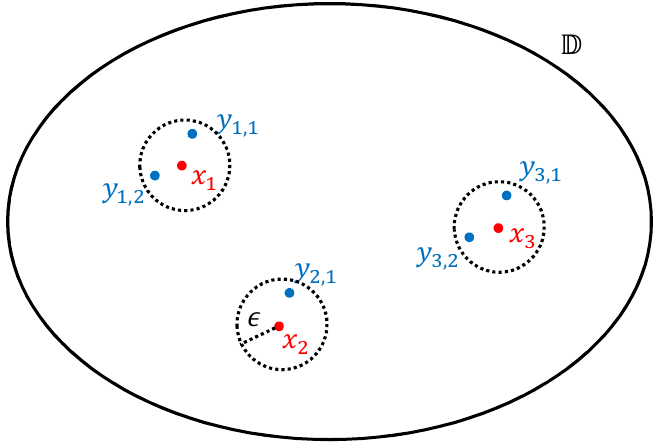}
\caption{Visualization of the problem setup. The probability measure $\mu$ is supported on the domain $\D\subseteq \mathbb{R}^n$ of a smooth function $f(\cdot)$.  Here, $N=3$ and $X=\{x_i\}_{i=1}^N$ are drawn independently from $\mu$. For sufficiently small $\epsilon$, we let $\B_{x_i,\epsilon}$ denote the $\epsilon$-neighborhood of each $x_i$ and set $\B_{X,\epsilon}=\cup_{i=1}^N \B_{x_i,\epsilon}$. On $\B_{X,\epsilon}$, $\mu$ induces the conditional measure $\mu_{X,\epsilon}$, from which $N_{X,\epsilon}$ points are independently drawn and collected in $Y_{X,\epsilon}=\{y_{ij}\}_{i,j}$. Here, $x_1$ has $N_{x_1,\epsilon}=2$ neighbors in $Y_{X,\epsilon}$ and we set  $Y_{x_1,\epsilon}=\{y_{1,j} \}_{j=1}^{N_{x_1,\epsilon}}$.  Similarly, $Y_{x_2,\epsilon}$ and $Y_{x_3,\epsilon}$ are formed. Note that $Y_{X,\epsilon}=\cup_{i=1}^N Y_{x_i,\epsilon} $. Our objective is to estimate the second-moment matrix of $f(\cdot)$ (with respect to the  probability measure $\mu$) given $\{x_i,f(x_i)\}$ and $\{y_{ij},f(y_{ij})\}$.\label{fig:diagram} }
\end{center}
\end{figure}

Since we furthermore  assume that only the point values of $f(\cdot)$ are at our disposal (rather than its gradients), it is not possible to directly calculate $\dot{\Sigma}_X$ as in~\eqref{eq:def of Sigma dot intro}. Thus, one might resort to using finite difference approximations of the partial derivatives, as we sketch here and formalize in Section~\ref{sec:problem statement}. Our procedure for estimating the second-moment matrix of $f(\cdot)$ will in fact rely not only on $\{x_i,f(x_i)\}_{i=1}^N$ but also on a supplementary set of points (also drawn randomly) nearby those in $X$.  In particular, for a sufficiently small $\epsilon>0$ and arbitrary $x$, let $\B_{x,\epsilon}$ denote  the Euclidean ball of radius $\epsilon$ about $x$, and set
\begin{equation*}
\B_{X,\epsilon} = \bigcup_{x\in X} \B_{x,\epsilon}.
\end{equation*}
Let also $\mu_{X,\epsilon}$ be the conditional probability measure on $\B_{X,\epsilon}$ induced by $\mu$. Consider $N_{X,\epsilon}$ random points drawn independently from $\mu_{X,\epsilon}$ and stored as the columns of $Y_{X,\epsilon}\in \R^{n\times N_{X,\epsilon}}$. Then partition $Y_{X,\epsilon}$ according to $X$ by setting $Y_{x,\epsilon}=Y_{X,\epsilon}\cap \B_{x,\epsilon}$, so that $Y_{x,\epsilon}\in\R^{n\times N_{x,\epsilon}}$ contains all $\epsilon$-neighbors of $x$ in $Y_{X,\epsilon}$. This setup is illustrated in Figure \ref{fig:diagram}.

For every $x\in X$, consider $\dot{\nabla}_{Y_{x,\epsilon}} f(x) \in \mathbb{R}^n$ as an estimate of the true gradient $\nabla f(x)$, where
\begin{equation}
\dot{\nabla}_{Y_{x,\epsilon}} f(x) := \frac{n}{N_{x,\epsilon}} \sum_{y\in Y_{x,\epsilon}}\frac{f(y)-f(x)}{\l\| y-x\r\|_2} \cdot
\frac{y-x}{\|y-x\|_2};
\end{equation}
the scaling with $n$ will be shortly justified. Then we could naturally consider $\dot{\Sigma}_{X,Y_{X,\epsilon}}\in\mathbb{R}^{n\times n}$ as an estimate of $\dot{\Sigma}_{X}$ in \eqref{eq:def of Sigma dot intro}, and in turn an estimate of  $\Sigma_{\mu}$ in \eqref{eq:def of Sigma intro}, where
\begin{equation}
\dot{\Sigma}_{X,Y_{X,\epsilon}}:=
\frac{1}{N} \sum_{x\in X}
\dot{\nabla}_{Y_{x,\epsilon}} f(x) \cdot \dot{\nabla}_{Y_{x,\epsilon}} f(x)^*.
\label{eq:Sigma single dot intro}
\end{equation}
In general, however, $\dot{\Sigma}_{X,Y_{X,\epsilon}}$ is a biased estimator of $\dot{\Sigma}_X$ and the importance of consistency in statistical learning motivated us to search for a better estimator, not unlike the approach taken in covariance matrix estimation \cite{lounici2014high,friedman2008sparse}, see also Section~\ref{sec:contribution}. Algorithm~\ref{alg:alg}, in fact, introduces a estimate of $\dot{\Sigma}_{X}$, denoted throughout by $\ddot{\Sigma}_{X,Y_{X,\epsilon}}$, which has a smaller bias than $\dot{\Sigma}_{X,Y_{X,\epsilon}}$. Indeed, Theorem~\ref{thm:bias} in Section~\ref{sec:results}, roughly speaking,\footnote{{ In order to simplify this overview, we suppress the less important terms here before turning to the details later.}} establishes that
\begin{equation}
\l\| \E\l[\ddot{\Sigma}_{X,Y_{X,\epsilon}}\r]- {\Sigma}_{\mu} \r\|_F
 \lesssim 
 \bias_{\mu,\epsilon}+\epsilon n^{3/2},
\label{eq:rough bias result}
\end{equation}
where the expectation is over $X,Y_{X,\epsilon}$ and $\|\cdot\|_F$ stands for the Frobenius norm. Throughout, we will use $\lesssim$ and similarly $\gtrsim,\approx$ to suppress universal constants and simplify the presentation. Above, the quantity $\bias_{\mu,\epsilon}$ depends in a certain way on the regularity of the measure $\mu$ and function $f(\cdot)$, with the dependence on $f(\cdot)$ suppressed  as usual. Moreover, loosely speaking, it  holds true that
\begin{equation}
\l\| \ddot{\Sigma}_{X,Y_{X,\epsilon}} - \Sigma_{\mu} \r\|_F
\lesssim \bias_{\mu,\epsilon}+
\epsilon n^2 +
\frac{n}{\sqrt{N_{X,\epsilon}}},
\label{eq:rough finite sample result}
\end{equation}
with high probability, as described in Theorem~\ref{thm:cvg rate} in Section~\ref{sec:results}. {  As a rule of thumb, \eqref{eq:rough bias result} and \eqref{eq:rough finite sample result} hold when $N_{X,\epsilon}\approx N\log^2 N$. Thus, it suffices to take only $O(\log^2 N)$ samples within the $\epsilon$-neighborhood of each of the $N$ data points in $X$. As we discuss in Remark~\ref{rem:maininterpretation}, the resulting convergence rate (as a function of the total number of samples) is then nearly the same as one achieves when perfect knowledge of the gradients is available and two-stage sampling is not required.}



\begin{figure}[!h]
\centering
\subfloat[$N_{X,\text{min},\epsilon}=50$]{
\label{fig:sub0}
\includegraphics[width=0.42\textwidth]{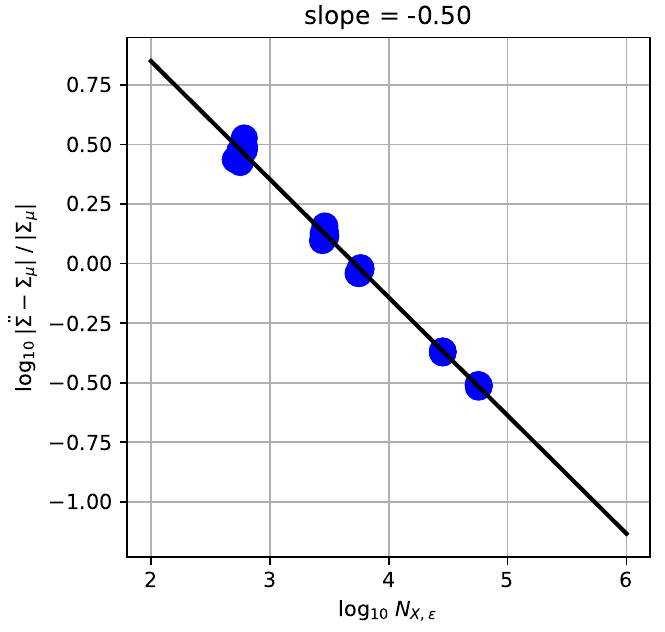}%
}
\hfil
\subfloat[$N_{X,\text{min},\epsilon}=200$]{
\label{fig:sub1}
\includegraphics[width=0.42\textwidth]{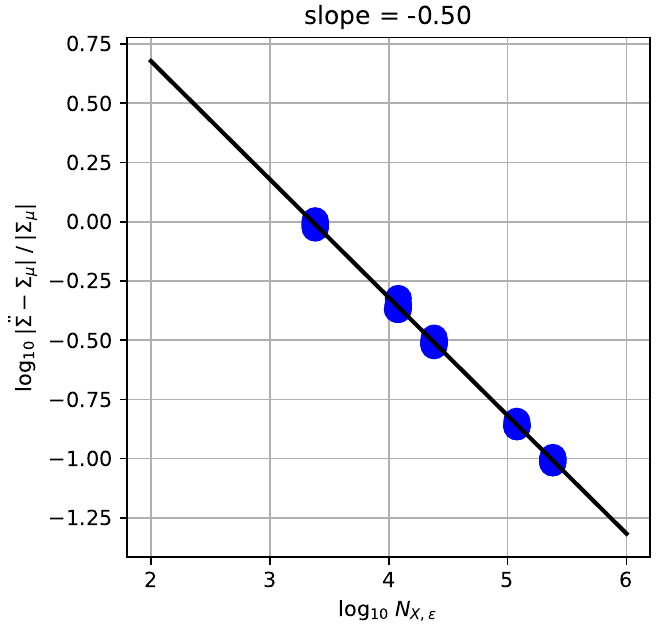}%
}
\\
\subfloat[$N_{X,\text{min},\epsilon}=400$]{
\label{fig:sub2}
\includegraphics[width=0.42\textwidth]{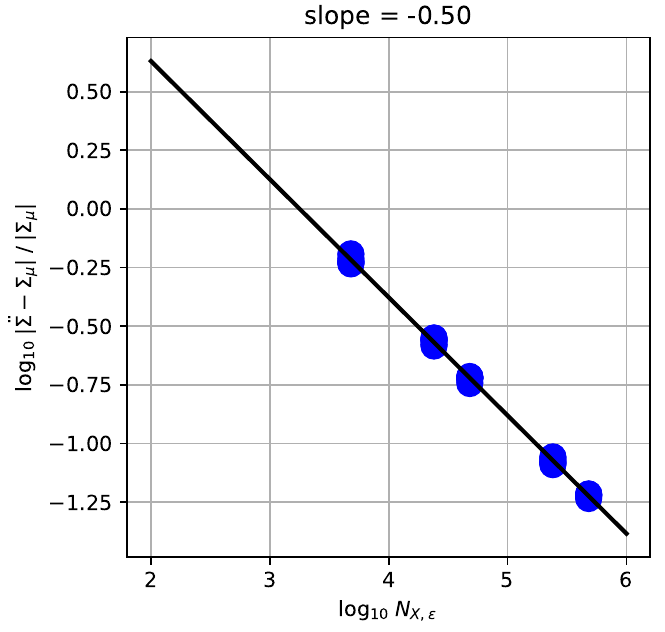}%
}
\hfil
\subfloat[$N_{X,\text{min},\epsilon}=550$]{
\label{fig:sub3}
\includegraphics[width=0.42\textwidth]{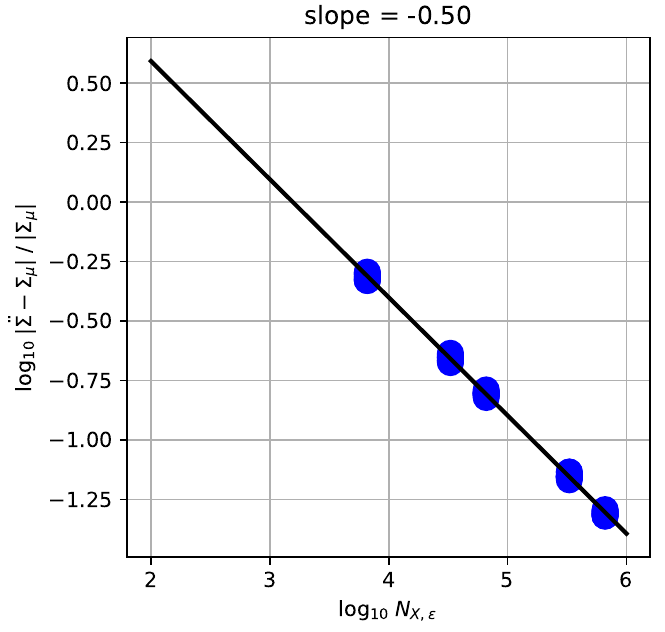}%
}
\caption{A simulation study using a 500-dimensional ($n=500$) quadratic function for the relative error in the approximation \eqref{eq:sigmahathat 0} as a function of the number $N_{X,\epsilon}$ of { total} samples. All simulations use $\epsilon=10^{-4}$. Each subfigure uses a different value for { the minimum number $N_{X,\text{min},\epsilon}$ of points in the $\epsilon$-neighborhood of each center point in $X$, and varies the number $N$ of centers.} Each of the five groups of blue dots in each subfigure { indicates a} different number $N$. Within each group, there are ten replications. The slope of each line is $-1/2$, which verifies the expected convergence rate from \eqref{eq:rough finite sample result}.}
\label{fig:0}
\end{figure}

To verify the convergence rate from \eqref{eq:rough finite sample result}, we consider the following numerical example. For $x\in [-1,1]^{500}$ (i.e., $n=500$) and $\mu$ a uniform probability distribution on the hypercube, let $f(x) = \frac{1}{2}x^\ast A x + b^\ast x$ for a known symmetric matrix $A\in\mathbb{R}^{500\times 500}$ and a known vector $b\in\mathbb{R}^{500}$. A quick calculation shows that $\Sigma_{\mu}=\frac{1}{3}A^2 + bb^\ast$. For the simulation, we generate $A$ and $b$ randomly, and we estimate the relative error (in the Frobenius norm) in the approximation $\ddot{\Sigma}_{X,Y_{X,\epsilon}}$ from Algorithm~\ref{alg:0}. All reported results use $\epsilon=10^{-4}$; using $10^{-2}$ and $10^{-6}$ produced similar results. Each subfigure in Figure \ref{fig:0} shows results using a different value for $N_{X,\text{min},\epsilon}$ from the set $\{50, 200, 400, 550\}$, i.e., the minimum number of samples in each ball. Note that the first three values of $N_{X,\text{min},\epsilon}$ tested are significantly less than the dimension $n$ of the space. In other words, each gradient approximation uses significantly fewer than the $n+1$ samples that would be needed for a finite difference approximation. The experiments in each subfigure use the number of centers $N\in\{10, 50, 100, 500, 1000\}$, and for each value of $N$ there are 10 independent replications. The slope of the line is $-1/2$ in each case, which verifies the expected convergence rate in \eqref{eq:rough finite sample result}{, in this case to a bias of approximately zero (see Remark~\ref{rem:discussion thm bias})}.

\begin{center}
\begin{algorithm}
\caption{for estimating the second-moment matrix of the function $f(\cdot)$ with respect to the measure $\mu$\label{alg:alg}}
\vspace{5pt}

\textbf{Input:}
\begin{itemize}
\item Open set $\D\subseteq \R^n$, equipped with probability measure $\mu$.
\item An oracle that returns $f(x)$ for a query point $x\in\D$.
\item Neighborhood radius $\epsilon>0$, sample sizes $N$, $N_{X,\epsilon}$,  and integer $N_{X,\min,\epsilon}\le N_{X,\epsilon}$.
\end{itemize}

\textbf{Output:}
\begin{itemize}
\item $\ddot{\Sigma}_{X,Y_{X,\epsilon}}$,  as an estimate of $\Sigma_\mu$.
\end{itemize}

\textbf{Body:}
\begin{itemize}
\item Draw $N$ random points independently from $\mu$ and store them as the columns of $X\in\R^{n\times N}$.
\item Draw $N_{X,\epsilon}$ random points independently from $\mu_{X,\epsilon}$ and store them as the columns of $Y_{X,\epsilon}\in\R^{n\times N_{X,\epsilon}}$. Here, $\mu_{X,\epsilon}$ is the conditional probability measure induced by $\mu$ on $\B_{X,\epsilon}=\cup_{x\in X} \B_{x,\epsilon}$. In turn, $\B_{x,\epsilon}\subset\R^n$ is the Euclidean ball of radius $\epsilon$ about $x$. Partition $Y_{X,\epsilon}$ according to $X$ by setting $Y_{x,\epsilon}=Y_{X,\epsilon}\cap \B_{x,\epsilon}$, so that $Y_{x,\epsilon}\in\R^{n\times N_{x,\epsilon}}$ contains all $\epsilon$-neighbors of $x$ in $Y_{X,\epsilon}$.
\item Compute and return
\begin{align}
\ddot{\Sigma}_{X,Y_{X,\epsilon}} 
:=
\frac{1}{N} \l(1+\frac{1-\frac{2}{n}}{1+\frac{2}{n} } \cdot \Nave^{-1}\r)^{-1}
\nonumber\\
\cdot
\l(
\sum_{N_{x,\epsilon}\ge N_{X,\min,\epsilon}}\dot{\nabla}_{Y_{x,\epsilon}}f(x) \cdot  \dot{\nabla}_{Y_{x,\epsilon}}f(x)^* -
 \frac{
 \l\|\dot{\nabla }_{Y_{x,\epsilon}} f(x) \r\|_2^2}{\l(1+\frac{2}{n} \r)\Nave+n+1-\frac{2}{n}}  \cdot I_n
\r),
\label{eq:sigmahathat 0}
\end{align}
where $I_n$ denotes the $n \times n$ identity matrix, and
\begin{equation}
\dot{\nabla}_{Y_{x,\epsilon}}f(x):=\frac{n}{N_{x,\epsilon}}\sum_{y\in Y_{x,\epsilon}}
\frac{f(y)-f(x)}{\|y-x\|_2}
\cdot
\frac{y-x}{\|y-x\|_2}
\in\R^n.
\label{eq:grad est 0}
\end{equation}
\end{itemize}


\label{alg:0}
\end{algorithm}
\end{center}


\subsection{Contribution and Organization}
\label{sec:contribution}

The main contribution of this paper is the design and analysis of a simple algorithm to estimate the second-moment matrix $\Sigma_\mu$ of a smooth function $f(\cdot)$ from its point samples; see \eqref{eq:def of Sigma intro} and Algorithm~\ref{alg:alg}.  As argued earlier and also in Section~\ref{sec:related work},  $\Sigma_\mu$ is a key quantity in ridge approximation and a number of related problems.

The key distinction of this work is the lack of any structural assumptions (such as small rank or sparsity) on $\Sigma_\mu$; mild assumptions on $f$ are specified at the beginning of Section~\ref{sec:problem statement}. Imposing a specific structure on $\Sigma_\mu$ can lead to more efficient algorithms as we discuss in Section~\ref{sec:related work}.

At a very high level, there is indeed a parallel between estimating the second-moment matrix of a function from random point samples and estimating the covariance matrix of a random vector; Algorithm~\ref{alg:alg} in a sense produces an analogue of the \emph{sample covariance matrix}, adjusted to handle missing data~\cite{lounici2014high}. In this context, more efficient algorithms are available for estimating, for example,  the covariance matrix with a sparse inverse~\cite{friedman2008sparse}.  In this sense, we feel that this work fills an important gap in the literature of ridge approximation and perhaps dimensionality reduction by addressing the problem in more generality.

The rest of this paper is organized as follows. The problem of learning the second-moment matrix of a function is formalized in Section~\ref{sec:problem statement}. Our approach to this problem, stated more formally, along with the theoretical guarantees, are described in Section~\ref{sec:results}. In Section~\ref{sec:related work}, we sift through a large body of literature and summarize the relevant prior work. Proofs and technical details are deferred to Section~\ref{sec:theory} and  the appendices.

\section{Problem Statement and Approach}
\label{sec:problem statement}

In this section, we formalize the problem outlined in Section~\ref{sec:intro}.   Consider an open  set $\D\subseteq\mathbb{R}^{n}$, equipped with  subspace Borel $\sigma$-algebra and  probability measure $\mu$.
 We assume throughout that $f:\D\rightarrow\mathbb{R}$  is  twice differentiable on
$\D$, and that
\begin{equation}
\L_f := \sup_{x\in \D} \left\| \nabla f(x) \right\|_2 <\infty,
\label{eq:Lf}
\end{equation}
\begin{equation}
\H_f := \sup_{x\in \D} \left\| \nabla^2 f(x)   \right\|_2 < \infty,
\label{eq:Hf}
\end{equation}
where $\nabla f(x)\in\mathbb{R}^n$ and $\nabla^2 f(x)\in\R^{n\times n}$ are the gradient and Hessian of $f(\cdot)$ at $x\in\D$, respectively, and we use the notation $\| \cdot \|_2$ to denote both the $\ell_2$-norm of vectors and the spectral norm of matrices. Moreover, for $\epsilon>0$, let $\D_\epsilon\subset\D$ denote the $\epsilon$-interior of $\D$, namely $\D_\epsilon=\{x\in\D\,:\, \B_{x,\epsilon}\subseteq\D\}$. Throughout, $\B_{x,\epsilon}\subset\mathbb{R}^n$ denotes the (open) Euclidean ball of radius $\epsilon$ centered at $x$.

Consider $\Sigma_{\mu}\in\R^{n\times n}$ defined as
\begin{equation}
\Sigma_{\mu}:=\mathbb{E}_{x}\left[\nabla f(x )\cdot \nabla f(x)^{*}\right]=\int_{\D}\nabla f(x)\cdot \nabla f(x)^{*}\,\mu(dx),\label{eq:main}
\end{equation}
where $\E_x$ computes the expectation with respect to $x\sim \mu$.
Our objective in this work is to estimate $\Sigma_\mu$.
To that end, consider $N$ random points drawn
independently from $\mu$ and stored as the columns of  $X\in\R^{n\times N}$. Then, as noted in Section~\ref{sec:approach}, it is easy to verify that
\begin{equation}
\dot{\Sigma}_{X}:=\frac{1}{N}\sum_{x\in X}\nabla f(x)\cdot \nabla f(x)^{*},\label{eq:emp}
\end{equation}
is an unbiased estimator for  $\Sigma_{\mu}$ in (\ref{eq:main}). To interpret \eqref{eq:emp}, recall also that we  treat matrices and sets interchangeably throughout, slightly abusing the standard notation. In particular, $x\in X$ can also be interpreted as $x$ being a column of $X\in\R^{n\times N}$. The following result quantifies how well $\dot{\Sigma}_X$ approximates $\Sigma_\mu$. Its proof is included in Appendix \ref{sec:proof of proposition Paul} for completeness; see~\cite{constantine2014computing} for related results concerning the accuracy of $\dot{\Sigma}_{X}$ as an estimate of $\Sigma_{\mu}$.
\begin{proposition}\label{prop:Paul}
Let $X\in\mathbb{R}^{n\times N}$ contain $N$ independent samples drawn from the probability measure $\mu$. Then, $\dot{\Sigma}_{X}$  is an unbiased estimator for $\Sigma_{\mu}\in\mathbb{R}^{n\times n}$, see (\ref{eq:main}) and (\ref{eq:emp}). Moreover, except for a probability of at most $n^{-1}$, it holds that
\begin{equation}
\l\|\dot{\Sigma}_{X}-\Sigma_{\mu} \r\|_F \lesssim  \frac{ \L_f^2\log n}{\sqrt{N}}.
\end{equation}
\end{proposition}

Since only point values of $f(\cdot)$ are at our disposal, we cannot compute $\dot{\Sigma}_{X}$ directly. Instead, we will systematically generate random points near the point cloud  $X$ and then estimate $\dot{\Sigma}_{X}$ by aggregating  local information, as detailed next.

Given the point cloud $X\subset \D$, fix $\epsilon>0$, small enough so that $X$ is a $2\epsilon$-separated point cloud that belongs to the $\epsilon$-interior of $\D$. Formally, fix $\epsilon \le \epsilon_X$, where
\begin{equation}
\label{eq:no overlap}
\epsilon_X := \sup\l\{ \epsilon'\,:\,
X\subset \D_{\epsilon'}
\mbox{ and }
\l\|x-x' \r\|_2\ge 2\epsilon',\,\, \forall x,x'\in X,\,\, x \ne x'
 \r\}.
\end{equation}
Let
\begin{equation}
\B_{X,\epsilon} := \bigcup_{x\in X} \B_{x,\epsilon} \subseteq \D
\label{eq:def of BX}
\end{equation}
denote the $\epsilon$-neighborhood of the point cloud $X$. Consider the conditional probability measure on $\B_{X,\epsilon}$ described as
\begin{equation}
\label{eq:mu_X}
\mu_{X,\epsilon} =
\begin{cases}
\mu / \mu\l( \B_{X,\epsilon} \r),
& \mbox{inside } \B_{X,\epsilon},\\
0, & \mbox{outside } \B_{X,\epsilon}.
\end{cases}
\end{equation}
For an integer $N_{X,\epsilon}$, draw $N_{X,\epsilon}$ independent random points from $\mu_{X,\epsilon}$ and store them as the columns of $Y_{X,\epsilon}\in\mathbb{R}^{n\times N_{X,\epsilon}}$.
Finally, an estimate of $\dot {\Sigma}_{X}$   and in turn of $\Sigma_\mu$ as a function of $X,Y_{X,\epsilon}\subset \D$ and evaluations of $f(\cdot)$ at these points is proposed by $\ddot{\Sigma}_{X,Y_{X,\epsilon}}$  in Algorithm~\ref{alg:alg}.

\section{Theoretical Guarantees}
\label{sec:results}

Recalling \eqref{eq:main} and \eqref{eq:emp}, how well does $\ddot{\Sigma}_{X,Y_{X,\epsilon}}$ in Algorithm~\ref{alg:alg} approximate $\dot{\Sigma}_{X}$ and in turn $\Sigma_\mu$? Parsing the answer requires introducing additional  notation and imposing  a certain regularity assumption on $\mu$. All these we set out to do now, before stating the results in Section~\ref{sec:Theoretical guarantees subsec}.

For each $x\in X$, let the columns of $Y_{x,\epsilon}\in \mathbb{R}^{n\times N_{x,\epsilon}}$ contain the $\epsilon$-neighbors of $x$ in $Y_{X,\epsilon}$. In our notation, this can be written as
\begin{equation} \label{eq:def of neighborhoods}
Y_{x,\epsilon} := Y_{X,\epsilon} \cap \B_{x,\epsilon},
\qquad
\#Y_{x,\epsilon} =  N_{x,\epsilon}.
\end{equation}
Because $\epsilon\le \epsilon_X$ is small, see \eqref{eq:no overlap}, these neighborhoods do not intersect, that is
\begin{equation*}
Y_{x,\epsilon} \cap Y_{x',\epsilon} = \emptyset,\quad
\forall x,x'\in X, \quad x \ne x';
\end{equation*}
therefore, $Y_{X,\epsilon}$ is simply partitioned into $\#X=N$ subsets $\{Y_{x,\epsilon}\}_{x\in X}$.
Observe also that, conditioned on $x\in X$ and $N_{x,\epsilon}$, each neighbor $y\in Y_{x,\epsilon}$ follows the conditional probability measure described as follows:
\begin{equation}
y|x,N_{x,\epsilon} \sim \mu_{x,\epsilon}:=
\begin{cases}
\mu/{\mu\l( \B_{x,\epsilon}\r)}, & \mbox{inside } \B_{x,\epsilon},\\
0,& \mbox{outside } \B_{x,\epsilon}.
\end{cases}
\label{eq:cond mu}
\end{equation}

\subsection{Regularity of $\mu$}
\label{sec:assumptions}

In order to  introduce the regularity condition imposed on $\mu$ here, consider first the special case where the domain $\mathbb{D}\subset \mathbb{R}^n$ is bounded and   $\mu$ is the uniform probability measure on  $\D$. Then, for  $\epsilon>0$ and arbitrary $\epsilon$-interior point  $x\in\D_\epsilon$,   the conditional measure $\mu_{x,\epsilon}$   too is the uniform  measure on  $\B_{x,\epsilon}$, see \eqref{eq:cond mu}. Draw $y$ from $\mu_{x,\epsilon}$, that is, $y|x\sim \mu_{x,\epsilon}$ in our notation. Then it is easy to verify that  $y-x$ is an \emph{isotropic} random vector, namely
\begin{equation*}
\mathbb{E}_{y|x}\l[ (y-x)(y-x)^*\r] = C \cdot I_n,
\end{equation*}
for some factor $C$.\footnote{A simple calculation shows that $C=1/n$. See Appendix \ref{sec:proof of lemma bias pointwise}.} Above, $I_n\in\mathbb{R}^{n\times n}$ is the identity matrix and $\E_{y|x}[\cdot]=\E_y[\cdot|x]$ stands for conditional expectation, given $x$. A similar property plays an important role in this paper, as captured by Assumption~\ref{def:moments} below.

\begin{assumption}\textbf{\emph{(Local near-isotropy of $\mu$)}} \label{def:moments}
Throughout this paper, we assume that there exist $\epsilon_\mu,\K_\mu>0$ such that for all $\epsilon \le \epsilon_\mu$, the following requirement holds for any arbitrary $\epsilon$-interior point $x\in \D_{\epsilon}$.

Given $x$, draw $y$ from the conditional measure on the $\epsilon$-neighborhood of $x$, namely  $y|x\sim \mu_{x,\epsilon}$ with $\mu_{x,\epsilon}$  defined in (\ref{eq:cond mu}). Then, for every $\Cl[gam]{beta}\ge 0$ and arbitrary (but fixed) $v\in\mathbb{R}^n$, it holds that
 \begin{equation}
 \Pr{}_{y|x}\l[ \l\|P_{x,y} \cdot v \r\|_2^2 >  \Cr{beta} \cdot  \frac{\|v\|_2^2}{n}\r] \lesssim  e^{-\K_\mu \Cr{beta}},
 \label{eq:projconcentration}
 \end{equation}
where
\[
P_{x,y}:=  \frac{(y-x)(y-x)^*}{\|y-x\|_2^2} \in\mathbb{R}^{n\times n}
\]
is  the orthogonal projection onto the direction of $y-x$. Above, $\Pr_{y|x}[\cdot]=\Pr_{y}[\cdot|x]$ stands for conditional probability.
\end{assumption}

Roughly speaking,   under Assumption  \ref{def:moments}, $\mu$ is locally isotropic. Indeed, this assumption is met when $\mu$ is the uniform probability measure on $\D$, as shown in Appendix \ref{sec:uni meas satisiefs assumption}.  Moreover, Assumption \ref{def:moments} is not  too restrictive. One would expect that a  probability measure $\mu$,  if dominated by the uniform measure on $\D$ and with a smooth Radon-Nikodym derivative, satisfies Assumption \ref{def:moments} when restricted to sufficiently small neighborhoods.

Assumption \ref{def:moments} also controls the growth of the moments of $\|P_{x,y}v\|_2$~\cite[Lemma 5.5]{vershynin2010introduction}. Finally, given the point cloud $X$, we also conveniently set
\begin{equation}
\epsilon_{\mu,X} := \min\l[\epsilon_\mu,\epsilon_X \r].
\qquad \mbox{(see Assumption \ref{def:moments} and  \eqref{eq:no overlap})}
\label{eq:eps mu X}
\end{equation}
We are now in position to present the main results.

\subsection{Performance of Algorithm~\ref{alg:alg}}
\label{sec:Theoretical guarantees subsec}
With the setup detailed in Section~\ref{sec:problem statement}, we now quantify the performance of Algorithm~\ref{alg:alg}. In Theorems \ref{thm:bias} and \ref{thm:cvg rate} below,  for a fixed point cloud $X$, we focus on how well the output of Algorithm~\ref{alg:alg}, namely $\ddot{\Sigma}_{X,Y_{X,\epsilon}}$, approximates $\dot{\Sigma}_{X}$. Then, in the ensuing remarks, we remove the conditioning on $X$, using Proposition~\ref{prop:Paul} to see how well $\ddot{\Sigma}_{X,Y_{X,\epsilon}}$ approximates $\Sigma_\mu$. We now turn to the details.

Theorem~\ref{thm:bias} below  states that $ {\ddot{\Sigma}}_{X,Y_{X,\epsilon}}$  can be a nearly unbiased estimator of $\dot{\Sigma}_{X}$ given $X$, see  \eqref{eq:emp}. The proof is given in  Section  \ref{sec:proof of thm bias}.
Throughout,  $\E_{z_1|z_2}[\cdot]=\E_{z_1}[\cdot|z_2]$ stands for conditional expectation over $z_1$ and conditioned on $z_2$ for random variables $z_1,z_2$.
\begin{theorem} \label{thm:bias}
\textbf{\emph{(Bias)}}
Consider an open set $\D\subseteq\R^n$ equipped with probability measure $\mu$ satisfying Assumption~\ref{def:moments}, and consider a twice differentiable function $f:\D\rightarrow\R$ satisfying (\ref{eq:Lf},\ref{eq:Hf}).  Assume that the columns of (fixed) $X\in\R^{n\times N}$ belong to  $\D$, namely  $X\subset \D$ in our notation. Fix also $\epsilon \in (0,\epsilon_{\mu,X}]$, see (\ref{eq:eps mu X}).
For an integer $N$ and integers $N_{X,\epsilon} \ge N$ and $N_{X,\min,\epsilon}\le N_{X,\epsilon}$,  assume  also that
\begin{equation}
N_{X,\epsilon} \ge \max \l(\frac{N_{X,\min,\epsilon}N}{\rho_{\mu,X,\epsilon}}, n^{\frac{1}{20}} \r)
\qquad \text{and} \qquad
N_{X,\min,\epsilon} \gtrsim  \log^2 N,
\label{eq:cnd bias}
\end{equation}
where
\begin{equation}
\label{eq:def of rhoX thm}
\rho_{\mu,X,\epsilon} := N \cdot \min_{x\in X} \frac{\mu\l(\B_{x,\epsilon} \r)}{\mu\l(\B_{X,\epsilon} \r)}.
\end{equation}
Then the output of Algorithm~\ref{alg:alg}, namely the estimator $\ddot{\Sigma}_{X,Y_{X,\epsilon}}$ defined in  (\ref{eq:sigmahathat 0}), satisfies
\begin{align}
 & \l\| \E_{Y_{X,\epsilon}|X}\l[ {\ddot{\Sigma}}_{X,Y_{X,\epsilon}} \r]- \dot{\Sigma}_{X} \r\|_F\nonumber\\
&  \lesssim
 \bias_{\mu,\epsilon}
+
n^{2}\L_f^2 N^{-10}
+
\epsilon^2 \H_f^2 n^2
+
\epsilon \L_f \H_f n^{3/2} \max(\K_\mu^{-1/2},1) \log^{\frac{1}{2}} N_{X,\epsilon},
\label{eq:bias thm}
\end{align}
where $\bias_{\mu,\epsilon}$ is  given explicitly in (\ref{eq:companions}).
\end{theorem}

{In this theorem statement and throughout the paper, we use the notation $\log^a b$ as shorthand for $(\log b)^a$.} A few remarks are in order.
\begin{rem}\label{rem:discussion thm bias}\textbf{\emph{(Discussion)}}
\emph{ Theorem~\ref{thm:bias} describes how well $\ddot{\Sigma}_{X,Y_{X,\epsilon}}$ approximates $\dot{\Sigma}_{X}$, in expectation. To form a better understanding of this result, let us first study the conditions listed in (\ref{eq:cnd bias}).
\begin{itemize}
\item The quantity $\rho_{\mu,X,\epsilon}$, defined in~\eqref{eq:def of rhoX thm}, reflects the non-uniformity of $\mu$ over the set $\D$. In particular, if $\D\subset\mathbb{R}^n$ is bounded and $\mu$ is the uniform probability measure on $\D$, then $\rho_{\mu,X,\epsilon}$ {achieves its maximum possible value of $1$}. Non-uniform measures could yield $\rho_{\mu,X,\epsilon} < 1$.
\item The requirements on $N_{X,\epsilon}$ and $N_{X,\min,\epsilon}$ in (\ref{eq:cnd bias}) ensure that every $x\in X$ has sufficiently many neighbors in $Y_{X,\epsilon}$, that is $N_{x,\epsilon}$ is  large enough for all $x$.  For example, if $\mu$ is the uniform probability measure on $\D$ and $\rho_{\mu,X,\epsilon}=1$, we might take $N_{X,\min,\epsilon}\approx \log^2 N$ so that \eqref{eq:bias thm} holds with a total of $N_{X,\epsilon}=O(N \log^2 N)$ samples.
\item The requirement that $N_{X,\epsilon} \ge n^{\frac{1}{20}}$ in (\ref{eq:cnd bias}) is very mild and will be automatically satisfied in cases of interest, as we discuss below.
\end{itemize}
}
\emph{
Let us next interpret the bound on the bias in (\ref{eq:bias thm}).
\begin{itemize}
\item The first term on the right-hand side of (\ref{eq:bias thm}), namely $\bias_{\mu,\epsilon}$, is given explicitly in (\ref{eq:companions}); it depends on both the probability measure $\mu$ and the function $f(\cdot)$, and it can also be viewed as a measure of the non-uniformity of $\mu$. In fact, as explained in the proof of Theorem~\ref{thm:bias}, in the special case where $\mu$ is the uniform probability measure on a bounded and open set $\D$ and every $x\in X$ has the same number of neighbors $N_{x,\epsilon}=N_{X,\epsilon}/N$ within $Y_{X,\epsilon}$, then conditioned on this event, (\ref{eq:bias thm}) can in fact be sharpened by replacing the definition of $\operatorname{B}_{\mu,\epsilon}$ in (\ref{eq:companions}) simply with $\operatorname{B}_{\mu,\epsilon}=0$. In general, the more isotropic $\mu$ is in the sense described in Assumption \ref{def:moments}, the smaller $\operatorname{B}_{\mu,\epsilon}$ will be.
\item The second term on the right-hand side of (\ref{eq:bias thm}) is negligible, as we will generally have $N$ growing at least with $n^2$, as explained below.
\item The third and fourth terms on the right-hand side of (\ref{eq:bias thm}) can be made arbitrarily small by choosing the neighborhood radius $\epsilon$ appropriately small (as a function of $L_f$, $H_f$, $n$, $\K_\mu$, and $N_{X,\epsilon}$). In computational applications, however, choosing $\epsilon$ too small could raise concerns about numerical precision.
\item To get a sense of when the bias in (\ref{eq:bias thm}) is small relative to the size of $\Sigma_\mu$, it may be appropriate to normalize (\ref{eq:bias thm}). A reasonable choice would be to divide both sides of (\ref{eq:bias thm}) by $\L_f^2$, where $\L_f$ bounds $\left\| \nabla f(x) \right\|_2$ on $\D$; see~\eqref{eq:Lf}. In particular, such a normalization accounts for the possible scaling behavior of $\|\Sigma_\mu\|_F$ if one were to consider a sequence of problems with $n$ increasing. For example, in the case where $n$ increases but the new variables in the domain of $f(\cdot)$ do not affect its value, then $\L_f^2$ and $\|\Sigma_\mu\|_F$ are both constant. On the other hand, in the case where $n$ increases and $f(\cdot)$ depends uniformly on the new variables, then $\L_f^2$ and $\|\Sigma_\mu\|_F$ both increase with $n$. In any case, one can show that $\|\Sigma_\mu\|_F \le L_f^2$. With this choice of normalization, the second, third, and fourth terms on the right-hand side of (\ref{eq:bias thm}) can still be made arbitrarily small as described above. In the special case where $\mu$ is uniform on $\D$ and every $x\in X$ has the same number of neighbors $N_{x,\epsilon}=N_{X,\epsilon}/N$, the first term on the right-hand side of (\ref{eq:bias thm}) remains zero, as also described above. More generally, however, $\bias_{\mu,\epsilon}/\L_f^2$ will contain a term that scales like $\sqrt{n}/ N_{X,\min,\epsilon}$, and to control this term it is necessary to choose $N_{X,\min,\epsilon} \gtrsim  n^{1/2} \log^2 N$ so that (\ref{eq:cnd bias}) is also satisfied.  Notably, though, this method can be implemented when fewer than $n$ neighbors are available for each $x \in X$, whereas estimating the local gradients via a conventional finite difference approximation would require $n$ neighbors per point using deterministic queries. For Algorithm~\ref{alg:alg}, we revisit the impact of $n$ on the choices of $N$ and $N_{X,\epsilon}$ after presenting Theorem~\ref{thm:cvg rate} below.
\end{itemize}
}
\end{rem}

\begin{rem} \textbf{\emph{(Sampling strategy)}}
\emph{In Algorithm~\ref{alg:alg}, $N_{X,\epsilon}$ points are independently drawn from the conditional probability measure on the $\epsilon$-neighborhood of the point cloud $X$ and then stored as the columns of $Y_{X,\epsilon}$, namely
\begin{equation}
Y_{X,\epsilon} \overset{\operatorname{i.i.d.}}{\sim} \mu_{X,\epsilon}.
\qquad \mbox{(see (\ref{eq:mu_X}))}
\label{eq:why not uniform pre}
\end{equation}
This sampling strategy  appears to best fit our fixed budget of  $N_{X,\epsilon}$ samples, as it ``prioritizes'' the areas of $\D$ with larger ``mass.'' For example, suppose that $\mu(dx) \gg \mu(dx')$ and $\nabla f(x)\approx \nabla f(x')$ for a pair $x,x'\in\D$. Then,  $\nabla f(x) \nabla f(x)^* \mu(dx) \gg \nabla f(x') \nabla f(x') ^* \mu(dx')$,  suggesting that  a larger weight should be placed on $x$ rather than $x'$ when estimating $\Sigma_\mu$ (see  (\ref{eq:main})).   In the same scenario, assume naturally that   $\mu(\B_{x,\epsilon})\gg \mu(\B_{x',\epsilon})$, so that it is more likely to sample from the neighborhood of $x$ than $x'$. Then, given a fixed budget  of  $N_{X,\epsilon}$ samples, it is highly likely that $N_{x,\epsilon} \gg N_{x',\epsilon}$. That is, $x$ likely has far more $\epsilon$-neighbors in $Y_{X,\epsilon}$ compared to $x'$. Loosely speaking then, the contribution of $x$ to $\ddot{\Sigma}_{X,Y_{X,\epsilon}}$ is calculated more accurately than that of $x'$. In other words, the sampling strategy used in Algorithm~\ref{alg:alg} indeed assigns more weight to areas of $\D$ with larger mass. }

\emph{In some applications, however, sampling points according to the distribution $\mu_{X,\epsilon}$ may be a challenge. A rejection sampling strategy---where points are drawn i.i.d.\ from $\mu$ on $\D$ and those falling outside $\cup_{x \in X} \B_{x,\epsilon}$ are discarded---is one possibility but is not feasible in high dimensions. As an alternative, one can consider a two-phase approach where first a ball $\B_{x,\epsilon}$ with $x \in X$ is selected with probability proportional to $\mu(\B_{x,\epsilon})$, and second a point is selected from the {\em uniform} measure within this ball. Such {\em locally uniform} sampling is an approximation to sampling from the distribution $\mu_{X,\epsilon}$. We expect that similar performance bounds hold for this locally uniform sampling strategy---especially when $\epsilon$ is small---but we do not quantify this here.}
\end{rem}

\begin{rem}\textbf{\emph{(Proof strategy)}}
\emph{
At a high level, the analysis handles the possible  non-uniformity of the measure $\mu$ and  higher order terms in $f(\cdot)$ by introducing  quantities that are simpler to work with but are similar to $\ddot{\Sigma}_{X,Y_{X,\epsilon}}$. Moreover, if $N_{X,\epsilon}$ is sufficiently large,  then  each $x\in X$ has many neighbors in $Y_{X,\epsilon}$ and this observation aids the analysis. The rest of the calculations, in effect, remove the estimation bias of $\dot{\Sigma}_{X,Y_{X,\epsilon}}$ in (\ref{eq:Sigma single dot intro}) to arrive at $\ddot{\Sigma}_{X,Y_{X,\epsilon}}$.
}
\end{rem}

Our second result, proved in Section~\ref{sec:proof of theorem cvg rate}, is a finite-sample bound for $\ddot{\Sigma}_{X,Y_{X,\epsilon}}$.

\begin{theorem}\label{thm:cvg rate} \textbf{\emph{(Finite-sample bound)}} Under the same setup as in Theorem~\ref{thm:bias} including the conditions in~(\ref{eq:cnd bias}), and under the mild assumptions that $\log(n) \ge 1$, $N \ge \log(n)$, and $N_{X,\epsilon} \ge n$, it holds that
\begin{align}
& \l\|  \ddot{\Sigma}_{X,Y_{X,\epsilon}}- \dot{\Sigma}_{X} \r\|_F \nonumber\\
& \lesssim \epsilon^2 \H_f^2 n^2 + \epsilon \L_f \H_f n^{2} + \bias_{\mu,\epsilon} \nonumber \\
& \quad + \log^4(N_{X,\epsilon}) \cdot  \frac{n \sqrt{\log n} }{\sqrt{\rho_{\mu,X,\epsilon} N_{X,\epsilon}}} \cdot \max[\K_\mu^{-1},\K_\mu^{-2}]  \L_f^2 + n^2 L_f^2 N_{X,\epsilon}^{-3}
 \label{eq:finite rate bound thm}
\end{align}
except with a probability of $O\left(n^{-3} + N^{-3}\right)$. Here, $O(\cdot)$ is the standard Big-$O$ notation, the probability is with respect to the selection of $Y_{X,\epsilon}$ conditioned on the fixed set $X$, and $\bias_{\mu,\epsilon}$ is  given explicitly in (\ref{eq:companions}).
\end{theorem}

\begin{rem}\textbf{\emph{(Discussion)}}\label{rem:discussion} \emph{Theorem~\ref{thm:cvg rate} states that $\ddot{\Sigma}_{X,Y_{X,\epsilon}}$ can reliably estimate $\dot{\Sigma}_{X}$ with high probability. We offer several remarks to help interpret this result.
\begin{itemize}
\item The conditions in~\eqref{eq:cnd bias} were discussed in Remark~\ref{rem:discussion thm bias}. The requirement that $N_{X,\epsilon} \ge n^{\frac{1}{20}}$ in (\ref{eq:cnd bias}) has been strengthened to $N_{X,\epsilon} \ge n$ in the statement of Theorem~\ref{thm:cvg rate}. However, this will again be automatically satisfied in cases of interest, as we discuss below.
\item Let us now dissect the estimation error, namely the right-hand side of (\ref{eq:finite rate bound thm}). As discussed in Remark~\ref{rem:discussion thm bias}, $\bias_{\mu,\epsilon}$ in effect captures the non-uniformity of measure $\mu$. In particular, the right-hand side of (\ref{eq:finite rate bound thm}) can be sharpened by setting $\bias_{\mu,\epsilon}=0$ in the setting described in that remark.
\item Similar to  Remark~\ref{rem:discussion thm bias}, the terms involving $\epsilon$ on the right hand side of~\eqref{eq:finite rate bound thm} can be made negligible by choosing $\epsilon$ to be suitably small. We omit these terms in the discussion below.
\item The fourth term on the right hand side of~\eqref{eq:finite rate bound thm} can be controlled by making $N_{X,\epsilon}$ suitably large. We discuss this point further below.
\item The final term on the right hand side of~\eqref{eq:finite rate bound thm} is negligible compared to the fourth, and we omit this in our discussion below.
\end{itemize}}
\end{rem}

\begin{rem}\textbf{\emph{(Estimating $\Sigma_\mu$)}} \label{rem:maininterpretation}
\emph{
{   Let
$$\rho_{\mu,\epsilon} := \frac{\min_{x\in \mathbb{D} } \mu(\mathbb{B}_{x,\epsilon})}{ \max_{x'\in \mathbb{D}} \mu(\mathbb{B}_{x',\epsilon})},
$$
and note that $\rho_{\mu,X,\epsilon}\ge \rho_{\mu,\epsilon}$ for any $X$; see \eqref{eq:def of rhoX thm}. Now}
combining Theorem~\ref{thm:cvg rate} with Proposition~\ref{prop:Paul}{, removing the conditioning on $X$,} and omitting the negligible terms yields
\begin{align}
& \l\|\ddot{\Sigma}_{X,Y_{X,\epsilon}}-\Sigma_\mu\r\|_F\nonumber\\
& \lesssim
\bias_{\mu,\epsilon}+
\log^4(N_{X,\epsilon}) \cdot  \frac{n \sqrt{\log n} }{\sqrt{\rho_{\mu,\epsilon} N_{X,\epsilon}}} \cdot \max[\K_\mu^{-1},\K_\mu^{-2}]  \L_f^2
+ \frac{\operatorname{L}_f^2\log n}{\sqrt{N}},
 \label{eq:overall}
\end{align}
with high probability when both $X$ and $Y_{X,\epsilon}$ are selected randomly, therefore quantifying how well the full algorithm in Algorithm~\ref{alg:alg} estimates the second-moment matrix of $f(\cdot)$. {This conclusion holds for any value of $\epsilon$ small enough that ($i$) $\epsilon_X \ge \epsilon$ (see~\eqref{eq:no overlap}) with high probability over the random draw of $X$ and ($ii$) the terms involving $\epsilon$ on the right hand side of~\eqref{eq:finite rate bound thm} are made negligible.}}
\emph{As suggested in Remark~\ref{rem:discussion thm bias}, we can normalize this bound by dividing both sides by $\L_f^2$:
\begin{equation}
\frac{\l\|\ddot{\Sigma}_{X,Y_{X,\epsilon}}-\Sigma_\mu\r\|_F}{L_f^2} \lesssim
\frac{\bias_{\mu,\epsilon}}{L_f^2} +
\log^4(N_{X,\epsilon}) \cdot  \frac{n \sqrt{\log n} }{\sqrt{\rho_{\mu,\epsilon} N_{X,\epsilon}}} \cdot \max[\K_\mu^{-1},\K_\mu^{-2}]
+ \frac{\log n}{\sqrt{N}}.
\label{eq:overallRenormalized}
\end{equation}
We discuss the terms appearing on the right hand side of~\eqref{eq:overallRenormalized}:
\begin{itemize}
\item As described in Remark~\ref{rem:discussion thm bias}, in some settings $\bias_{\mu,\epsilon}/\L_f^2$ will be zero, while in other settings controlling $\bias_{\mu,\epsilon}/\L_f^2$  will require choosing $N_{X,\min,\epsilon} \gtrsim  \sqrt{n} \log^2 N$.
\item The second and third terms in~\eqref{eq:overallRenormalized} dictate the convergence rate of the error as the number of samples increases. In particular, setting $N_{X,\epsilon}$ proportional to $N\log^2(N)$ gives $N \approx N_{X,\epsilon}/\log^2(N)$ and an overall convergence rate (perhaps to a nonzero bias $\bias_{\mu,\epsilon}/\L_f^2$) of $\log^4(N_{X,\epsilon})/\sqrt{N_{X,\epsilon}}$ as the number $N_{X,\epsilon}$ of secondary samples (which dominates the total) increases. Up to logarithmic terms, this is the same as the convergence rate appearing in Proposition~\ref{prop:Paul} where perfect knowledge of gradients was available.
\item As a function of the ambient dimension $n$, the second term in~\eqref{eq:overallRenormalized} will dominate the third. Controlling the second term in~\eqref{eq:overallRenormalized} will require ensuring that $N_{X,\epsilon}$ (and thus the overall number of samples) scales like $n^2$,  neglecting logarithmic factors.
\end{itemize}
}
\end{rem}

\begin{rem}\emph{\textbf{(Proof strategy)}} \emph{The estimation error here is decomposed into  ``diagonal'' and ``off-diagonal'' terms. The diagonal term, we find, can be written as a sum of independent random matrices and controlled by applying a standard Bernstein inequality. The off-diagonal term, however, is a second-order chaos (a certain sum of products of random variables) and requires additional care. }
\end{rem}

\begin{rem}\emph{\textbf{(Possible improvements)} In combination with  Weyl's inequality~\cite{bernstein2009matrix}, \eqref{eq:overall} might be used to control the distance between the spectrum of $\ddot{\Sigma}_{X,Y_{X,\epsilon}}$ and that of $\Sigma_\mu$. Likewise, given an integer $r\le n$, standard perturbation results \cite{wedin1972perturbation} might be deployed to measure the principal angle between the span of the leading $r$ eigenvectors of $\ddot{\Sigma}_{X,Y_{X,\epsilon}}$ and an $r$-dimensional active subspace of $f(\cdot)$. To obtain the sharpest bounds, both these improvements would require controlling the spectral norm of $\ddot{\Sigma}_{X,Y_{X,\epsilon}}-\Sigma_\mu$ rather than its Frobenius norm (which is bounded in Theorem~\ref{thm:cvg rate} above). { A detailed argument favoring the spectral norm in the context of active subspaces is also given in \cite{lam2018multifidelity}.} Controlling the spectral norm of the error however appears to be considerably more difficult. As an aside, let us point out that the spectrum of $\Sigma_\mu$ in relation to $f(\cdot)$ has been studied in \cite{fornasier2012learning,tyagi2014learning}. {  Another interesting question for future work is the extension of these results to the case where $f(\cdot)$ is vector-valued, rather than scalar-valued; see \cite{zahm2018gradient} and the references therein.}
}
\end{rem}

\section{Related Work}
\label{sec:related work}

As argued in Section~\ref{sec:intro}, the second-moment matrix (or its leading eigenvectors) is of particular relevance in the context of ridge approximation. A ridge function $f(\cdot)$ is one for which $f(x) = h(A^* x)$ for all $x \in \mathbb{D}$, where $A$ is an $n \times r$ matrix with $r < n$ and $h:\mathbb{R}^r\rightarrow\mathbb{R}$. Such a function varies only along the $r$-dimensional subspace spanned by the columns of $A$ and is constant along directions in the $(n-r)$-dimensional orthogonal complement of this subspace.  A large body of work exists in the literature of approximation theory on learning ridge functions from point samples~\cite{pinkus2015ridge,devore2011approximation,cohen2012capturing,stone1985additive,gaiffas2007optimal,juditsky2009nonparametric,golubev1992asymptotic,novak2010tractability,candes1998ridgelets,keiper2015analysis}. Most of these works focus on finding an approximation to the underlying function $h$ and/or the dimensionality-reducing matrix $A$ (or its column span). When $f(\cdot)$ is a ridge function, the $r$-dimensional column span of $A$ coincides with the span of the eigenvectors of $\Sigma_\mu$, which will have rank $r$. This illuminates the connection between ridge approximation and second-moment matrices.

In \cite{fornasier2012learning}, the authors develop an algorithm to learn the column span of $A$ when its basis vectors are (nearly) sparse. The sparsity assumption was later removed in \cite{tyagi2014learning,bogunovic2015active} and replaced with an assumption that this column span is low-dimensional ($r$ is small). For learning such a low-dimensional subspace, these models allow for algorithms with better sample complexities compared to Theorem~\ref{thm:cvg rate} which, in contrast, provides a guarantee on learning the entire second-moment matrix $\Sigma_\mu$ and holds without any assumption (such as low rank) on $\Sigma_\mu$. In this sense, the present work fills a gap in the literature of ridge approximation; see also Section~\ref{sec:contribution}. For completeness, we note that it is natural to ask whether the results in~\cite{tyagi2014learning} could simply be applied in the ``general case'' where the subspace dimension $r$ approaches the ambient dimension $n$ (thus relaxing the critical structural assumption in that work). As detailed in Section 5 of~\cite{tyagi2014learning}, however, the sampling complexity in this general case will scale with $n^5$ (ignoring log factors). In contrast, our bound~\eqref{eq:overall} requires only that the total number of function samples $N + N_{X,\epsilon}$ scale with $n^2$.

A ridge-like function is one for which $f(x) \approx h(A^* x)$. The framework of {\em active subspaces} provides a mechanism for detecting ridge-like structure in functions and reducing the dimensionality of such functions~\cite{constantine2014computing,constantine2015active,constantine2016near}. For example, in scientific computing $f(x)$ may represent the scalar-valued output of some complicated simulation that depends on a high-dimensional input parameter $x$. By finding a suitable $r \times n$ matrix $A$, one can reduce the complexity of parameter studies by varying inputs only in the $r$-dimensional column space of $A$. The term {\em active subspace} refers to the construction of $A$ via the $r$ leading eigenvectors of $\Sigma_\mu$.

In high-dimensional statistics and machine learning, similar structures arise in the task of regression, where given a collection of data pairs $(x_i, z_i)$, the objective is to construct a function $z = f(x)$ that is a model for the relationship between $x$ and $z$. One line of work in this area is \emph{projection pursuit} where, spurred by the interest in \emph{generalized additive models} \cite{hastie1990generalized}, the aim is to construct $f(\cdot)$ using functions of the form $\sum_i h_i(a_i^*x)$~\cite{friedman1981projection,huber1985projection,donoho1989projection}. Further connections with neural networks are studied in~\cite{pinkus1999approximation},\cite[Chapter 11]{friedman2001elements}. See also \cite{vivarelli1999discovering,tripathy2016gaussian} for connections with Gaussian process regression and uncertainty quantification. \emph{Sufficient dimension reduction} and related topics \cite{li1991sliced,yin2011sufficient,cook1994using,xia2002adaptive,fukumizu2004dimensionality,samarov1993exploring,hristache2001structure,glaws2017inverse} are still other lines of related work in statistics. In this context, a collection of data pairs $(x_i, z_i)$ are observed having been drawn independently from some unknown joint density. The assumption is that $z$ is conditionally independent of $x$, given $A^* x$ for some $n \times r$ matrix $A$. The objective is then to estimate the column span of $A$, known as the \emph{effective subspace for regression} in this literature.

Finding the second-moment matrix of a function is also closely related to covariance estimation (see \eqref{eq:def of Sigma intro}), which is  widely studied in modern statistics often under various structural assumptions on the covariance matrix, e.g., sparsity of its inverse \cite{cai2015rop,chen2015exact,dasarathy2015sketching,kolar2012consistent,ravikumar2011high}. In this context, it appears that
\cite{azizyan2015extreme,krishnamurthy2014subspace,anaraki2014memory,pourkamali2015estimation}
are the most relevant to the present work, in part because of their lack of any structural assumptions. For the sake of brevity, we focus on \cite{azizyan2015extreme}, which offers an unbiased estimator for the covariance matrix of a random vector $x$ given few measurements of multiple realizations of $x$ in the form of $\{\Phi_i  x_i\}_i$ for low-dimensional (and uniformly random) orthogonal projection matrices $\{\Phi_i\}_i$. It is important to point out that, by design, the estimator in \cite{azizyan2015extreme} is not applicable to our setup.\footnote{The use of finite differences will effectively replace $\Phi_t x_t$ in $\widehat{\Sigma}_1$ in \cite[Section 3]{azizyan2015extreme} with a sum of rank-$1$ projections of $x_t$.}  Our framework might be interpreted as sum of rank-$1$ projections. To further complicate matters, the probability measure $\mu$ on $\D$ is not necessarily uniform; we cannot hope to explicitly determine the distribution of the crucial components of the estimator. Instead, we rely on the standard  tools in empirical processes to control the bounds. It is also worth including a few other works~\cite{lounici2014high,loh2011high,gonen2014sample} which also involve covariance estimation from partially observed random vectors.

Yet another related field is matrix completion and recovery \cite{recht2011simpler,recht2010guaranteed,eftekhari2016mc} and subspace estimation from data with erasures \cite{eftekhari2016expect}, where typically a low-rank structure is imposed.
 Lastly, in  numerical linear algebra, random projections are increasingly used to facilitate matrix operations \cite{sarlos2006improved,halko2011finding,liberty2007randomized}. As a result, a very similar mathematical toolbox is used in that line of research.

\section{Theory}
\label{sec:theory}

This section contains the proofs of the  two main results of this paper.

\subsection{Proof of Theorem~\ref{thm:bias}}
\label{sec:proof of thm bias}

Let us begin by outlining the proof strategy.
\begin{itemize}
\item First, we introduce a new quantity: $\dddot{\Sigma}_{X,Y_{X,\epsilon}}\in\mathbb{R}^{n \times n}$. Conditioned on a certain ``good'' event $\Cl[event]{Q}$, $\dddot{\Sigma}_{X,Y_{X,\epsilon}}$ is easier to work with than $\ddot{\Sigma}_{X,Y_{X,\epsilon}}$.
\item Then, for fixed $X\subset\D_\epsilon$, we define another ``good'' event $\Cl[event]{good}$ where each $x\in X$ has sufficiently many neighbors in $Y_{X,\epsilon}$. Lemma~\ref{lem:like a good neighbor} below shows that $\Cr{good}$ is very likely to happen if $N_{X,\epsilon}=\#Y_{X,\epsilon}$ is large enough. Conditioned on the event $\Cr{good}$, Lemma~\ref{lem:bias} below shows that ${\dddot{\Sigma}}_{X,Y_{X,\epsilon}}$ is a nearly unbiased estimator of $\dot{\Sigma}_{X}$:
\begin{equation}
\E_{Y_{X,\epsilon}|\Cr{good},X} \l[{\dddot{\Sigma}}_{X,Y_{X,\epsilon}} \r] \approx \dot{\Sigma}_{X}.
\end{equation}
\item Lastly, we remove the conditioning on $\Cr{Q}\cap \Cr{good}$ to complete the proof of Theorem~\ref{thm:bias}.
\end{itemize}

We now turn to the details and introduce $\dddot{\Sigma}_{X,Y_{X,\epsilon}}\in \mathbb{R}^{n \times n}$:
\begin{align}
\label{eq:sigmahathat}
\dddot{\Sigma}_{X,Y_{X,\epsilon}}
:=
\frac{1}{N} \l(1+\frac{1-\frac{2}{n}}{1+\frac{2}{n} } \cdot \Nave^{-1}\r)^{-1}
\nonumber\\
\cdot \l(
\sum_{N_{x,\epsilon}\ge N_{X,\min,\epsilon}}\ddot{\nabla}_{Y_{x,\epsilon}}f(x) \cdot  \ddot{\nabla}_{Y_{x,\epsilon}}f(x)^* -
 \frac{\sum_{N_{x,\epsilon}\ge N_{X,\min,\epsilon}}\l\|\ddot{\nabla }_{Y_{x,\epsilon}} f(x) \r\|_2^2}{\l(1+\frac{2}{n} \r)\Nave+n+1-\frac{2}{n}}  \cdot I_n
\r),
\end{align}
Here,
\begin{equation}\label{eq:grad est}
\ddot{\nabla}_{Y_{x,\epsilon}}f(x):=\frac{n}{N_{x,\epsilon}}\sum_{y\in Y_{x,\epsilon}}
P_{x,y} \cdot \nabla f(x)
\in\R^n,
\end{equation}
and $P_{x,y}\in\mathbb{R}^{n\times n}$ is the orthogonal projection onto the direction of $y-x$. In order to relate $\dddot{\Sigma}_{X,Y_{X,\epsilon}}$ to $\ddot{\Sigma}_{X,Y_{X,\epsilon}}$, we invoke the following result, proved in Appendix \ref{sec:proof of lemma ddd to dd bias}.
\begin{lemma}
\label{lem:ddd to dd bias}
Fix $X$ and $\epsilon\in(0,\epsilon_{\mu,X}]$. It holds that
\begin{equation}
\label{eq:ddd to dd bias unconditional}
\l\|  \ddot{\Sigma}_{X,Y_{X,\epsilon}}- \dddot{\Sigma}_{X,Y_{X,\epsilon}} \r\|_F
\le \frac{1}{2} \epsilon^2 \H_f^2 n^2 + 2\epsilon \L_f \H_f n^{2}.
\end{equation}
Moreover,  consider the event
\begin{equation}
\Cr{Q} := \l\{ \max_{x\in X}\max_{y\in Y_{x,\epsilon}} \l\| P_{x,y} \cdot \nabla f(x) \r\|_2^2 \le \frac{Q_{X,\epsilon}\L_f^2}{n}
 \r\},
 \label{eq:event Q def lemma}
\end{equation}
for $Q_{X,\epsilon}>0$ to be set later. Then, conditioned on the event $\Cr{Q}$, it holds that
\begin{equation}
\l\|  \ddot{\Sigma}_{X,Y_{X,\epsilon}}- \dddot{\Sigma}_{X,Y_{X,\epsilon}} \r\|_F
\le \frac{1}{2} \epsilon^2 \H_f^2 n^2 + 2\epsilon \L_f \H_f Q_{X,\epsilon}^{1/2} n^{3/2}.
\label{eq:ddd to dd conditional}
\end{equation}
\end{lemma}
Thanks to Assumption \ref{def:moments}, the event $\Cr{Q}$ is very likely to happen for the right choice of $Q_{X,\epsilon}$. Indeed, if we set $Q_{X,\epsilon}=\Cl[gam]{events}\log N_{X,\epsilon}$ for $\Cr{events}\ge1$, then
\begin{equation}
\Pr_{Y_{X,\epsilon}|X} \l[  \Cr{Q}^C \r]
\lesssim
N_{X,\epsilon}^{1-\K_\mu\Cr{events}},
\label{eq:pre invoke}
\end{equation}
which follows from \eqref{eq:projconcentration} and an application of the union bound (similar to the slightly more general result in Lemma~\ref{lem:bnd on Q}).

Roughly speaking, in light of Lemma~\ref{lem:ddd to dd bias}, $\ddot{\Sigma}_{X,Y_{X,\epsilon}}\approx \dddot{\Sigma}_{X,Y_{X,\epsilon}}$. It therefore suffices to study the  bias of $\dddot{\Sigma}_{X,Y_{X,\epsilon}}$ in the sequel.
 As suggested earlier, if $\# Y_{X,\epsilon}=N_{X,\epsilon}$ is sufficiently large, then every $x\in X$ will likely have many neighbors in $Y_{X,\epsilon}$, namely   $\#Y_{x,\epsilon}=N_{x,\epsilon}\gg 1$ for every $x\in X$.   This claim is formalized below and proved in Appendix \ref{sec:proof of like a good neighbor}.
\begin{lemma} \label{lem:like a good neighbor}
Fix $X$ and $\epsilon\in(0,\epsilon_{X}]$.
With $\Cr{nb1}\ge 1$, assume that
\begin{equation}
N_{X,\epsilon}  \gtrsim \frac{\Cr{nb1}^2 \log^2 N \cdot \mu\l(\B_{X,\epsilon}\r)}{\min_{x\in X} \mu\l(\B_{x,\epsilon}\r)}.
\end{equation}
Then, except with a probability of at most $N^{1-\Cr{nb1}}$, it holds that
\begin{equation}
\frac{1}{2} \cdot \frac{\mu\l(\B_{x,\epsilon}\r)}{\mu\l(\B_{X,\epsilon} \r)} N_{X,\epsilon} \le N_{x,\epsilon} \le \frac{3}{2} \cdot \frac{\mu\l(\B_{x,\epsilon}\r)}{\mu\l(\B_{X,\epsilon} \r)} N_{X,\epsilon}, \qquad \forall x\in X.
\end{equation}
\end{lemma}
To use Lemma~\ref{lem:like a good neighbor} here, we proceed as follows.
 For $\Cl[gam]{nb1}\ge 1 $, suppose that
\begin{equation}\label{eq:good N_min}
N_{X,\min,\epsilon}\gtrsim   \Cr{nb1}^2 \log^2 N,
\end{equation}
and consider the  event
\begin{equation}\label{eq:good event def}
\Cr{good} :=  \bigcap_{x\in X}\l\{ N_{x,\epsilon}
\ge
\frac{1}{2}\cdot \frac{\mu\l(\B_{x,\epsilon} \r)}{\mu\l( \B_{X,\epsilon}\r)} N_{X,\epsilon}
\ge N_{X,\min,\epsilon}  \r\},
\end{equation}
where, in particular, each $x\in X$ has at least $N_{X,\min,\epsilon}$ neighbors in $Y_{X,\epsilon}$. In light of Lemma~\ref{lem:like a good neighbor}, $\Cr{good}$ is very likely to happen. To be specific,
\begin{equation}
\Pr_{Y_{X,\epsilon}|X}\l[ \Cr{good}^C \r] \le  N^{1-\Cr{nb1}},
\label{eq:good event is likely}
\end{equation}  	
provided that
\begin{equation}
N_{X,\epsilon} \gtrsim \frac{ N_{X,\min,\epsilon} \cdot \mu\l( \B_{X,\epsilon}\r)}{{\min_{x\in X} \mu\l(\B_{x,\epsilon}\r)}}
= \frac{N_{X,\min,\epsilon}N}{\rho_{\mu,X,\epsilon}}
,
\qquad \mbox{(see \eqref{eq:good N_min})}
\label{eq:min on NX}
\end{equation}
where we conveniently defined
\begin{equation}
\rho_{\mu,X,\epsilon}=N\cdot \min_{x\in X}\frac{\mu(\B_{x,\epsilon})}{\mu(\B_{X,\epsilon})}.
\label{eq:def of rhoXeps 1}
\end{equation}
Conditioned on the event $\Cr{good}$, ${\dddot{\Sigma}}_{X,Y_{X,\epsilon}}$ in  \eqref{eq:sigmahathat} takes the following simplified form:
\begin{align}
\label{eq:sigmahathat extended}
{\dddot{\Sigma}}_{X,Y_{X,\epsilon}}
=
\frac{1}{N} \l(1+\frac{1-\frac{2}{n}}{1+\frac{2}{n} } \cdot \Nave^{-1}\r)^{-1} \nonumber\\
\cdot \l(
\sum_{x\in X}\ddot{\nabla}_{Y_{X,\epsilon}}f(x) \ddot{\nabla}_{Y_{X,\epsilon}}f(x)^* -
 \frac{\sum_{x\in X}\l\|\ddot{\nabla }_{Y_{x,\epsilon}} f(x) \r\|_2^2}{\l(1+\frac{2}{n} \r)\Nave+n+1-\frac{2}{n}}  \cdot I_n
\r).
\end{align}
Using the above simplified form, we will prove the following result in Appendix \ref{sec:proof of lemma bias}. Roughly speaking it  states that, conditioned on the event $\Cr{good}$, ${\dddot{\Sigma}}_{X,Y_{X,\epsilon}}$ is a nearly-unbiased estimator of  $\dot{\Sigma}_{X}$.
\begin{lemma}\label{lem:bias}
Fix $X$ and $\epsilon\in(0,\epsilon_{\mu,X}]$. Then, it holds that
\begin{align}
\l\|
\mathbb{E}_{Y_{X,\epsilon}|\Cr{good},X} \l[\dddot{{\Sigma}}_{X,Y_{X,\epsilon}}\r]-
\dot{\Sigma}_{X}
\r\|_F
\le \bias_{\mu,\epsilon},
\end{align}
where
\begin{equation}
\bias_{\mu,\epsilon} := \frac{2\bias''_{\mu,\epsilon}}{N_{X,\min,\epsilon}} +4 \bias'_{\mu,\epsilon}\l(\bias'_{\mu,\epsilon}+1 \r)\L_f^2 +
 \frac{2 \L_f^2  \left(1+\sqrt{n}\right)}{N_{X,\min,\epsilon}},
 \label{eq:companions}
\end{equation}
\begin{equation*}
\bias'_{\mu,\epsilon} :=  n  \cdot \sup_{x\in \D_\epsilon} \left\|
\mathbb{E}_{y|x} \left[P_{x,y}\right]- \frac{I_n}{n}
 \right\|_2, \qquad \l(y|x \sim \mu_{x,\epsilon}\r)
\end{equation*}
\begin{align*}
&  \bias''_{\mu,\epsilon} := n^{2} \cdot
\sup_{x\in \D_\epsilon}
\left\|
\mathbb{E}_{y|x} \left[
P_{x,y} \nabla f(x) \nabla f(x)^* P_{x,y}
\right] \r. \nonumber\\
& \qquad \qquad \qquad  \l. -
\left(
\frac{2\nabla f(x)\nabla f(x)^*}{n(n+2)}+
\frac{\left\|\nabla f(x) \right\|_2^2 }{n(n+2)}\cdot I_n
\right)
\right\|_F,\qquad  \l(y|x \sim \mu_{x,\epsilon} \r).
\end{align*}
Moreover, suppose that  $\mu$ is the uniform probability measure on $\D$, and that $N_{x,\epsilon}=N_{x',\epsilon}$ for every pair $x,x'\in X$. Then, conditioned on $\Cr{good}$, one can replace $\bias_{\mu,\epsilon}$ with $0$, and thus ${\dddot{\Sigma}}_{X,Y_{X,\epsilon}}$ is an unbiased estimator of $\dot{\Sigma}_{X}$.
\end{lemma}
Next, we remove the conditioning on the event $\Cr{good}$, with the aid of the following bounds:
\begin{equation*}
\l\| \dot{\Sigma}_{X} \r\|_F \le \L_f^2, \qquad \mbox{(see \eqref{eq:emp} and \eqref{eq:Lf})}
\end{equation*}
\begin{equation*}
\l\| \ddot{\nabla}_{Y_{x,\epsilon}} f(x) \r\|_2
\le n \L_f, \qquad \forall x\in X,
\qquad \mbox{(see \eqref{eq:grad est} and \eqref{eq:Lf})}
\end{equation*}
\begin{equation}
\left\| {\dddot{\Sigma}}_{X,Y_{X,\epsilon}} \r\|_F \lesssim n^2 \L_f^2.
\qquad \mbox{(see \eqref{eq:sigmahathat} and \eqref{eq:Lf})}
\label{eq:useful in removing good eve}
\end{equation}
Then, we write that
\begin{align}
& \l\| \E_{Y_{X,\epsilon}|X}\l[ {\dddot{\Sigma}}_{X,Y_{X,\epsilon}} \r]- \dot{\Sigma}_{X} \r\|_F \nonumber\\
&  =  \l\| \E_{Y_{X,\epsilon}|\Cr{good},X}\l[ {\dddot{\Sigma}}_{X,Y_{X,\epsilon}} \r] \cdot \Pr_{Y_{X,\epsilon}|X}\l[\Cr{good} \r]+ \E_{Y_{X,\epsilon}|\Cr{good}^C,X}\l[{\dddot{\Sigma}}_{X,Y_{X,\epsilon}} \r] \cdot \Pr_{Y_{X,\epsilon}|X}\l[\Cr{good}^C \r]
- \dot{\Sigma}_{X} \r\|_F\nonumber\\
&  \le \l\| \E_{Y_{X,\epsilon}|\Cr{good},X}\l[ {\dddot{\Sigma}}_{X,Y_{X,\epsilon}} \r]
- \dot{\Sigma}_{X} \r\|_F \cdot  \Pr_{Y_{X,\epsilon}|X}\l[\Cr{good}\r] \nonumber\\
& \qquad + \l\| \E_{Y_{X,\epsilon}|\Cr{good}^C,X} \l[{\dddot{\Sigma}}_{X,Y_{X,\epsilon}}\r] -  \dot{\Sigma}_{X} \r\|_F  \cdot \Pr_{Y_{X,\epsilon}|X}\l[\Cr{good}^C \r]
\nonumber\\
& \le \l\| \E_{Y_{X,\epsilon}|\Cr{good},X}\l[ \dddot{{\Sigma}}_{X,Y_{X,\epsilon}} \r]   - \dot{\Sigma}_{X}\r\|_F  \cdot  \Pr_{Y_{X,\epsilon}|X}\l[\Cr{good} \r] \nonumber\\
& \qquad + \l( \sup \l\| {\dddot{\Sigma}}_{X,Y_{X,\epsilon}} \r\|_F + \sup \l\| \dot{\Sigma}_{X} \r\|_F  \r) \cdot \Pr_{Y_{X,\epsilon}|X}\l[\Cr{good}^C \r]  \nonumber\\
& \lesssim \l\| \E_{Y_{X,\epsilon}|\Cr{good},X}\l[ {\dddot{\Sigma}}_{X,Y_{X,\epsilon}} \r]   - \dot{\Sigma}_{X}\r\|_F
+
n^{2}\L_f^2  \cdot N^{1-\Cr{nb1}} \qquad \mbox{(see
\eqref{eq:useful in removing good eve} and
\eqref{eq:good event is likely})}\nonumber\\
& \le \bias_{\mu,\epsilon} +
n^{2}\L_f^2 \cdot N^{1-\Cr{nb1}}, \qquad
\mbox{(see Lemma~\ref{lem:bias} and \eqref{eq:emp})}
\label{eq:bias pre final}
\end{align}
which, to reiterate, holds with $N_{X,\min,\epsilon} \gtrsim \Cr{nb1}^2 \log^2 N$ and under \eqref{eq:min on NX}.
Lastly, we reintroduce $\ddot{\Sigma}_{X,Y_{X,\epsilon}}$  by invoking Lemma~\ref{lem:ddd to dd bias}  as follows:
\begin{align}
&
\l\| \E_{Y_{X,\epsilon}|X}\l[ {\ddot{\Sigma}}_{X,Y_{X,\epsilon}} - \dddot{\Sigma}_{X,Y_{X,\epsilon}}\r] \r\|_F
\nonumber\\
& \le
\l\| \E_{Y_{X,\epsilon}|\Cr{Q},X}\l[ {\ddot{\Sigma}}_{X,Y_{X,\epsilon}} - \dddot{\Sigma}_{X,Y_{X,\epsilon}}\r] \r\|_F
\cdot \Pr_{Y_{X,\epsilon}|X}\l[ \Cr{Q}\r] \nonumber\\
& \qquad + \l( \sup \l\|  {\ddot{\Sigma}}_{X,Y_{X,\epsilon}}\r\|_F + \sup \l\| \dddot{\Sigma}_{X,Y_{X,\epsilon}}  \r\|_F  \r)
\cdot \Pr_{Y_{X,\epsilon}|X}\l[ \Cr{Q}^C\r]
\qquad \mbox{(similar to \eqref{eq:bias pre final})}
\nonumber\\
&
\lesssim
\epsilon^2 \H_f^2 n^2 + \epsilon \L_f \H_f Q_{X,\epsilon}^{1/2} n^{3/2}
\nonumber\\
& \qquad +
\l(
\sup \l\|\ddot{\Sigma}_{X,Y_{X,\epsilon}}\r\|_F
+
\sup \l\|\dddot{\Sigma}_{X,Y_{X,\epsilon}}\r\|_F
\r)
\cdot
\Pr_{Y_{X,\epsilon}|X}\l[ \Cr{Q}^C\r]
\qquad \l(\mbox{see  \eqref{eq:ddd to dd conditional}} \r)
\nonumber\\
&
\lesssim
\epsilon^2 \H_f^2 n^2 + \epsilon \L_f \H_f \l(\Cr{events} \log N_{X,\epsilon} \r)^{\frac{1}{2}} n^{3/2}
\nonumber\\
& \qquad
+
\l(
\sup \l\|\ddot{\Sigma}_{X,Y_{X,\epsilon}}\r\|_F
+\sup \l\|\dddot{\Sigma}_{X,Y_{X,\epsilon}}\r\|_F
\r)
\cdot
N_{X,\epsilon}^{1-\K_\mu \Cr{events}}
\qquad
\l(Q_{X,\epsilon}\mbox{ in  \eqref{eq:pre invoke} } \r)
\nonumber\\
& \le
\epsilon^2 \H_f^2 n^2 + \epsilon \L_f \H_f \l(\Cr{events} \log N_{X,\epsilon} \r)^{\frac{1}{2}} n^{3/2}
\nonumber\\
&
\qquad
+ \l(
\sup \l\|\ddot{\Sigma}_{X,Y_{X,\epsilon}}-
\dddot{\Sigma}_{X,Y_{X,\epsilon}}
\r\|_F
+
2\sup \l\|\dddot{\Sigma}_{X,Y_{X,\epsilon}}\r\|_F
\r)
\cdot
N_{X,\epsilon}^{1-\K_\mu\Cr{events}}
\,\, \mbox{(triangle ineq.)}
\nonumber\\
& \lesssim
\epsilon^2 \H_f^2 n^2 + \epsilon \L_f \H_f \l(\Cr{events} \log N_{X,\epsilon} \r)^{\frac{1}{2}} n^{3/2} \nonumber\\
& \qquad
+
\l(
\epsilon^2 \H_f^2 n^2 + \epsilon \L_f \H_f n^{2}
+
\L_f^2 n^2
\r)
\cdot
N_{X,\epsilon}^{1-\K_\mu\Cr{events}}.
\qquad \mbox{(see \eqref{eq:ddd to dd bias unconditional} and \eqref{eq:useful in removing good eve})}
\label{eq:bias pre final 2}
\end{align}
Combining the above bound with \eqref{eq:bias pre final} yields that
\begin{align}
&\l\| \E_{Y_{X,\epsilon}|X}\l[ {\ddot{\Sigma}}_{X,Y_{X,\epsilon}} \r]- \dot{\Sigma}_{X} \r\|_F
\nonumber\\
& \le
\l\| \E_{Y_{X,\epsilon}|X}\l[ {\dddot{\Sigma}}_{X,Y_{X,\epsilon}} \r]- \dot{\Sigma}_{X} \r\|_F
+
\l\| \E_{Y_{X,\epsilon}|X}\l[ {\ddot{\Sigma}}_{X,Y_{X,\epsilon}} - \dddot{\Sigma}_{X,Y_{X,\epsilon}}\r] \r\|_F
\,\, \mbox{(triangle ineq.)} \nonumber\\
& \lesssim
\bias_{\mu,\epsilon}
+
n^{2}\L_f^2 \cdot N^{1-\Cr{nb1}}
+
\epsilon^2 \H_f^2 n^2 + \epsilon \L_f \H_f \l(\Cr{events} \log N_{X,\epsilon} \r)^{\frac{1}{2}} n^{3/2} \nonumber \\
& \qquad
+
\l(\epsilon^2 \H_f^2 n^2 + \epsilon \L_f \H_f n^{2} + \L_f^2 n^2\r) \cdot N_{X,\epsilon}^{1-\K_\mu\Cr{events}}
\qquad \mbox{(see \eqref{eq:bias pre final} and \eqref{eq:bias pre final 2})} \nonumber \\
& =
\bias_{\mu,\epsilon}
+
n^{2}\L_f^2 \l( N^{1-\Cr{nb1}} + N_{X,\epsilon}^{1-\K_\mu\Cr{events}}\r)
+
\epsilon^2 \H_f^2 n^2 \l(1 + N_{X,\epsilon}^{1-\K_\mu\Cr{events}}\r)  \nonumber \\
& \qquad
+
\epsilon \L_f \H_f n^{2} \l( \l( \frac{\Cr{events} \log N_{X,\epsilon}}{n} \r)^{\frac{1}{2}} + N_{X,\epsilon}^{1-\K_\mu\Cr{events}} \r) \nonumber \\
& \lesssim
\bias_{\mu,\epsilon}
+
n^{2}\L_f^2 \l( N^{-10} + N_{X,\epsilon}^{-10}\r)
+
\epsilon^2 \H_f^2 n^2 \l(1 + N_{X,\epsilon}^{-10}\r)  \nonumber \\
& \qquad
+
\epsilon \L_f \H_f n^{2} \l( \l( \frac{\max(\K_\mu^{-1},1) \log N_{X,\epsilon}}{n} \r)^{\frac{1}{2}} + N_{X,\epsilon}^{-10} \r)
\nonumber\\
& \qquad \qquad  \mbox{(setting $\Cr{events} = 11\max(\K_\mu^{-1},1)$ and $\Cr{nb1} = 11$)} \nonumber \\
& \lesssim
\bias_{\mu,\epsilon}
+
n^{2}\L_f^2 N^{-10}
+
\epsilon^2 \H_f^2 n^2 \nonumber\\
& \qquad +
\epsilon \L_f \H_f n^{2} \l( \l( \frac{\max(\K_\mu^{-1},1) \log N_{X,\epsilon}}{n} \r)^{\frac{1}{2}} + N_{X,\epsilon}^{-10} \r)
\qquad \mbox{($N_{X,\epsilon} \ge N \ge 1$)} \nonumber \\
& \lesssim
\bias_{\mu,\epsilon}
+
n^{2}\L_f^2 N^{-10}
+
\epsilon^2 \H_f^2 n^2 \nonumber\\
& \qquad 
+
\epsilon \L_f \H_f n^{3/2} \max(\K_\mu^{-1/2},1) \log^{\frac{1}{2}} N_{X,\epsilon} .
\qquad \mbox{($N_{X,\epsilon} \ge n^{\frac{1}{20}}$)}
\end{align}
This completes the proof of Theorem~\ref{thm:bias}.

\subsection{Proof of Theorem~\ref{thm:cvg rate}}\label{sec:proof of theorem cvg rate}

At a high level, the proof strategy here matches that of Theorem~\ref{thm:bias}. First, we replace $\ddot{\Sigma}_{X,Y_{X,\epsilon}}$ with the simpler quantity $\dddot{\Sigma}_{X,Y_{X,\epsilon}}$ defined in \eqref{eq:sigmahathat}. More specifically, in light of Lemma~\ref{lem:ddd to dd bias}, it suffices to study $\dddot{\Sigma}_{X,Y_{X,\epsilon}}$ in the sequel.

Next, for $N_{X,\min,\epsilon}>0$ to be set later, recall the ``good'' event $\Cr{good}$ in \eqref{eq:good event def} whereby every $x\in X$ has at least $N_{X,\min,\epsilon}$ neighbors in $Y_{X,\epsilon}$. Conditioned on the event $\Cr{good}$, $\dddot{{\Sigma}}_{X,Y_{X,\epsilon}}$ takes the simpler form of \eqref{eq:sigmahathat extended}, using which we prove the following result in Appendix~\ref{sec:proof of main result}.
\begin{lemma}\label{lem:main result}
Fix $X$ and $\epsilon\in (0,\epsilon_{\mu,X}]$. If $\log(n) \ge 1$, $N \ge \log(n)$, and $\log(N_{X,\epsilon}) \ge \log(n)$, then conditioned on $\Cr{good}$ and $X$, it holds that
\begin{align}\label{eq:mainresultlemmaRHS}
\left\Vert {\dddot{\Sigma}}_{X,Y_{X,\epsilon}}-\dot{\Sigma}_{X}\right\Vert_F & \lesssim
\bias_{\mu,\epsilon} + \Cr{g2} \Cr{events}^2 \log^4(N_{X,\epsilon}) \cdot  \frac{n \sqrt{\log n} }{\sqrt{\rho_{\mu,X,\epsilon} N_{X,\epsilon}}} \cdot \max[\K_\mu^{-1},\K_\mu^{-2}]  \L_f^2 \nonumber\\
& \qquad + 4 n^2 L_f^2 N_{X,\epsilon}^{(1-\Cr{events}\log(N_{X,\epsilon}))},
\end{align}
for $\Cr{g2} \ge 1$ and $\Cr{events} \ge 3$, except with a probability $\lesssim  e^{-\Cr{g2}} + n^{2-\log \Cr{g2}} + N_{X,\epsilon}^{(1-\Cr{events}\log(N_{X,\epsilon}))}$.
\end{lemma}

We next remove the conditioning on the event $\Cr{good}$ by letting $R$ denote the right hand side of~\eqref{eq:mainresultlemmaRHS} and by writing that
\begin{align}
& \Pr_{Y_{X,\epsilon}|X}\l[  \l\|\dddot{{\Sigma}}_{X,Y_{X,\epsilon}} -\dot{\Sigma}_{X}\r\|_F \gtrsim R \r] \nonumber\\
& \le \Pr_{Y_{X,\epsilon}|\Cr{good},X}\l[  \l\|{\dddot{\Sigma}}_{X,Y_{X,\epsilon}} -\dot{\Sigma}_{X}\r\|_F \gtrsim R \r]+ \Pr_{Y_{X,\epsilon}|X}\l[ \Cr{good}^C\r]
\, \mbox{(see \eqref{eq:useful ineq})}
\nonumber\\
&
\lesssim
e^{-\Cr{g2}} +  n^{2-\log \Cr{g2}} + N_{X,\epsilon}^{(1-\Cr{events}\log(N_{X,\epsilon}))}
+ N^{1-\Cr{nb1}},
\quad \mbox{(Lemma~\ref{lem:main result} and
\eqref{eq:good event is likely})}
\label{eq:fail pr pre}
\end{align}
under \eqref{eq:good N_min}. Lastly, we reintroduce $\ddot{\Sigma}_{X,Y_{X,\epsilon}}$ by invoking Lemma~\ref{lem:ddd to dd bias}: it holds that
\begin{align}
& \l\|  \ddot{\Sigma}_{X,Y_{X,\epsilon}}- \dot{\Sigma}_{X} \r\|_F \nonumber\\
& \le
\l\|  \ddot{\Sigma}_{X,Y_{X,\epsilon}}- \dddot{\Sigma}_{X,Y_{X,\epsilon}} \r\|_F
+ \l\|  \dddot{\Sigma}_{X,Y_{X,\epsilon}}- \dot{\Sigma}_{X} \r\|_F
\qquad \mbox{(triangle inequality)}\nonumber\\
&
\lesssim  \frac{1}{2} \epsilon^2 \H_f^2 n^2 + 2\epsilon \L_f \H_f n^{2} + R
 \qquad \mbox{(see Lemma~\ref{lem:ddd to dd bias})}
\end{align}
with a failure probability of the order of
\begin{equation}
e^{-\Cr{g2}} +  n^{2-\log \Cr{g2}} + N_{X,\epsilon}^{(1-\Cr{events}\log(N_{X,\epsilon}))} + N^{1-\Cr{nb1}} \quad \mbox{(see \eqref{eq:fail pr pre})},
\end{equation}
assuming \eqref{eq:good N_min} holds and that $\log(n) \ge 1$, $N \ge \log(n)$, and $\log(N_{X,\epsilon}) \ge \log(n)$. Setting $\Cr{events} = 4$, $\Cr{nb1} = 4$, and $\Cr{g2} = 149\log(N)$ and noting that $N_{X,\epsilon} \ge N$ completes the proof of Theorem~\ref{thm:cvg rate}.

\section{Acknowledgements}

The authors thank Rachel Ward for her helpful discussions during the preparation of this work. At the time, author AE was a Peter O'Donnell, Jr. Postdoctoral Fellow at UT Austin, mentored by Rachel Ward. AE would also like to thank Hemant Tyagi for many interesting conversations regarding ridge approximation.


\providecommand{\href}[2]{#2}
\providecommand{\arxiv}[1]{\href{http://arxiv.org/abs/#1}{arXiv:#1}}
\providecommand{\url}[1]{\texttt{#1}}
\providecommand{\urlprefix}{URL }

\appendix

\section{Toolbox}

In this section, we list a few  results that are repeatedly used in the rest of appendices.
Recall the following  inequalities for a random variable $z$ and event $\mathcal{A}$ (with complement $\mathcal{A}^C$):\footnote{To see why the  first inequality holds, note that
\begin{align*}
\E^p_z[z] & = \E^p_z\l[z \cdot 1_{\mathcal{A}}(z)+z \cdot 1_{\mathcal{A}^C}(z) \r] \nonumber\\
& \le  \E^p_z\l[z\cdot 1_{\mathcal{A}}(z)\r] + \sup |z| \cdot \E^p\l[z \cdot 1_{\mathcal{A}^C}(z)\r],
\qquad \mbox{(triangle inequality)}
\end{align*}
where $1_{\mathcal{A}}(\cdot)$ is the indicator function for the event $\mathcal{A}$. It is easily verified that
\begin{equation}
\E^p_z\l[z\cdot 1_{\mathcal{A}}(z) \r]
\le \E^p_{z|\mathcal{A}}\l[ z\r], \qquad
\E^p_z\l[z\cdot 1_{\mathcal{A}^C}(z) \r]
\le \sup |z| \cdot \Pr_z[\mathcal{A}^C ]^{\frac{1}{p}},
\end{equation}
from which \eqref{eq:useful ineq} follows immediately.
}
\begin{equation*}
\E^p_z [z] \le \E^p_{z| \mathcal{A}} \l[ z \r] + \sup |z|\cdot  \l( \Pr_z\l[\mathcal{A}^C \r]\r)^{\frac{1}{p}},
\qquad
\l(\mbox{if } \sup |z|< \infty \r),
\end{equation*}
\begin{equation}\label{eq:useful ineq}
\Pr_z\l[ z > z_0 \r] \le \Pr_{z|\mathcal{A}}\l[ z >z_0 \r] + \Pr_z\l[\mathcal{A}^C \r],
\qquad \forall z_0.
\end{equation}
We also recall the Bernstein inequality \cite{gross2011recovering}.
\begin{proposition}\label{prop:Bernstein recall}
\textbf{\emph{(Bernstein inequality)}}
Let $\{A_i\}_i$ be a finite sequence of zero-mean independent random matrices, and set
\begin{equation}
b:=\max_i \l\|A_i\r\|_F,
\end{equation}
\begin{equation}
\sigma^2 := \sum_i \E \l\| A_i \r\|_F^2.
\end{equation}
Then, for $\gamma\ge 1$ and except with a probability of at most $e^{-\gamma}$, it holds that
\begin{equation}
\l\| \sum_i A_i \r\|_F \lesssim  \gamma \cdot \max[b,\sigma].
\end{equation}
\end{proposition}

\section{Proof of Proposition~\ref{prop:Paul}}
\label{sec:proof of proposition Paul}
Recalling the definition of $\dot{\Sigma}_{X}$ from \eqref{eq:emp}, we write that
\begin{align}
\mathbb{E}_X \l[\dot{\Sigma}_{X} \r]
& = \frac{1}{N} \sum_{x\in X} \mathbb{E}_X \l[ \nabla f(x) \nabla f(x)^*\r]
\qquad \mbox{(see (\ref{eq:emp}))}
\nonumber\\
&  = \mathbb{E}_x \l[ \nabla f(x) \nabla f(x)^*\r]
\qquad \l(\#X = N \r)\nonumber\\
& = \Sigma_{\mu}, \qquad \mbox{(see \eqref{eq:main})}
\end{align}
which proves the first claim. To control the deviation about the mean, we will invoke  the standard Bernstein inequality, recorded in Proposition~\ref{prop:Bernstein recall} for the reader's convenience. Note that
\begin{align}
 \dot{\Sigma}_{X}-\Sigma_{\mu}
 &
 = \dot{\Sigma}_{X}-\mathbb{E}_X\l[ \dot{\Sigma}_{X}\r] \nonumber\\ 	
 &= \frac{1}{N}\sum_{x\in X} \nabla f(x)\nabla f(x)^*  - \mathbb{E}_{x}\l[ \nabla f(x) \nabla f(x)^* \r] \nonumber\\
 & =:\sum_{x\in X} A_x,
 \label{eq:def of Ax}
\end{align}
where $\{A_x\}_x\subset \mathbb{R}^{n\times n}$ are independent and zero-mean random matrices. To apply the Bernstein inequality (Proposition~\ref{prop:Bernstein recall}), we compute the parameters
\begin{align*}
b & = \max_{x\in X} \l\| A_x \r\|_F \nonumber\\
& = \frac{1}{N} \max_{x\in X} \l\| \nabla f(x)\nabla f(x)^* -\mathbb{E}_x\l[ \nabla f(x)\nabla f(x)^* \r]\r\|_F \qquad \mbox{(see \eqref{eq:def of Ax})}\nonumber\\
& \le \frac{1}{N} \max_{x\in X} \l\| \nabla f(x)\nabla f(x)^*\r\|_F + \frac{1}{N}\mathbb{E}_x\l\|  \nabla f(x)\nabla f(x)^* \r\|_F
\,\, \mbox{(triangle, Jensen's ineqs.)}
\nonumber\\
& \le \frac{2}{N} \sup_{x\in \D} \l\| \nabla f(x)\nabla f(x)^*\r\|_F \nonumber\\
& =  \frac{2}{N} \sup_{x\in \D} \l\| \nabla f(x)\r\|_2^2 \nonumber\\
& = \frac{2\L_f^2}{N} \qquad \mbox{(see \eqref{eq:Lf})}
\end{align*}
and
\begin{align*}
\sigma^2 & = \sum_{x\in X} \mathbb{E}_x \l\|A_x \r\|_F^2 \nonumber\\
& = \frac{1}{N}\mathbb{E}_x \l\|\nabla f(x) \nabla f(x)^* - \mathbb{E}_x\l[\nabla f(x)\nabla f(x)^* \r] \r\|_F^2
\qquad \l(\mbox{see \eqref{eq:def of Ax} and } \#X=N  \r) \nonumber\\
& \le \frac{1}{N} \mathbb{E}_x \l\| \nabla f(x)\nabla f(x)^* \r\|_F^2
\qquad \l(\E \| Z-\E[Z]\|_F^2 \le \E \|Z\|_F^2 \mbox{ for a random matrix }Z\r)
\nonumber\\
& = \frac{1}{N} \mathbb{E}_x \l\| \nabla f(x)\r\|_2^4 \nonumber\\
& \le \frac{\L_f^4}{N},\qquad \mbox{(see \eqref{eq:Lf})} \nonumber\\
\end{align*}
and thus
\begin{equation}
\max[b,\sigma] \le \frac{2\L_f^2}{\sqrt{N}}.
\end{equation}
Therefore, for $\Cl[gam]{gPaul}\ge 1$ and except with a probability of at most $e^{-\Cr{gPaul}}$, Proposition~\ref{prop:Bernstein recall} dictates that
\begin{align*}
\l\| \dot{\Sigma}_{X}-{\Sigma}_{\mu}\r\|_F & = \l\| \sum_{x\in X}A_x \r\|_F
\qquad \mbox{(see \eqref{eq:def of Ax})}
\nonumber\\
& \lesssim \Cr{gPaul} \cdot \max[b,\sigma]\nonumber\\
& \lesssim \Cr{gPaul} \cdot \frac{\L_f^2}{\sqrt{N}},
\end{align*}
which  completes the proof of Proposition~\ref{prop:Paul} when we take $\Cr{gPaul}=\log n$.

\section{Uniform Measure Satisfies Assumption \ref{def:moments}}
\label{sec:uni meas satisiefs assumption}

We verify in this appendix that the uniform probability measure on $\D$ satisfies Assumption \ref{def:moments}.  Fix arbitrary $\epsilon>0$ and  $x$ in the $\epsilon$-interior of $\D\subseteq \R^n$, namely $x \in\D_{\epsilon}$, assuming that $\D_\epsilon\ne \emptyset$. The conditional measure in the neighborhood $\B_{x,\epsilon}$ too is uniform, so that $y|x\sim \mbox{uniform}(\B_{x,\epsilon})$. Then, for fixed $v\in\mathbb{R}^n$ with $\|v\|_2=1$, observe that
\begin{equation}
\l\| P_{x,y} v \r\|_2^2
\sim \operatorname{beta}\l(\frac{1}{2},\frac{n-1}{2} \r).
\label{eq:uni dist}
\end{equation}
To study the tail bound of the random variable $\|P_{x,y}v\|_2^2$, we proceed as follows. We note that~\eqref{eq:projconcentration} trivially holds for any $\Cr{beta} > n$ since $\l\| P_{x,y}v \r\|_2^2 \le \l\| v \r\|_2^2$ because $P_{x,y}$ is an orthogonal projection. Thus, it suffices to consider fixed $\Cr{beta} \in (0,n]$. Recalling the moments of the beta distribution, write that
\begin{align}
& \Pr{}_{y|x}\l[ \l\| P_{x,y}v \r\|_2^2 > \frac{\Cr{beta}}{n} \r] \nonumber\\
&
= \Pr{}_{y|x}\l[ \l\| P_{x,y}v \r\|_{2}^{2\lambda} > \l(\frac{\Cr{beta}}{n}\r)^\lambda \r]
\qquad \l( \lambda >0 \r)
 \nonumber\\
 & \le \l(\frac{\Cr{beta}}{n} \r)^{-\lambda}  \mathbb{E}\l[ \l\| P_{x,y}v \r\|_{2}^{2\lambda} \r]
 \qquad \mbox{(Markov's inequality)}
 \nonumber\\
 & = \l(\frac{\Cr{beta}}{n} \r)^{-\lambda} \frac{B\l(\lambda+\frac{1}{2} ,\frac{n-1}{2}\r)}{B\l( \frac{1}{2}, \frac{n-1}{2}\r)},
 \label{eq:beta tail bnd 0}
\end{align}
where
\begin{equation}
B(a,b)=\int_0^1 t^{a-1} (1-t)^{b-1} dt = \frac{\Gamma(a)\Gamma(b)}{\Gamma(a+b)}
\label{eq:beta fcn}
\end{equation}
is the beta function. Above, $\Gamma(a) = \int_0^\infty t^{a-1} e^{-t} dt$ is the usual gamma function. In order to choose $\lambda$ above, we rewrite \eqref{eq:beta tail bnd 0} as
\begin{align}
\Pr{}_{y|x}\l[ \l\| P_{x,y} v \r\|_2^2  > \frac{\Cr{beta}}{n}\r]
& \le
e^{-\lambda \log\l( \frac{\Cr{beta}}{n}\r)+ \log\l( B\l( \lambda+\frac{1}{2},\frac{n-1}{2}\r) \r)-\log\l( B\l(\frac{1}{2},\frac{n-1}{2} \r)\r) } \nonumber\\
& =: e^{l\l(\lambda \r)}.
\end{align}
In order to minimize $l(\cdot)$, we compute its derivative:
\begin{align}
l'(\lambda)
& = - \log\l(\frac{\Cr{beta}}{n} \r)
+ \frac{d}{d\lambda} \log\l(B\l(\lambda+\frac{1}{2},\frac{n-1}{2} \r) \r) - \frac{d}{d\lambda} \log\l( B\l(\frac{1}{2},\frac{n-1}{2}  \r) \r)
\nonumber\\
& =
- \log\l( \frac{\Cr{beta}}{n} \r) + \frac{d}{d\lambda} \log\l( \Gamma\l(\lambda+\frac{1}{2} \r) \r) - \frac{d}{d\lambda} \log\l( \Gamma\l(\lambda+\frac{n}{2} \r) \r)
\qquad \mbox{(see \eqref{eq:beta fcn})}
\nonumber\\
& =
- \log\l( \frac{\Cr{beta}}{n} \r) + \frac{ \Gamma'\l(\lambda+\frac{1}{2} \r)}{\Gamma\l(\lambda+\frac{1}{2} \r)} -
\frac{ \Gamma'\l(\lambda+\frac{n}{2} \r)}{\Gamma\l(\lambda+\frac{n}{2} \r)}
\nonumber\\
& = - \log\l(\frac{\Cr{beta}}{n} \r)
+ \psi\l( \lambda+\frac{1}{2} \r) - \psi \l(\lambda+\frac{n}{2} \r),
\label{eq:l derivative}
\end{align}
where $\psi(a) = \frac{\Gamma'(a)}{\Gamma(a)}$ is the ``digamma'' function. It is well-known that $\psi(a)\approx \log(a)$ for large $a$ (see, for example, \cite{olver2010nist}). To guide our choice of $\lambda$, note that if $n$ is sufficiently large and we take $\lambda$ such that $1\ll \lambda \ll n$, we have that
\begin{align}
l'(\lambda) &
= - \log\l( \frac{\Cr{beta}}{n}\r) + \psi\l( \lambda+ \frac{1}{2} \r)
- \psi \l(\lambda+\frac{n}{2} \r)
\qquad \mbox{(see \eqref{eq:l derivative})}
 \nonumber\\
& \approx
 - \log\l(\frac{\Cr{beta}}{n} \r) + \log \lambda - \log \l(\frac{n}{2}\r) \nonumber\\
 & = - \log\l( \frac{2\lambda}{\Cr{beta}}\r),
\end{align}
thereby suggesting the choice of $\lambda = \Cr{beta}/2$.
With this choice, we find that
\begin{align}
&\Pr{}_{y|x}\l[ \l\| P_{x,y}v \r\|_2^2 > \frac{\Cr{beta}}{n} \r]
\nonumber\\
 &
\le
\l( \frac{\Cr{beta}}{n} \r) ^{-\frac{\Cr{beta}}{2}}
\frac{B\l( \frac{\Cr{beta}+1}{2},\frac{n-1}{2} \r)}{B\l(\frac{1}{2},\frac{n-1}{2} \r)}
\qquad \mbox{(see \eqref{eq:beta tail bnd 0})}
\nonumber\\
& = \l( \frac{\Cr{beta}}{n} \r) ^{-\frac{\Cr{beta}}{2}}
 \frac{\Gamma\l( \frac{\Cr{beta}+1}{2}\r)\Gamma\l( \frac{n}{2}\r)} {\Gamma\l( \frac{1}{2} \r)\Gamma\l(  \frac{n+\Cr{beta}}{2}\r)}
 \qquad \mbox{(see \eqref{eq:beta fcn})}
\nonumber\\
& \lesssim
\l( \frac{\Cr{beta}}{n} \r) ^{-\frac{\Cr{beta}}{2}}
\frac{ \l(\frac{\Cr{beta}+1}{2} \r)^{\frac{\Cr{beta}}{2}} e^{-\frac{\Cr{beta}+1}{2}} \l(\frac{n}{2} \r)^{\frac{n-1}{2}} e^{-\frac{n}{2}} }{\l( \frac{n+\Cr{beta}}{2} \r)^{\frac{n+\Cr{beta}-1}{2}} e^{-\frac{n+\Cr{beta}}{2}}}
\quad \l(
1
<
\frac{a^{\frac{1}{2}-a} e^a}{\sqrt{2\pi}}
\Gamma\l(a\r)
< e^{\frac{1}{12a}}
,\,\, \forall a >0
\r)\nonumber\\
& \lesssim \l( \frac{n}{n+\Cr{beta}}\r)^{\frac{n+\Cr{beta}-1}{2}}\nonumber\\
& \le \l( \frac{n}{n+\Cr{beta}}\r)^{\frac{n-1}{2}}
\qquad \l( \Cr{beta}>0 \r)\nonumber\\
& = \l( 1+ \frac{\Cr{beta}}{n}\r)^{-\frac{n-1}{2}}\nonumber\\
& \le e^{-\frac{\Cr{beta}}{n}\cdot \frac{n-1}{2}}
\qquad \l( 1+a\le a^a \r)\nonumber\\
& \le  e^{-\frac{\Cr{beta}}{2}+\frac{1}{2}}.
\qquad \l( \Cr{beta} \le n \r)
\end{align}
Therefore, Assumption \ref{def:moments} holds for the uniform probability measure with $\epsilon_\mu=\infty$ and $\K_\mu = 1/2$.

\section{Estimating $\nabla f(x)$}
\label{sec:proof of lemma bias pointwise}

For fixed $x\in \D$,  by drawing samples from the neighborhood of $x$ and then applying  the method of finite differences, we may estimate $\nabla f(x)$. This is described below for the sake of completeness.
\begin{proposition}\label{lem:bias pointwise}
Fix $x\in \D$ and take $\epsilon>0$ small enough so that $x$ belongs to $\epsilon$-interior of $\D$, namely $x\in \D_\epsilon$.  Draw $y$ from the conditional measure on the neighborhood  $\B_{x,\epsilon}$, namely   $y|x \sim \mu_{x,\epsilon}$ (see (\ref{eq:cond mu})). For an integer $N_{x,\epsilon}$, let $Y_{x,\epsilon} \subset \B_{x,\epsilon}$ contain $N_{x,\epsilon}$ independent copies of $y$. Then, it holds that
\begin{align}
\left\|
\mathbb{E}_{Y_{x,\epsilon}|N_{x,\epsilon},x}
\left[\dot{\nabla}_{Y_{x,\epsilon}} f(x)
\right]
 -
\nabla f(x)
 \right\|_2
 \le  \bias_{\mu,\epsilon}'  \L_f + \frac{\epsilon \H_f n}{2}.
\end{align}
where
\begin{equation}
\dot{\nabla}_{Y_{x,\epsilon}}f(x):=\frac{n}{N_{x,\epsilon}}\sum_{y\in Y_{x,\epsilon}}
\frac{f(y)-f(x)}{\|y-x\|_2} \cdot \frac{y-x}{\|y-x\|_2}\in\R^n,
\end{equation}
\begin{equation}\label{eq:Bf}
\bias_{\mu,\epsilon}' :=  n \cdot \sup_{x\in \D_\epsilon} \left\|
\mathbb{E}_{y|x} \left[P_{x,y}\right]- \frac{I_n}{n}
 \right\|. \qquad \l( y|x \sim \mu_{x,\epsilon}\r)
\end{equation}
In particular, if $\mu$ is the uniform probability measure on $\D$, then $\bias_{\mu,\epsilon}'=0$.
\end{proposition}

\begin{proof}
First, we replace $\dot{\nabla}_{Y_{x,\epsilon}}f(x)$  with the simpler quantity $\ddot{\nabla}_{Y_{x,\epsilon}} f(x)$, defined as
\begin{equation}
\ddot{\nabla}_{Y_{x,\epsilon}} f(x)
:= \frac{n}{N_{x,\epsilon}} \sum_{y\in Y_{x,\epsilon}} P_{x,y} \cdot \nabla f(x) \in\R^n,
\end{equation}
where $P_{x,y}\in\R^{n\times n}$ is the orthogonal projection onto the direction of $y-x$. By definition, the two quantities are related as follows:
\begin{align}
\label{eq:Hess 1}
& \l\| \dot{\nabla}_{Y_{x,\epsilon}} f(x)- \ddot{\nabla }_{Y_{x,\epsilon}}  f(x)\r\|_2
\nonumber\\
&
= \frac{n}{N_{x,\epsilon}} \l\| \sum_{y\in Y_{x,\epsilon}} \frac{y-x}{\|y-x\|_2^2}\r. \nonumber\\
& \qquad \qquad \l.
\cdot \l( {f(y)-f(x)- (y-x)^*\nabla f(x)} \r) \r\|_2
\qquad \l( P_{x,y}= \frac{(y-x)(y-x)^*}{\|y-x\|_2^2}\r)\nonumber\\
& \le \frac{n}{N_{x,\epsilon}}\sum_{y\in Y_{x,\epsilon}} \l\| \frac{y-x}{\|y-x\|_2^2}\l( f(y)-f(x)-(y-x)^*\nabla f(x)\r)  \r\|_2
\qquad \mbox{(triangle inequality)}
\nonumber\\
& = \frac{n}{N_{x,\epsilon}}\sum_{y\in Y_{x,\epsilon}} \frac{\l|  f(y)-f(x)-(y-x)^*\nabla f(x) \r|}{\|y-x\|_2}\nonumber\\
& \le n \cdot \sup_{y\in \B_{x,\epsilon}}
\frac{\l| f(y)-f(x)-(y-x)^*\nabla f(x) \r|}{\|y-x\|_2}
\qquad \l( \#Y_{x,\epsilon} = N_{x,\epsilon} \r)\nonumber\\
& \le n \cdot \sup_{y\in \B_{x,\epsilon}}
\frac{\H_f \|y-x\|_2 }{2}
\qquad \mbox{(Taylor's expansion and  \eqref{eq:Hf})} \nonumber\\
& \le n \cdot \frac{\H_f  \cdot  \epsilon}{2}.
\qquad \l( y\in \B_{x,\epsilon} \r)
\end{align}
Loosely speaking then,  $\dot{\nabla}_{Y_{x,\epsilon}} f(x) \approx \ddot{\nabla}_{Y_{x,\epsilon}f(x)}$ and it therefore suffices to study the estimation bias of $\ddot{\nabla}_{Y_{x,\epsilon}}f(x)$. To that end, we simply note that
\begin{align}\label{eq:pntwise unbiased}
& \left\|
\E_{Y_{x,\epsilon}|N_{x,\epsilon},x}\left[ \ddot{\nabla}_{Y_{x,\epsilon}}f(x)\right]
- \nabla f(x)
\right\|_2 \nonumber\\
& = \left\| n\cdot \E_{y|x}\left[ P_{x,y} \nabla f(x)\right]
- \nabla f(x)
\right\|_2
 \qquad \left( y|x\sim \mu_{x,\epsilon}\right)\nonumber\\
& =
\left\|
n\cdot \E_{y|x}\left[ P_{x,y} \right]\cdot \nabla f(x)
- \nabla f(x)
\right\|_2
\nonumber\\
& \le n \cdot
\sup_{x\in \D} \left\|
\mathbb{E}_{y|x} \left[P_{x,y} \right]
- \frac{I_n}{n}
 \right\|
\cdot \sup_{x\in \D} \left\|\nabla f(x)\right\|_2
\nonumber \\
& =: \bias_{\mu,\epsilon}' \cdot \L_f,  \qquad
\mbox{(see \eqref{eq:Lf})}
\end{align}
which, in turn, implies that
\begin{align}
& \l\| \E_{Y_{x,\epsilon}|N_{x,\epsilon},x}\l[ \dot{\nabla}_{Y_{x,\epsilon}}f(x)
\r]
- \nabla f(x)  \r\|_2
\nonumber\\
& \le \l\| \E_{Y_{x,\epsilon}|N_{x,\epsilon},x}\l[ \dot{\nabla}_{Y_{x,\epsilon}}f(x)
-
\ddot{\nabla}_{Y_{x,\epsilon}} f(x) \r]\r\|_2 \nonumber\\
& \qquad + \l\|
\E_{Y_{x,\epsilon}|N_{x,\epsilon},x}\l[
\ddot{\nabla}_{Y_{x,\epsilon}} f(x)
\r]
- \nabla f(x)\r\|_2
\qquad \mbox{(triangle inequality)} \nonumber\\
& \le \frac{n \H_f \epsilon }{2} +  \bias_{\mu,\epsilon}' \L_f.
\qquad \mbox{(see \eqref{eq:Hess 1} and \eqref{eq:pntwise unbiased})}
\end{align}
In particular, when $\mu$ is the uniform probability measure on $\mathbb{D}$, $P_{x,y}$ is an isotropic random matrix (for fixed $x\in\D$). Therefore, $\mathbb{E}_{y|x} [P_{x,y}] = C \cdot I_n$ for some scalar $C$. To find $C$, we note that
$$
\mbox{trace}\left[\E_{y|x} \left[ P_{x,y}\right]\right] = \E_{y|x}\left[ \mbox{trace}\left[P_{x,y}\right]\right] = 1 = C \cdot \mbox{trace} [I_n] = C\cdot n\Longrightarrow C = \frac{1}{n},
$$
where we used the fact that $P_{x,y}$ is a rank-$1$ orthogonal projection. Consequently, when $\mu$ is the uniform measure, $\bias_{\mu,\epsilon}'=0$. This completes the proof of Proposition~\ref{lem:bias pointwise}.
\end{proof}

\section{Proof of Lemma~\ref{lem:ddd to dd bias}}
\label{sec:proof of lemma ddd to dd bias}

We only verify  the second claim, as the other proof is similar. Conditioned on the event $\Cr{Q}$, note that
\begin{align}
\label{eq:Q comes to help}
\l\| \ddot{\nabla}_{Y_{x,\epsilon}}f(x) \r\|_2
& \le \frac{n}{N_{x,\epsilon}}\sum_{y\in Y_{x,\epsilon}}
\l\| P_{x,y} \cdot \nabla f(x) \r\|_2
\qquad \mbox{(see \eqref{eq:grad est})}
\nonumber\\
& \le n \cdot  \max_{x\in X}\max_{y\in Y_{x,\epsilon}}
\l\| P_{x,y} \cdot \nabla f(x) \r\|_2
\qquad \l( \#Y_{x,\epsilon} = N_{x,\epsilon} \r)
\nonumber\\
& \le n \cdot \sqrt{\frac{{Q_{X,\epsilon}}\L_f^2}{n}}.
\qquad
\mbox{(see \eqref{eq:event Q def lemma})}
\end{align}
Using the inequality $\l\| aa^*-bb^* \r\|_2 \le \|a-b\|( \|a\|_2+\|b\|_2)$ for any $a,b\in\mathbb{R}^n$ in the third line below, it follows that
\begin{align}
& \frac{1}{N} \l\|
\sum_{N_{x,\epsilon}>N_{X,\min,\epsilon}}
\dot{\nabla}_{Y_{x,\epsilon}} f(x) \cdot \dot{\nabla}_{Y_{x,\epsilon}}f(x)^*
-
\ddot{\nabla}_{Y_{x,\epsilon}} f(x) \cdot \ddot{\nabla}_{Y_{x,\epsilon}}f(x)^*
\r\|_F
\nonumber\\
& \le
\frac{1}{N}
\sum_{N_{x,\epsilon}>N_{X,\min,\epsilon}}
\l\|
\dot{\nabla}_{Y_{x,\epsilon}} f(x) \cdot \dot{\nabla}_{Y_{x,\epsilon}}f(x)^* \r. \nonumber\\
& \l. \qquad \qquad \qquad \qquad \qquad -
\ddot{\nabla}_{Y_{x,\epsilon}} f(x) \cdot \ddot{\nabla}_{Y_{x,\epsilon}}f(x)^*
\r\|_F
\qquad
\mbox{(triangle inequality)}
\nonumber\\
&
\le \frac{1}{N}\sum_{N_{x,\epsilon}>N_{X,\min,\epsilon}} \l\| \dot{\nabla}_{Y_{x,\epsilon}} f(x) - \ddot{\nabla}_{Y_{x,\epsilon}}f(x)  \r\|_2
\l( \l\| \dot{\nabla}_{Y_{x,\epsilon}} f(x)\r\|_2+
\l\| \ddot{\nabla}_{Y_{x,\epsilon}} f(x)\r\|_2
\r)
\nonumber\\
& \le \max_{x\in X}
 \l\| \dot{\nabla}_{Y_{x,\epsilon}} f(x) - \ddot{\nabla}_{Y_{x,\epsilon}} f(x)  \r\|_2 \nonumber\\
& \qquad \cdot \l( \max_{x\in X} \l\| \dot{\nabla}_{Y_{x,\epsilon}} f(x)\r\|_2+
\max_{x\in X}\l\| \ddot{\nabla}_{Y_{x,\epsilon}} f(x)\r\|_2
\r)
\qquad \l( \#X = N \r)
\nonumber\\
& \le \max_{x\in X}
 \l\| \dot{\nabla}_{Y_{x,\epsilon}} f(x) - \ddot{\nabla}_{Y_{x,\epsilon}} f(x)  \r\|_2 \nonumber\\
 & \qquad \cdot 
\l( \max_{x\in X} \l\| \dot{\nabla}_{Y_{x,\epsilon}} f(x)- \ddot{\nabla}_{Y_{x,\epsilon}} f(x)\r\|_2+
2 \max_{x\in X}\l\| \ddot{\nabla}_{Y_{x,\epsilon}} f(x)\r\|_2
\r)
\quad \mbox{(triangle ineq.)}
\nonumber\\
& \le
\frac{ \epsilon
\H_f n}{2} \l( \frac{\epsilon \H_f n}{2} + 2  \sqrt{Q_{X,\epsilon}\L_f^2 n}  \r)
\qquad \mbox{(see \eqref{eq:Hess 1} and \eqref{eq:Q comes to help})}\nonumber\\
& \le \frac{1}{4} \epsilon^2 \H_f^2 n^2 + \epsilon \L_f \H_f Q_{X,\epsilon}^{1/2} n^{3/2},
\label{eq:ddd to dd pre}
\end{align}
which, in turn, immediately implies that
\begin{align}
&  \frac{1}{N}
\l| \sum_{N_{x,\epsilon}>N_{X,\min,\epsilon}}
\l\|
\dot{\nabla}_{Y_{x,\epsilon}} f(x)
\r\|_2^2
-  \l\|
\ddot{\nabla}_{Y_{x,\epsilon}} f(x)
\r\|_2^2
 \r|
 \nonumber\\
& = \frac{1}{N}
\l| \sum_{N_{x,\epsilon}>N_{X,\min,\epsilon}}
\mbox{trace}
\l[
\dot{\nabla}_{Y_{x,\epsilon}} f(x)\cdot
\dot{\nabla}_{Y_{x,\epsilon}} f(x)^*
-  \ddot{\nabla}_{Y_{x,\epsilon}} f(x)\cdot
\ddot{\nabla}_{Y_{x,\epsilon}} f(x)^*
\r]
 \r|
 \nonumber\\
 & \le
 \frac{\sqrt{n}}{N} \l\| \sum_{N_{x,\epsilon}>N_{X,\min,\epsilon}}
\dot{\nabla}_{Y_{x,\epsilon}} f(x) \cdot \dot{\nabla}_{Y_{x,\epsilon}}f(x)^*
-
\ddot{\nabla}_{Y_{x,\epsilon}} f(x) \cdot \ddot{\nabla}_{Y_{x,\epsilon}}f(x)^*
\r\|_F
  \nonumber\\
  & \le \frac{1}{4} \epsilon^2 \H_f^2 n^{5/2} + \epsilon \L_f \H_f Q_{X,\epsilon}^{1/2} n^{2},
 \qquad
\mbox{(see \eqref{eq:ddd to dd pre})}
\label{eq:ddd to dd pre pre}
\end{align}
where the third line above uses the fact that $|\mbox{trace}(A) | \le \sqrt{n}\|A\|_F$ for any $A\in \R^{n\times n}$. Recall the definitions of $\ddot{\Sigma}_{X,Y_{X,\epsilon}}$ and $\dddot{\Sigma}_{X,Y_{X,\epsilon}}$ in \eqref{eq:sigmahathat 0} and \eqref{eq:sigmahathat}, respectively. Then, by combining \eqref{eq:ddd to dd pre} and \eqref{eq:ddd to dd pre pre}, it follows that
\begin{align}
& \l\| \ddot{\Sigma}_{X,Y_{X,\epsilon}} - \dddot{\Sigma}_{X,Y_{X,\epsilon}}  \r\|_F \nonumber\\
& \le \frac{1}{N}
\l\|
\sum_{N_{x,\epsilon}\ge N_{X,\min,\epsilon}}
\dot{\nabla}_{Y_{x,\epsilon}}f(x) \dot{\nabla}_{Y_{x,\epsilon}}f(x)^* -
\ddot{\nabla}_{Y_{x,\epsilon}}f(x) \ddot{\nabla}_{Y_{x,\epsilon}}f(x)^*
\r\|_F \nonumber\\
& \qquad +
\frac{1}{N}
\l|
 \sum_{N_{x,\epsilon}>N_{X,\min,\epsilon}}
\l\|
\dot{\nabla}_{Y_{x,\epsilon}} f(x)
\r\|_2^2
-  \l\|
\ddot{\nabla}_{Y_{x,\epsilon}} f(x)
\r\|_2^2
 \r| \cdot \frac{\|I_n\|_F}{n}
\qquad \mbox{(see (\ref{eq:sigmahathat 0},\ref{eq:sigmahathat}))}
\nonumber\\
& \le \frac{1}{2} \epsilon^2 \H_f^2 n^2 + 2 \epsilon \L_f \H_f Q_{X,\epsilon}^{1/2} n^{3/2}.
\qquad \mbox{(see  (\ref{eq:ddd to dd pre},\ref{eq:ddd to dd pre pre}))}
\end{align}
This completes the proof of Lemma~\ref{lem:ddd to dd bias}.

\section{Proof of Lemma~\ref{lem:like a good neighbor}}\label{sec:proof of like a good neighbor}

Our objective is to establish that, given $X$ and neighborhood radius $\epsilon$,  each $x\in X$ has many neighbors in $Y_{X,\epsilon}$ provided that  $N_{X,\epsilon}=\# Y_{X,\epsilon}$ is sufficiently large.  To that end, we proceed as follows. Recall that $\mu_{X,\epsilon}$ is the conditional distribution on the $\epsilon$-neighborhood of the point cloud $X$ (see \eqref{eq:mu_X}).
With $y\sim \mu_{X,\epsilon}$  and for fixed $x\in X$,   observe that $y$ belongs to the $\epsilon$-neighborhood of $x$ (namely, $y\in \B_{x,\epsilon}$) with the following probability:
\begin{equation}
\Pr{}_{y|x}\l[ y\in \B_{x,\epsilon} \r] = \frac{\mu\l( \B_{x,\epsilon}\r)}{\mu\l( \B_{X,\epsilon} \r)}.
\end{equation}
Equivalently, the indicator function $1_{y\in \B_{x,\epsilon}}$
follows a Bernoulli distribution:
\begin{equation}
1_{y\in Y_{x,\epsilon}}|x\sim \operatorname{Bernoulli}\l(\frac{\mu\l(\B_{x,\epsilon}\r)}{\mu\l(\B_{X,\epsilon}\r)}\r).
\label{eq:Bernolli dist fixed x}
\end{equation}
Then,
\begin{equation}\label{eq:exp of ind}
\E_{Y_{X,\epsilon}|X} \l[ N_{x,\epsilon} \r] = \frac{\mu\l(\B_{x,\epsilon}\r)}{\mu\l(\B_{X,\epsilon}\r)}\cdot \# Y_{X,\epsilon} = \frac{\mu\l(\B_{x,\epsilon}\r)}{\mu\l( \B_{X,\epsilon}\r)}\cdot N_{X,\epsilon},
\end{equation}
and, to investigate the concentration of $N_{x,\epsilon}$ about its expectation, we write that
\begin{align}\label{eq:Bernie prep neighbor}
N_{x,\epsilon} - \frac{\mu\l(\B_{x,\epsilon}\r)}{\mu\l( \B_{X,\epsilon}\r)} \cdot N_{X,\epsilon}
& = N_{x,\epsilon} - \E_{Y_{X,\epsilon}|X}\l[ N_{x,\epsilon} \r]
\qquad \mbox{(see \eqref{eq:exp of ind})}
\nonumber\\
& = \sum_{y\in Y_{X,\epsilon}} \l( 1_{y\in \B_{x,\epsilon}}- \E_{Y_{X,\epsilon}|X}  \l[1_{y\in \B_{x,\epsilon}}\r]\r)\nonumber\\
& = \sum_{y\in Y_{X,\epsilon}} \l( 1_{y\in \B_{x,\epsilon}}- \frac{\mu\l(\B_{x,\epsilon}\r)}{\mu\l( \B_{X,\epsilon}\r)}\r) \nonumber\\
& =: \sum_{y\in Y_{X,\epsilon}} a_{y},
\end{align}
where $\{a_y\}_y$ are independent zero-mean random variables (for fixed $x\in X$). In order to apply the Bernstein's inequality (Proposition~\ref{prop:Bernstein recall}) to the last line of \eqref{eq:Bernie prep neighbor}, we write that
\begin{align}
b & = \max_y \l|a_y \r|\nonumber\\
& = \max_y \l|1_{y\in \B_{x,\epsilon}}- \frac{\mu\l(\B_{x,\epsilon}\r)}{\mu\l( \B_{X,\epsilon} \r)}  \r| \qquad
\mbox{(see \eqref{eq:Bernie prep neighbor})}
\nonumber\\
& \le 1,
\end{align}
\begin{align}
\sigma^2 & = \sum_{y\in Y_{X,\epsilon}} \E_{Y_{x,\epsilon}|x} \l[ a_y^2 \r] \nonumber\\
& = \sum_{y\in Y_{X,\epsilon}} \E_{Y_{x,\epsilon}|x} \l[ \l(1_{y\in \B_{x,\epsilon}} -
\frac{ \mu\l(\B_{x,\epsilon}\r)}{\mu\l(\B_{X,\epsilon}\r)} \r)^2 \r]
\qquad \mbox{(see \eqref{eq:Bernie prep neighbor})}
\nonumber\\
& = \sum_{y\in Y_{X,\epsilon}} \frac{\mu\l(\B_{x,\epsilon}\r)}{\mu\l(\B_{X,\epsilon}\r)} \l(1-\frac{\mu\l(\B_{x,\epsilon}\r)}{\mu\l(\B_{X,\epsilon}\r)} \r)
\qquad \mbox{(see \eqref{eq:Bernolli dist fixed x})}
\nonumber\\
& \le \sum_{y\in Y_{X,\epsilon}} \frac{\mu\l(\B_{x,\epsilon}\r)}{\mu\l(\B_{X,\epsilon}\r)} =  \frac{\mu\l(\B_{x,\epsilon}\r)}{\mu\l(\B_{X,\epsilon}\r)}\cdot   N_{X,\epsilon},
\qquad \l(\# Y_{x,\epsilon} = N_{x,\epsilon}\r)
\end{align}
\begin{equation}
\max\l[b,\sigma \r] = \frac{\mu\l(\B_{x,\epsilon}\r)}{\mu\l( \B_{X,\epsilon} \r)}   N_{X,\epsilon}.
\qquad \l(\mbox{if } N_{X,\epsilon}\ge \frac{\mu\l(\B_{X,\epsilon} \r)}{\mu\l(\B_{x,\epsilon}\r)} \r)
\label{eq:b n sigma for neighbors}
\end{equation}
From Proposition~\ref{prop:Bernstein recall}, then, it follows that
\begin{align}
\l| N_{x,\epsilon}- \frac{\mu\l(\B_{x,\epsilon}\r)}{\mu\l(\B_{X,\epsilon} \r)} N_{X,\epsilon}\r|
& \lesssim  \Cl[gam]{nb} \cdot \max\l[b,\sigma \r]\nonumber\\
& = \Cr{nb} \cdot \sqrt{\frac{\mu\l(\B_{x,\epsilon}\r)}{\mu\l(\B_{X,\epsilon}\r)} N_{X,\epsilon}},
\qquad \mbox{(see \eqref{eq:b n sigma for neighbors})}
\end{align}
for $\Cr{nb}\ge 1$  and except with a probability of at most $e^{-\Cr{nb}}$. Recall that $\# X = N$. Then, an application of the union bound with the choice of $\Cr{nb} = \Cr{nb1} \log N$ (with $\Cr{nb1}\ge 1$) yields that
\begin{equation}
\label{eq:neigh pre final}
\max_{x\in X} \l| N_{x,\epsilon}- \frac{\mu\l(\B_{x,\epsilon}\r)}{\mu\l(\B_{X,\epsilon}\r)} N_{X,\epsilon}\r| \lesssim  \Cr{nb1} \log N \cdot \sqrt{
\frac{\mu\l(\B_{x,\epsilon}\r)}{\mu\l(\B_{X,\epsilon}\r)} N_{X,\epsilon}},
\end{equation}
except with a probability of at most $N e^{-\Cr{nb1} \log N} = N^{1-\Cr{nb1}}$. For the bound above to hold, we assume that $N_{X,\epsilon}$ is sufficiently large (so that the requirement in  \eqref{eq:b n sigma for neighbors} hold for every $x\in X$). In fact, if
\begin{equation}
N_{X,\epsilon} \gtrsim \frac{\Cr{nb1}^2  \log^2 N \cdot \mu\l(\B_{X,\epsilon}\r)}{\min_{x\in X}\mu\l(\B_{x,\epsilon}\r)},
\end{equation}
then \eqref{eq:neigh pre final} readily yields that
\begin{equation}
\frac{1}{2} \cdot \frac{\mu\l(\B_{x,\epsilon}\r)}{\mu\l( \B_{X,\epsilon}\r)} N_{X,\epsilon} \le N_{x,\epsilon} \le \frac{3}{2} \cdot \frac{\mu\l(\B_{x,\epsilon}\r)}{\mu\l( \B_{X,\epsilon}\r)}  N_{X,\epsilon}, \qquad \forall x\in X,
\end{equation}
except with a probability of at most $N^{1-\Cr{nb1}}$. This completes the proof of Lemma~\ref{lem:like a good neighbor}.

\section{Proof of Lemma~\ref{lem:bias}}
\label{sec:proof of lemma bias}

Throughout, $X$ and $\epsilon\in(0,\epsilon_{\mu,X}]$ are fixed, and we further assume that the event $\Cr{good}$ holds (see \eqref{eq:good event def}). For now, suppose in addition that the neighborhood structure $\NN_{X,\epsilon}:=\{ N_{x,\epsilon}\}_{x\in X}$ is fixed too.
Recalling the definition of $\ddot{\nabla}_{Y_{x,\epsilon}}f(\cdot)$ from \eqref{eq:grad est}, we first set
\begin{equation}
\mathbb{R}^{n\times n}\ni
\ddddot{\Sigma}_{X,Y_{X,\epsilon}}
:= \frac{1}{N} \sum_{x\in X}\ddot{\nabla}_{Y_{x,\epsilon}}f(x)  \ddot{\nabla}_{Y_{x,\epsilon}}f(x) ^{*},
\label{eq:Sigma prime}
\end{equation}
for short, and then separate the ``diagonal'' and ``off-diagonal'' components of the expectation of $\ddddot{\Sigma}_{X,Y_{X,\epsilon}}$ as follows:
\begin{align}
 &\mathbb{E}_{Y_{X,\epsilon}|N_{X,\epsilon},X}\left[ \ddddot{\Sigma}_{X,Y_{X,\epsilon}}\right]
 \nonumber\\
 & =
\frac{1}{N}\cdot  \mathbb{E}_{Y_{X,\epsilon}|N_{X,\epsilon},X}\left[
 \sum_{x\in X} \ddot{\nabla}_{Y_{x,\epsilon}} f(x) \cdot
\ddot{\nabla}_{Y_{x,\epsilon}} f(x)^*
 \r]
 \qquad \mbox{(see \eqref{eq:Sigma prime})}\nonumber\\
 & =\frac{n^2}{N }\cdot \mathbb{E}_{Y_{X,\epsilon}|N_{X,\epsilon},X}
 \left[
\sum_{x\in X}
\frac{1}{N_{x,\epsilon}^2}
\sum_{y,y'\in Y_{x,\epsilon}}P_{x,y}\nabla f(x)\nabla f(x)^{*}P_{x,y'}\right] \qquad \mbox{(see \eqref{eq:grad est})}
\nonumber\\
 & = \frac{n^2}{N}\cdot \mathbb{E}_{Y_{X,\epsilon}|N_{X,\epsilon},X}\left[
 \sum_{x\in X}
\frac{1}{N_{x,\epsilon}^2}
 \sum_{y\in Y_{x,\epsilon}}P_{x,y}\nabla f(x)\nabla f(x)^{*}P_{x,y}\right]\nonumber\\
 & \qquad
+\frac{n^2}{N}\cdot \mathbb{E}_{Y_{X,\epsilon}|N_{X,\epsilon},X}\left[
 \sum_{x\in X}
\frac{1}{N_{x,\epsilon}^2}
 \sum_{y,y'\in Y_{x,\epsilon}} 1_{y\ne y'}\cdot P_{x,y}\nabla f(x)\nabla f(x)^{*}P_{x,y'}\right]
 \nonumber\\
 & =\frac{n^2}{N} \sum_{x\in X}
 \frac{1}{N_{x,\epsilon}^2}  \sum_{y\in Y_{x,\epsilon}}
  \mathbb{E}_{y|x}\left[P_{x,y}\nabla f(x)\nabla f(x)^{*}P_{x,y}\right]\nonumber\\
  & \,
+  \frac{n^2}{N} \sum_{x\in X}
\frac{1}{N_{x,\epsilon}^2}
 \sum_{y,y'\in Y_{x,\epsilon}}\E_{y,y'|x}\left[1_{y\ne y'}\cdot P_{x,y}\nabla f(x)\nabla f(x)^{*}P_{x,y'}\right]
\qquad \l(y,y'\sim \mu_{x,\epsilon} \r)
  \nonumber\\
 & =\frac{n^2}{N } \sum_{x\in X}
 \frac{1}{N_{x,\epsilon}}\cdot  \mathbb{E}_{y|x}\left[P_{x,y}\nabla f(x)\nabla f(x)^{*}P_{x,y}\right]
\nonumber\\
& \qquad \qquad +\frac{n^2}{N}\sum_{x\in X}
\frac{1}{N_{x,\epsilon}^2}
\sum_{y,y'\in Y_{x,\epsilon}}1_{y\ne y'}\cdot \mathbb{E}_{y|x}\left[P_{x,y}\nabla f(x)\right]\cdot \mathbb{E}_{y'|x}\left[\nabla f(x)^{*}P_{x,y'}\right].
\label{eq:long bias}
\end{align}
The last line above uses the fact that distinct elements of $Y_{x,\epsilon}$ are  statistically  independent. We next replace both the diagonal  and off-diagonal components  (namely, the first and second sums in the last line above) with simpler expressions. We approximate the diagonal term with another sum as follows:
\begin{align}
&
\left\|\frac{n^2}{N}\sum_{x\in X} \frac{1}{N_{x,\epsilon}}
\mathbb{E}_{y|x} \left[
P_{x,y} \nabla f(x) \nabla f(x)^* P_{x,y}
\right] \r. \nonumber\\
& \l. \qquad -
\frac{n^2}{N} \sum_{x\in X}
\frac{1}{N_{x,\epsilon}}
\left(
\frac{2\nabla f(x)\nabla f(x)^*}{n(n+2)}+
\frac{\left\|\nabla f(x) \right\|_2^2 }{n(n+2)}\cdot I_n
\right)
\right\|_F\nonumber\\
& \le
\frac{n^2}{\min_{x\in X} N_{x,\epsilon}} \cdot
\sup_{x\in \D}
\Bigg \|
\mathbb{E}_{y|x} \left[
P_{x,y} \nabla f(x) \nabla f(x)^* P_{x,y}
\right]  \nonumber\\
&  \qquad \qquad \qquad 
-
\left(
\frac{2\nabla f(x)\nabla f(x)^*}{n(n+2)}+
\frac{\left\|\nabla f(x) \right\|_2^2 }{n(n+2)}\cdot I_n
\right)
\Bigg\|_F
\quad \l( \# X= N\r)
\nonumber\\
& =:  \frac{\bias''_{\mu,\epsilon}}{\min_{x\in X} N_{x,\epsilon}} \nonumber\\
& \le \frac{\bias''_{\mu,\epsilon}}{N_{X,\min,\epsilon}}.  \qquad \mbox{(see \eqref{eq:good event def})}
\label{eq:2nd order reduction}
\end{align}
To replace the off-diagonal term  in the last line of \eqref{eq:long bias}, first recall the  inequality
\begin{align}\label{eq:conv ineq}
& \l\|ab^*-cd^* \r\|_F \nonumber\\
& \le 2  \max\l[\|a-c\|_2,\|b-d\|_2\r] \cdot \max\l[\|b\|_2,\|c\|_2 \r],
\qquad  a,b,c,d \in\R^n,
\end{align}
and then note  that
\begin{align}
&
\Bigg\|
\frac{n^2}{N} \sum_{x\in X} \frac{1}{N_{x,\epsilon}^2} \sum_{y,y'\in Y_{x,\epsilon}} 1_{y\ne y'} \cdot
 \mathbb{E}_{y|x} \left[P_{x,y}\nabla f(x) \right]\cdot \mathbb{E}_{y'|x} \left[ \nabla f(x)^* P_{x,y'}\right] \nonumber\\
& \qquad -\frac{1}{N} \sum_{x\in X} \frac{N_{x,\epsilon}-1}{N_{x,\epsilon}} \nabla f(x) \nabla f(x)^*
\Bigg\|_F \nonumber\\
& =\Bigg\|
\frac{n^2}{N} \sum_{x\in X} \frac{1}{N_{x,\epsilon}^2} \sum_{y,y'\in Y_{x,\epsilon}} 1_{y\ne y'} \nonumber\\
& \qquad \cdot 
\left( \mathbb{E}_{y|x} \left[P_{x,y}\nabla f(x) \right]\cdot \mathbb{E}_{y'|x} \left[ \nabla f(x)^* P_{x,y'}\right]
-
\frac{\nabla f(x) \nabla f(x)^*}{n^2}
\right)
\Bigg\|_F
\,\, \, \l( \# Y_{x,\epsilon} = N_{x,\epsilon}\r)
\nonumber\\
& \le n^2
\max_{x\in X} \max_{y,y'\in Y_{x,\epsilon}}
\Bigg\|
\mathbb{E}_{y|x} \left[P_{x,y}\nabla f(x) \right]\cdot \mathbb{E}_{y'|x} \left[ \nabla f(x)^* P_{x,y'}\right] \nonumber\\
& \qquad \qquad \qquad\qquad \qquad-
\frac{\nabla f(x) \nabla f(x)^*}{n^2}
\Bigg\|_F
\qquad \l(\#X = N, \#Y_{X,\epsilon}=N_{x,\epsilon} \r)
\nonumber\\
& \le 2n^2 \max_{x\in X}
\Bigg[
\left\|
\mathbb{E}_{y|x}\left[P_{x,y}\nabla f(x) \right]
- \frac{\nabla f(x)}{n}
\right\|_2 \nonumber\\
& \qquad \qquad \qquad 
\cdot
\max\left[
\left\|
\mathbb{E}_{y|x} \left[ P_{x,y} \nabla f(x)\right]
\right\|_2
,
\frac{\left\|\nabla f(x)\right\|_2}{n}
\right]
\Bigg]
\qquad \mbox{(see \eqref{eq:conv ineq})}
\nonumber\\
& \le 2n^2 \max_{x\in X}
\Bigg[
\left\|
\mathbb{E}_{y|x}\left[P_{x,y}\nabla f(x) \right]
- \frac{\nabla f(x)}{n}
\right\|_2 \nonumber\\
& \qquad \qquad \qquad \cdot
\l(
\left\|
\mathbb{E}_{y|x} \left[ P_{x,y} \nabla f(x)\right]
- \frac{\nabla f(x)}{n}
\right\|_2
+
\frac{\left\|\nabla f(x)\right\|_2}{n}
\r)
\Bigg] \quad \mbox{(triangle ineq.)}
\nonumber\\
& \le
2n^2 \l( \frac{\bias'_{\mu,\epsilon}}{n}\cdot \L_f \r)  \l( \frac{\bias'_{\mu,\epsilon}}{n} \cdot \L_f +\frac{\L_f}{n}\r)
\qquad \mbox{(see \eqref{eq:Lf} and \eqref{eq:Bf})}
\nonumber\\
& = 2 \bias'_{\mu,\epsilon}\l( \bias'_{\mu,\epsilon}+1\r) \L_f^2.
\label{eq:1st order reduction}
\end{align}
We may now replace the diagonal and off diagonal components in the last line of  \eqref{eq:long bias} with simpler expressions while incurring a typically small error. More specifically, in light of \eqref{eq:2nd order reduction} and \eqref{eq:1st order reduction}, \eqref{eq:long bias} now  implies that
\begin{align}
&
\Bigg\|
\mathbb{E}_{Y_{X,\epsilon}|N_{X,\epsilon},X} \left[ \ddddot{\Sigma}_{X,Y_{X,\epsilon}} \right]
-
\frac{1}{N} \sum_{x\in X}
\left( 1+ \frac{n-2}{N_{x,\epsilon}(n+2)} \right)
\nabla f(x) \nabla f(x)^* \nonumber\\
& \qquad \qquad\qquad \qquad  
- \frac{n}{N(n+2)} \sum_{x\in X} \frac{\left\|\nabla f(x) \right\|_2^2}{N_{x,\epsilon}}  \cdot I_n
\Bigg\|_F\nonumber\\
& =\Bigg\|
\mathbb{E}_{Y_{X,\epsilon}|N_{X,\epsilon},X} \left[ \ddddot{\Sigma}_{X,Y_{X,\epsilon}} \right]
-
\frac{n^2}{N} \sum_{x\in X} \frac{1}{N_{x,\epsilon}}
\left(
\frac{2\nabla f(x) \nabla f(x)^*}{n(n+2)}+\frac{\left\|\nabla f(x)\right\|_2^2}{n(n+2)}\cdot I_n
\right)
\nonumber\\
& \qquad\qquad\qquad   - \frac{1}{N} \sum_{x\in X} \frac{N_{x,\epsilon}-1}{N_{x,\epsilon}}  \nabla f(x) \nabla f(x)^*
\Bigg\|_F\nonumber\\
& \le  \frac{\bias''_{\mu,\epsilon}}{N_{X,\min,\epsilon}} +2 \bias'_{\mu,\epsilon}\l(\bias'_{\mu,\epsilon}+1 \r)\L_f^2.\qquad \mbox{(see \eqref{eq:2nd order reduction} and \eqref{eq:1st order reduction})}
\label{eq:bias 10}
\end{align}
We can further simplify the first line of \eqref{eq:bias 10} by replacing $N_{x,\epsilon}$
with $N_{X,\min,\epsilon}$ as follows.
By invoking \eqref{eq:emp} in the second line below, we note that
\begin{align}
&\Bigg\|
\mathbb{E}_{Y_{X,\epsilon}|N_{X,\epsilon},X} \left[ \ddddot{\Sigma}_{X,Y_{X,\epsilon}} \right]
-
\left( 1+ \frac{n-2}{ N_{X,\min,\epsilon}(n+2)} \right)
\dot{\Sigma}_{X} \nonumber\\
& \qquad \qquad \qquad \qquad - \frac{n}{N_{X,\min,\epsilon}(n+2)} \cdot \mbox{trace}\left[
\dot{\Sigma}_{X}
\right] \cdot I_n
\Bigg\|_F\nonumber\\
&
 = \Bigg\|
\mathbb{E}_{Y_{X,\epsilon}|N_{X,\epsilon},X} \left[ \ddddot{\Sigma}_{X,Y_{X,\epsilon}} \right]
-
\left( 1+ \frac{n-2}{ N_{X,\min,\epsilon}(n+2)} \right)\frac{1}{N}
\sum_{x \in X}
\nabla f(x) \nabla f(x)^*
\nonumber\\
& \qquad \qquad \qquad - \frac{n}{ N_{X,\min,\epsilon}(n+2)}\cdot  \frac{1}{N} \sum_{x \in X} \left\|\nabla f(x) \right\|_2^2 \cdot I_n
\Bigg\|_F
\nonumber\\
& \le \l(\frac{\bias''_{\mu,\epsilon}}{N_{X,\min,\epsilon}} +2 \bias'_{\mu,\epsilon}\l(\bias'_{\mu,\epsilon}+1 \r)\L_f^2\r) \nonumber\\
& \qquad \qquad +
\max_{x\in X}  \left| \frac{1}{N_{x,\epsilon}} -\frac{1}{N_{X,\min,\epsilon}}\right|
\cdot \max_{x\in X} \left\| \nabla f(x)\right\|_2^2 \cdot \left(1+\left\| I_n\right\|_F\right)
\qquad \mbox{(see \eqref{eq:bias 10})}\nonumber\\
& \le  \l(\frac{\bias''_{\mu,\epsilon}}{N_{X,\min,\epsilon}} +2 \bias'_{\mu,\epsilon}\l(\bias'_{\mu,\epsilon}+1 \r)\L_f^2\r) \nonumber\\
& \qquad \qquad +
\max_{x\in X} \l|\frac{1}{N_{x,\epsilon}}- \frac{1}{N_{X,\min,\epsilon}} \r| \cdot
 \L_f^2  \left(1+\sqrt{n}\right) \qquad \l( \mbox{see \eqref{eq:Lf}}\r)\nonumber\\
 & \le  \l(\frac{\bias''_{\mu,\epsilon}}{N_{X,\min,\epsilon}} +2 \bias'_{\mu,\epsilon}\l(\bias'_{\mu,\epsilon}+1 \r)\L_f^2\r) +
 \frac{
 \L_f^2  \left(1+\sqrt{n}\right)}{N_{X,\min,\epsilon}}
 \qquad \mbox{(see  \eqref{eq:good event def})}
 \nonumber\\
  & =: \frac{1}{2}\bias_{\mu,\epsilon}.
 \label{eq:bias 11}
\end{align}
Next, we replace $\mbox{trace}[\dot{\Sigma}_{X}]$ in the first line of  \eqref{eq:bias 11} with $\mbox{trace}[\ddddot{\Sigma}_{X,Y_{X,\epsilon}}]$. To that end, we first notice the following consequence of \eqref{eq:bias 11}:
\begin{align}
&
\Bigg| \mathbb{E}_{Y_{X,\epsilon}|N_{X,\epsilon},X} \left[\tr\left[ \ddddot{\Sigma}_{X,Y_{X,\epsilon}}\right] \right]
-
\left( 1+ \frac{n-2}{ N_{X,\min,\epsilon}(n+2)} \right)
\tr\left[\dot{\Sigma}_{X}\right] \nonumber\\
& \qquad \qquad \qquad \qquad 
- \frac{n^2}{N_{X,\min,\epsilon}(n+2)} \cdot \mbox{trace}\left[
\dot{\Sigma}_{X}
\right]
\Bigg|
\nonumber\\
& =
\Bigg| \tr\Bigg[
\mathbb{E}_{Y_{X,\epsilon}|N_{X,\epsilon},X} \left[ \ddddot{\Sigma}_{X,Y_{X,\epsilon}} \right]
-
\left( 1+ \frac{n-2}{N_{X,\min,\epsilon}(n+2)} \right)
\dot{\Sigma}_{X} \nonumber\\
& \qquad \qquad \qquad \qquad 
- \frac{n}{N_{X,\min,\epsilon}(n+2)} \cdot \mbox{trace}\left[
\dot{\Sigma}_{X}
\right] \cdot I_n
\Bigg]
\Bigg|
\nonumber\\
& \le \sqrt{n}
 \Bigg\|
\mathbb{E}_{Y_{X,\epsilon}|N_{X,\epsilon},X} \left[ \ddddot{\Sigma}_{X,Y_{X,\epsilon}} \right]
-
\left( 1+ \frac{n-2}{N_{X,\min,\epsilon}(n+2)} \right)
\dot{\Sigma}_{X} \nonumber\\
& \qquad \qquad \qquad \qquad 
- \frac{n}{N_{X,\min,\epsilon}(n+2)} \cdot \mbox{trace}\left[
\dot{\Sigma}_{X}
\right] \cdot I_n
\Bigg\|_F
\nonumber\\
& \le \frac{\sqrt{n}}{2}\bias_{\mu,\epsilon},
\qquad \mbox{(see \eqref{eq:bias 11})}
\label{eq:bias 12}
\end{align}
where the second line uses the fact that $\tr\left[I_n\right] =n$.
Also, the third line follows from the inequality $\l|\tr[A]\r| \le \sqrt{n}\|A\|_F$ for an arbitrary matrix $A\in\R^{n\times n}$. After rearranging, \eqref{eq:bias 12} immediately implies that
\begin{align}
& \l|
\l(1+\frac{n^2+n-2}{N_{X,\min,\epsilon}(n+2)} \r)^{-1}
\mathbb{E}_{Y_{X,\epsilon}|N_{X,\epsilon},X} \left[\tr\left[ \ddddot{\Sigma}_{X,Y_{X,\epsilon}}\right] \right]
-
\tr\left[\dot{\Sigma}_{X}\right]
\r| \nonumber\\
& \le
\l(1+\frac{n^2+n-2}{N_{X,\min,\epsilon}(n+2)} \r)^{-1} \frac{\sqrt{n}}{2} \bias_{\mu,\epsilon}.
\label{eq:bias 13}
\end{align}
The above inequality  enables us to remove $\tr[\dot{\Sigma}_{X} ]$ from the first line of \eqref{eq:bias 11}:
\begin{align}
&\left\|
\mathbb{E}_{Y_{X,\epsilon}|N_{X,\epsilon},X} \left[ \ddddot{\Sigma}_{X,Y_{X,\epsilon}} \right]
-
\left( 1+ \frac{n-2}{N_{X,\min,\epsilon}(n+2) } \right)
\dot{\Sigma}_{X}\right. \nonumber\\
& \left.
- \frac{n}{N_{X,\min,\epsilon}(n+2)} \cdot
\l(1+\frac{n^2+n-2}{N_{X,\min,\epsilon}(n+2)} \r)^{-1}
\mathbb{E}_{Y_{X,\epsilon}|N_{X,\epsilon},X}\l[
\tr\l[
\ddddot{\Sigma}_{X,Y_{X,\epsilon}}
\r]
\r]
  \cdot I_n
\right\|_F\nonumber\\
&
\le \frac{1}{2} \bias_{\mu,\epsilon}+ \frac{n}{N_{X,\min,\epsilon}(n+2)} \l(1+\frac{n^2+n
-2}{N_{X,\min,\epsilon}(n+2)} \r)^{-1} \frac{\sqrt{n}}{2}\bias_{\mu,\epsilon} \cdot \l\|I_n\r\|_F
\, \mbox{(\ref{eq:bias 11},\ref{eq:bias 13})}\nonumber\\
& =
\frac{1}{2}
\l( 1+ \frac{n^2}{N_{X,\min,\epsilon}(n+2) } \l(1+\frac{n^2+n
-2}{N_{X,\min,\epsilon}(n+2)} \r)^{-1} \r) \bias_{\mu,\epsilon}.
\qquad \l( \|I_n\|_F= \sqrt{n} \r)
\label{eq:bias 20}
\end{align}
Lastly, \eqref{eq:bias 20} can be rewritten as follows by introducing ${\dddot{\Sigma}}_{X,Y_{X,\epsilon}}\in\mathbb{R}^{n\times n}$:
\begin{align}
& \l\|
\mathbb{E}_{Y_{X,\epsilon}|\Cr{good},N_{X,\epsilon},X} \l[{\dddot{\Sigma}}_{X,Y_{X,\epsilon}}\r]-
\dot{\Sigma}_{X}
\r\|_F \nonumber\\
&
\le \frac{1}{2} \l( 1+\frac{n-2}{N_{X,\min,\epsilon}(n+2)}\r)^{-1} \nonumber\\
& \qquad 
\cdot \l( 1
+
\frac{n^2}{N_{X,\min,\epsilon}(n+2)} \l(1+\frac{n^2+n
-2}{N_{X,\min,\epsilon}(n+2)} \r)^{-1}\
\r) \bias_{\mu,\epsilon}\nonumber\\
& \le  \bias_{\mu,\epsilon},
\qquad \l( \mbox{the factor in front of } \bias_{\mu,\epsilon} \mbox{ does not exceed 1}  \r)
\label{eq:bias pre final wo scaling}
\end{align}
where, above, we set
\begin{align}
& {\dddot{\Sigma}}_{X,Y_{X,\epsilon}} \nonumber\\
& :=
\l(1+\frac{n-2}{N_{X,\min,\epsilon}(n+2)} \r)^{-1} \nonumber\\
&\,\, \cdot 
\l( \ddddot{\Sigma}_{X,Y_{X,\epsilon}}	- \frac{n}{N_{X,\min,\epsilon}(n+2)} \l(1+\frac{n^2+n-2}{N_{X,\min,\epsilon}(n+2)} \r)^{-1} \cdot \tr\l[\ddddot{\Sigma}_{X,Y_{X,\epsilon}}\r] \cdot I_n\r)\nonumber\\
& =
\l(1+\frac{1-\frac{2}{n}}{1+\frac{2}{n} } \cdot N_{X,\min,\epsilon}^{-1}\r)^{-1} \nonumber\\
& \cdot 
\l( \ddddot{\Sigma}_{X,Y_{X,\epsilon}}	-
\l(\l(1+\frac{2}{n} \r)N_{X,\min,\epsilon}+n+1-\frac{2}{n}\r)^{-1}
  \tr\l[\ddddot{\Sigma}_{X,Y_{X,\epsilon}}\r] \cdot I_n\r)\nonumber\\
& = \l(1+\frac{1-\frac{2}{n}}{1+\frac{2}{n} } \cdot N_{X,\min,\epsilon}^{-1}\r)^{-1} \nonumber\\
& \qquad \cdot
\Bigg(
\frac{1}{N}\sum_{x\in X}\ddot{\nabla}_{Y_{X,\epsilon}}f(x) \ddot{\nabla}_{Y_{X,\epsilon}}f(x)^* \nonumber\\
& \qquad \qquad \qquad 
-
\l( \l(1+\frac{2}{n} \r)N_{X,\min,\epsilon}+n+1-\frac{2}{n}\r)^{-1}  \frac{1}{N} \sum_{x\in X}\l\|\ddot{\nabla }_{Y_{x,\epsilon}} f(x) \r\|_2^2  \cdot I_n
\Bigg)
\nonumber\\
& = \frac{1}{N}\l(1+\frac{1-\frac{2}{n}}{1+\frac{2}{n} } \cdot N_{X,\min,\epsilon}^{-1}\r)^{-1}  \cdot
\Bigg(
\sum_{N_{x,\epsilon}\ge N_{X,\min,\epsilon}}\ddot{\nabla}_{Y_{X,\epsilon}}f(x) \ddot{\nabla}_{Y_{X,\epsilon}}f(x)^* \nonumber\\
& \qquad \qquad -
\l( \l(1+\frac{2}{n} \r)N_{X,\min,\epsilon}+n+1-\frac{2}{n}\r)^{-1}   \sum_{N_{x,\epsilon}\ge N_{X,\min,\epsilon}}\l\|\ddot{\nabla }_{Y_{x,\epsilon}} f(x) \r\|_2^2  \cdot I_n
\Bigg),
\end{align}
where the third identity uses \eqref{eq:Sigma prime} and the last line above follows from \eqref{eq:good event def}. Because $\bias_{\mu,\epsilon}$ does not depend on $N_{X,\epsilon}$, it is easy to remove the conditioning on $N_{X,\epsilon}$ in \eqref{eq:bias pre final wo scaling}:
\begin{align}
& \l\|
\mathbb{E}_{Y_{X,\epsilon}|\Cr{good},X} \l[{\dddot{\Sigma}}_{X,Y_{X,\epsilon}}\r]-
\dot{\Sigma}_{X}
\r\|_F \nonumber\\
& =
\l\|
\E\l[ \mathbb{E}_{Y_{X,\epsilon}|\Cr{good},N_{X,\epsilon},X} \l[{\dddot{\Sigma}}_{X,Y_{X,\epsilon}}\r] -
\dot{\Sigma}_{X} \r]
\r\|_F \nonumber\\
& \le
\E \l\|
 \mathbb{E}_{Y_{X,\epsilon}|\Cr{good},N_{X,\epsilon},X} \l[{\dddot{\Sigma}}_{X,Y_{X,\epsilon}}\r] -
\dot{\Sigma}_{X}
\r\|_F
\qquad \mbox{(Jensen's inequality)}\nonumber\\
& \le \E \bias_{\mu,\epsilon} \qquad \mbox{(see \eqref{eq:bias pre final wo scaling})} \nonumber\\
& = \bias_{\mu,\epsilon}.
\qquad \mbox{(see \eqref{eq:bias 11})}
\label{eq:bias main final wo scaling}
\end{align}

Consider also the following special case. Let $\mu$ be the uniform probability measure on $\mathbb{D}$ and fix $x$ within  the $\epsilon$-interior of $\D$, namely $x\in \D_\epsilon$. Also draw $y$ from $\mu_{x,\epsilon}$, namely  $y|x\sim \mu_{x,\epsilon}$ (see \eqref{eq:cond mu}). Then, as stated in Proposition \ref{lem:bias pointwise},  $\bias'_{\mu,\epsilon}=0$. Furthermore, it is known \cite{azizyan2015extreme}  that
\begin{equation}\label{eq:dist of proj}
P_{x,y}\cdot \nabla f(x) \overset{\operatorname{dist.}}{=} \omega \cdot  \nabla f(x) + \sqrt{\omega - \omega^2}
\left\|\nabla f(x)\right\|_2
\cdot  A  \alpha,
\end{equation}
where $\omega$ follows the beta distribution, $\alpha$ is uniformly distributed on the unit sphere in $\mathbb{R}^{n-1}$, and the two variables are independent, i.e.,
$$
\omega \sim \operatorname{beta}\left(\frac{1}{2},\frac{n-1}{2}\right),
\qquad \alpha\sim \operatorname{uniform}\left(\mathbb{S}^{n-2}\right),
\qquad
\omega \independent \alpha.
$$
Finally,  $A\in\mathbb{R}^{n\times (n-1)}$ in \eqref{eq:dist of proj} is an orthonormal basis for the directions orthogonal to $\nabla f(x)\in\mathbb{R}^n$, namely
\begin{equation}
A^* \nabla f(x) = 0, \qquad A^* A = I_{n-1}.
\label{eq:orthogonality}
\end{equation}
Using the expressions for the first and second moments of the beta distribution in the fourth line below, we write that
\begin{align}
& \mathbb{E}_{y|x}
\left[
P_{x,y}\nabla f(x)
\nabla f(x)^* P_{x,y}
 \right]
\nonumber\\
 & =
 \mathbb{E}
  \Bigg[
\left(
\omega \nabla f(x)+ \sqrt{\omega - \omega^2}
\left\| \nabla f(x)  \right\|_2
\cdot A\alpha
\right) \nonumber\\
& \qquad 
\cdot
\left(
\omega \nabla f(x)+ \sqrt{\omega - \omega^2}
\left\| \nabla f(x)  \right\|_2
\cdot A\alpha
\right)^*
\Bigg]
\qquad \mbox{(see \eqref{eq:dist of proj})}
\nonumber\\
& =
\mathbb{E} \left[\omega^2\right] \cdot  \nabla f(x) \nabla f(x)^* \nonumber\\
& \qquad \qquad 
+
\mathbb{E} \left[ \omega - \omega ^2  \right]
\left\| \nabla f(x) \right\|_2^2
\cdot
A \cdot
\mathbb{E}\left[  \alpha \alpha^*\right]
\cdot
A^*
\qquad
\left(\omega \independent \alpha, \quad
\mathbb{E} \alpha = 0
\right)\nonumber\\
&
= \frac{3}{n(n+2)} \cdot  	
\nabla f(x) \nabla f(x)^*
 \nonumber\\
 & \qquad  +
 \frac{n-1}{n(n+2)} \cdot
\left\|
 \nabla f(x)
 \right\|_2^2
 \cdot
 A\cdot  \frac{I_{n-1}}{n-1}
 \cdot A^*
 \qquad
 \left(
 \mathbb{E}\left[\alpha \alpha^*\right]
 = \frac{I_{n-1}}{n-1}
 \right)
\nonumber\\
& =
\frac{3}{n(n+2)} \cdot  	
\nabla f(x) \nabla f(x)^*
 +
 \frac{1}{n(n+2)} \cdot
\left\|
 \nabla f(x)
 \right\|_2^2
 \cdot
AA^* \nonumber\\
& = \frac{2}{n(n+2)} \cdot
\nabla f(x) \nabla f(x)^*\nonumber\\
& \qquad 
+ \frac{1}{n(n+2)} \cdot
\left\| \nabla f(x) \right\|_2^2
\cdot
\left(
\frac{\nabla f(x)}{\left\| \nabla f(x) \right\|_2}
\cdot
\frac{\nabla f(x)^*}{\left\| \nabla f(x)  \right\|_2}
+
AA^*
\right)
\nonumber\\
&
= \frac{2}{n(n+2)} \cdot
\nabla f(x) \nabla f(x)^*
+ \frac{1}{n(n+2)} \cdot
\left\| \nabla f(x) \right\|_2^2
\cdot I_n,
\qquad
\mbox{(see
\eqref{eq:orthogonality})}
\end{align}
and, consequently, $\bias_{\mu,\epsilon}''=0$.  Furthermore, assume that $N_{x,\epsilon}=N_{x',\epsilon}$ for every pair $x,x'\in X$. Then, we observe that the upper bound in \eqref{eq:bias 11} can be improved to $\bias_{\mu,\epsilon}=0$, namely  ${\dddot{\Sigma}}_{X,Y_{X,\epsilon}}$ is an unbiased estimator of $\dot{\Sigma}_{X}$, conditioned on the event $\Cr{good}$. This completes the proof of Lemma~\ref{lem:bias}.

\section{Proof of Lemma~\ref{lem:main result}}
\label{sec:proof of main result}

Throughout, $X$ is fixed and we assume that the event $\Cr{good}$ holds (see \eqref{eq:good event def}). We also consider $\NN_{X,\epsilon}=\{N_{x,\epsilon}\}_{x\in X}$ (see \eqref{eq:def of neighborhoods}) to be any fixed neighborhood structure consistent with $\Cr{good}$.

To bound the estimation error, we write that
\begin{align}\label{eq:triangleeqfordddot}
& \l\| {\dddot{\Sigma}}_{X,Y_{X,\epsilon}} - \dot{\Sigma}_{X} \r\|_F
\nonumber\\
& \le  \l\| {\dddot{\Sigma}}_{X,Y_{X,
\epsilon}} - \E_{Y_{X,\epsilon}|\Cr{good},\NN_{X,\epsilon},X}\l[{\dddot{\Sigma}}_{X,Y_{X,\epsilon}}\r] \r\|_F
+
\l\| \E_{Y_{X,\epsilon}|\Cr{good},\NN_{X,\epsilon},X}\l[{\dddot{\Sigma}}_{X,Y_{X,\epsilon}} \r]- \dot{\Sigma}_{X} \r\|_F
\nonumber\\
& \le   \l\| {\dddot{\Sigma}}_{X,Y_{X,\epsilon}} - \E_{Y_{X,\epsilon}|\Cr{good},\NN_{X,\epsilon},X}\l[{\dddot{\Sigma}}_{X,Y_{X,\epsilon}}\r] \r\|_F+ \bias_{\mu,\epsilon}.
\qquad \mbox{(see \eqref{eq:bias pre final wo scaling})}
\end{align}
It therefore suffices to study the concentration of ${\dddot{\Sigma}}_{X,Y_{X,\epsilon}}$ about its expectation. In fact, as we show next, it is  more convenient to first study the concentration of $\ddddot{\Sigma}_{X,Y_{X,\epsilon}}\in\mathbb{R}^{n\times n}$ instead, where
\begin{align}\label{eq:recall Sigma prime 1st}
\ddddot{\Sigma}_{X,Y_{X,\epsilon}}
& := \frac{1}{N} \sum_{x\in X} \ddot{\nabla}_{Y_{x,\epsilon}}f(x) \cdot \ddot{\nabla}_{Y_{x,\epsilon}}f(x)^*,
\end{align}
\begin{equation*}
\ddot{\nabla}_{Y_{x,\epsilon}}f(x):=\frac{n}{N_{x,\epsilon}}\sum_{y\in Y_{x,\epsilon}}P_{x,y}\cdot \nabla f(x) ~ \in \R^n , \qquad \forall x\in X.
\end{equation*}
Indeed, conditioned on $\Cr{good},\NN_{X,\epsilon},X$, the expression for $\dddot{\Sigma}_{X,Y_{X,\epsilon}}$ in \eqref{eq:sigmahathat} simplifies to
\begin{align}
{\dddot{\Sigma}}_{X,Y_{X,\epsilon}} & =
 \l(1+\frac{1-\frac{2}{n}}{1+\frac{2}{n} } \cdot \Nave^{-1}\r)^{-1} \nonumber\\
& \qquad \cdot \l(
\ddddot{\Sigma}_{X,Y_{X,\epsilon}}
-
 \frac{\tr\l[\ddddot{\Sigma}_{X,Y_{X,\epsilon}}\r]}{\l(1+\frac{2}{n} \r)\Nave+n+1-\frac{2}{n}}  \cdot I_n
\r).
\label{eq:sigmahathat simplified cvg}
\end{align}
Consequently, the deviation of ${\dddot{\Sigma}}_{X,Y_{X,\epsilon}}$ about its expectation can be bounded as:
\begin{align}
& \l\|{\dddot{\Sigma}}_{X,Y_{X,\epsilon}} - \E_{Y_{X,\epsilon}|\Cr{good},\NN_{X,\epsilon},X}\l[ {\dddot{\Sigma}}_{X,Y_{X,\epsilon}}\r] \r\|_F \nonumber\\
& \le
\l\|\ddddot{\Sigma}_{X,Y_{X,\epsilon}}- \E_{Y_{X,\epsilon}|\Cr{good},\NN_{X,\epsilon},X}\l[\ddddot{\Sigma}_{X,Y_X,\epsilon}\r] \r\|_F \nonumber\\
& \qquad +
\l(\l(1+\frac{2}{n} \r)\Nave+n+1-\frac{2}{n} \r)^{-1}\nonumber\\
& \qquad \qquad 
 \cdot
\l|\tr\l[\ddddot{\Sigma}_{X,Y_{X,\epsilon}}\r]
- \E_{Y_{X,\epsilon}|\Cr{good},\NN_{X,\epsilon},X}\l[\tr\l[\ddddot{\Sigma}_{X,Y_{X,\epsilon}}\r]\r] \r| \cdot \l\|I_n \r\|_F
\nonumber\\
& \le \l\|\ddddot{\Sigma}_{X,Y_{X,\epsilon}}- \E_{Y_{X,\epsilon}|\Cr{good},\NN_{X,\epsilon},X}\l[\ddddot{\Sigma}_{X,Y_X,\epsilon}\r] \r\|_F \nonumber\\
& \qquad +
\l(\l(1+\frac{2}{n} \r)\Nave+n+1-\frac{2}{n} \r)^{-1}\nonumber\\
& \qquad  \cdot \sqrt{n}
\l\|\ddddot{\Sigma}_{X,Y_{X,\epsilon}}
- \E_{Y_{X,\epsilon}|\Cr{good},\NN_{X,\epsilon},X}\l[\ddddot{\Sigma}_{X,Y_{X,\epsilon}}\r] \r\|_F \cdot \sqrt{n}
\qquad   \l(\l\|I_n\r\|_F=\sqrt{n} \r)
\nonumber\\
& = \l( 1+ \frac{n}{\l( 1+\frac{2}{n} \r)\Nave+n+1-\frac{2}{n}} \r)
\l\|\ddddot{\Sigma}_{X,Y_{X,\epsilon}}
- \E_{Y_{X,\epsilon}|\Cr{good},\NN_{X,\epsilon},X}\l[\ddddot{\Sigma}_{X,Y_{X,\epsilon}}\r] \r\|_F
\nonumber\\
& \le 2 \l\|\ddddot{\Sigma}_{X,Y_{X,\epsilon}}
- \E_{Y_{X,\epsilon}|\Cr{good},\NN_{X,\epsilon},X}\l[\ddddot{\Sigma}_{X,Y_{X,\epsilon}}\r] \r\|_F.
\qquad \mbox{(factor above $\le$ 2)}
\label{eq:Sigma double hat to single hat}
\end{align}
Above, the first inequality uses \eqref{eq:sigmahathat simplified cvg}. We also used the linearity of trace and the inequality $|\tr[A]|\le \sqrt{n}\|A\|_F$ for arbitrary $A\in\R^{n\times n}$. Thanks to  \eqref{eq:Sigma double hat to single hat}, it  suffices to study the concentration of $\ddddot{\Sigma}_{X,Y_{X,\epsilon}}$ about its expectation. The following result is proved in Appendix \ref{sec:proof of lemma com of Sigma prome}.
\begin{lemma}\label{lem:com of Sigma prime}
Fix $X$ and $\epsilon\in(0,\epsilon_{\mu,X}]$. If $\log(n) \ge 1$, $N \ge \log(n)$, and $\log(N_{X,\epsilon}) \ge \log(n)$, then conditioned on $\Cr{good}, \NN_{X,\epsilon}, X$,
\begin{align}
& \l\|\ddddot{\Sigma}_{X,Y_{X,\epsilon}} - \mathbb{E}_{Y_{X,\epsilon}|\Cr{good},\NN_{X,\epsilon},X}\l[
\ddddot{\Sigma}_{X,Y_{X,\epsilon}}
\r] \r\|_F \nonumber\\
& \lesssim
\Cr{g2} \Cr{events}^2 \log^4(N_{X,\epsilon}) \cdot  \frac{n \sqrt{\log n} }{\sqrt{\rho_{\mu,X,\epsilon} N_{X,\epsilon}}} \cdot \max[\K_\mu^{-1},\K_\mu^{-2}]  \L_f^2 \nonumber \\
& \quad + 4 n^2 L_f^2 N_{X,\epsilon}^{(1-\Cr{events}\log(N_{X,\epsilon}))},
\end{align}
for $\Cr{g2} \ge 1$ and $\Cr{events} \ge 3$, except with a probability of at most
\begin{equation}
 e^{-\Cr{g2}} + n^{2-\log \Cr{g2}} + N_{X,\epsilon}^{(1-\Cr{events}\log(N_{X,\epsilon}))}.
\label{eq:lem5prob}
\end{equation}
\end{lemma}
Combining \eqref{eq:triangleeqfordddot}, \eqref{eq:Sigma double hat to single hat}, and Lemma~\ref{lem:com of Sigma prime} tells us that if $\log(n) \ge 1$, $N \ge \log(n)$, and $\log(N_{X,\epsilon}) \ge \log(n)$, then conditioned on $\Cr{good}, \NN_{X,\epsilon}, X$,
\begin{align*}
& \l\| {\dddot{\Sigma}}_{X,Y_{X,\epsilon}} - \dot{\Sigma}_{X} \r\|_F \lesssim \bias_{\mu,\epsilon} + \Cr{g2} \Cr{events}^2 \log^4(N_{X,\epsilon}) \cdot  \frac{n \sqrt{\log n} }{\sqrt{\rho_{\mu,X,\epsilon} N_{X,\epsilon}}} \cdot \max[\K_\mu^{-1},\K_\mu^{-2}]  \L_f^2 \nonumber\\
& \qquad\qquad \qquad \qquad  + 4 n^2 L_f^2 N_{X,\epsilon}^{(1-\Cr{events}\log(N_{X,\epsilon}))},
\end{align*}
except with the probability appearing in~\eqref{eq:lem5prob}. We observe that this expression and probability do not depend on $\NN_{X,\epsilon}$, and so the same statement holds with the same probability when we condition only on $\Cr{good}, X$. This completes the proof of Lemma~\ref{lem:main result}.

\section{Proof of Lemma~\ref{lem:com of Sigma prime}}
\label{sec:proof of lemma com of Sigma prome}

Throughout, $X$ is fixed and the event $\Cr{good}$ holds. We will also use $\overline{N}_{X,\epsilon}= \{ N_{x,\epsilon}\}_{x\in X}$ to summarize the neighborhood structure of data (see \eqref{eq:def of neighborhoods}). As in Appendix \ref{sec:proof of lemma bias},  we  again decompose $\ddddot{\Sigma}_{X,Y_{X,\epsilon}}$ into ``diagonal'' and ``off-diagonal'' components:
\begin{align}
\label{eq:recall Sigma prime}
\ddddot{\Sigma}_{X,Y_{X,\epsilon}}
& = \frac{1}{N} \sum_{x\in X} \ddot{\nabla}_{Y_{x,\epsilon}}f(x) \cdot \ddot{\nabla}_{Y_{x,\epsilon}}f(x)^*
\qquad \mbox{(see \eqref{eq:recall Sigma prime 1st})}
\nonumber\\
& = \frac{n^2}{N} \sum_{x\in X} \frac{1}{N_{x,
\epsilon}^2} \sum_{y,y'\in Y_{x,\epsilon}}P_{x,y} \nabla f(x) \nabla f(x)^* P_{x,y'}
\qquad \mbox{(see \eqref{eq:grad est})}\nonumber\\
& = \frac{n^2}{N} \sum_{x\in X} \frac{1}{N_{x,\epsilon}^2} \sum_{y\in Y_{x,\epsilon}}P_{x,y} \nabla f(x) \nabla f(x)^* P_{x,y} \nonumber\\
& \qquad 
+ \frac{n^2}{N} \sum_{x\in X} \frac{1}{N_{x,\epsilon}^2} \sum_{y,y'\in Y_{x,\epsilon}} 1_{y\ne y'}\cdot P_{x,y} \nabla f(x) \nabla f(x)^* P_{x,y'}
\nonumber\\ 
& =: \ddddot{\Sigma}_{X,Y_{X,\epsilon}}^d + \ddddot{\Sigma}_{X,Y_{X,\epsilon}}^o.
\end{align}
This decomposition, in turn, allows us to break down the error into the contribution of the diagonal and off-diagonal components:
\begin{align}
&\l\|\ddddot{\Sigma}_{X,Y_{X,\epsilon}} - \mathbb{E}_{Y_{X,\epsilon}|\Cr{good},\NN_{X,\epsilon},X}\l[
\ddddot{\Sigma}_{X,Y_{X,\epsilon}}
\r] \r\|_F \nonumber\\
& \le \l\|\ddddot{\Sigma}^d_{X,Y_{X,\epsilon}} - \mathbb{E}_{Y_{X,\epsilon}|\Cr{good},\NN_{X,\epsilon},X}\l[
\ddddot{\Sigma}^d_{X,Y_{X,\epsilon}}
\r] \r\|_F  \nonumber\\
& \qquad \qquad +
\l\|\ddddot{\Sigma}^o_{X,Y_{X,\epsilon}} - \mathbb{E}_{Y_{X,\epsilon}| \Cr{good},\NN_{X,\epsilon},X}\l[
\ddddot{\Sigma}^o_{X,Y_{X,\epsilon}}
\r] \r\|_F. 
\label{eq:decomp pre}
\end{align}
We bound the norms on the right-hand side above separately in Appendices \ref{sec:proof of lemma diagonal result} and \ref{sec:proof of chaos result}, respectively, and report the results below.

\begin{lemma} \label{lem:diagonal result}
Fix $X$ and $\epsilon\in(0,\epsilon_{\mu,X}]$. Consider the event
\begin{equation}
\Cr{Q} := \l\{ \max_{x\in X}\max_{y\in Y_{X,\epsilon}} \l\| P_{x,y}\nabla f(x) \r\|_2^2 \le \frac{Q_{X,\epsilon}\L_f^2}{n}
 \r\},
 \label{eq:eventE1def}
\end{equation}
for $Q_{X,\epsilon}>\K_\mu^{-1}$ to be set later. Then, conditioned on $\Cr{good}, \NN_{X,\epsilon}, X$, it holds that
\begin{align}
& \l\|\ddddot{\Sigma}^d_{X,Y_{X,\epsilon}} - \mathbb{E}_{Y_{X,\epsilon}|\Cr{good},\NN_{X,\epsilon},X}\l[
\ddddot{\Sigma}^d_{X,Y_{X,\epsilon}}
\r] \r\|_F\nonumber\\
&  \lesssim  \Cl[gam]{gd} \cdot  \frac{Q_{X,\epsilon} \L_f^2 n}{\sqrt{\rho_{\mu,X,\epsilon}   N_{X,\epsilon} }} + 2 n^2 L_f^2 \Pr_{Y_{X,\epsilon}|\Cr{good},\NN_{X,\epsilon},X}\l[ \Cr{Q}^C\r] ,
\end{align}
for $\Cr{gd}\ge 1$ and except with a probability of at most $e^{-\Cr{gd}}+ \Pr_{Y_{X,\epsilon}|\Cr{good},\NN_{X,\epsilon},X}[ \Cr{Q}^C]$.
\end{lemma}

\begin{lemma}\label{lem:chaos result}
Fix $X$ and $\epsilon\in(0,\epsilon_{\mu,X}]$.
Let $\widetilde{Y}_{X,\epsilon}$ contain $Y_{X,\epsilon}$ and three independent copies of it. That is, $\widetilde{Y}_{X,\epsilon}= \cup_{x\in X} \widetilde{Y}_{x,\epsilon}$, where each $\widetilde{Y}_{x,\epsilon}$ contains $Y_{x,\epsilon}$ and three independent copies of it. Consider the event $\Cr{Q}$ defined in~\eqref{eq:eventE1def} for $Q_{X,\epsilon}>\K_\mu^{-1}$ to be set later.
Consider also the event
\begin{align}
\Cl[event]{max} & :=
 \l\{ \max_{ x\in X}\max_{y\in \widetilde{Y}_{x,\epsilon}}\max_{i\in [1:n]}
\l\| P_{x,y} e_i  \r\|_2^2 \le \frac{Q_{X,\epsilon}}{n}\r\} \nonumber\\
& \qquad \qquad  \bigcap
 \l\{ \max_{ x\in X}
\max_{y\in \widetilde{Y}_{x ,\epsilon}}
\l\| P_{x,y} \nabla f(x)  \r\|_2^2 \le \frac{Q_{X,\epsilon}\L_f^2}{n}\r\}.
\end{align}
Here, $e_i\in\mathbb{R}^n$ is the $i$th canonical vector.
Assume that
\begin{equation}\label{eq:chaos event likely}
 \Pr_{\widetilde{Y}_{X,\epsilon}|\NN_{X,\epsilon},\Cr{Q},\Cr{good},X}\l[ \Cr{max}^C \r] \lesssim
 \l( \frac{\log n}{N_{X,\min,\epsilon}\rho_{\mu,X,\epsilon} N_{X,\epsilon}} \r)^{\frac{\log n}{2}},
\end{equation}
and $1 \le \log n \le N$. Then, conditioned on $\Cr{good}, \NN_{X,\epsilon}, X$, it holds that
\begin{align}
& \l\| \ddddot{\Sigma}^o_{X,Y_{X,\epsilon}} - \mathbb{E}_{Y_{X,\epsilon}|\Cr{good},\NN_{X,\epsilon},X}\l[
\ddddot {\Sigma}^o_{X,Y_{X,\epsilon}}
\r] \r\|_F \nonumber\\
& \lesssim  \Cl[gam]{g2} \cdot \sqrt{\log n} \cdot \frac{n \cdot \max[Q_{X,\epsilon},Q_{X,\epsilon}^2] \cdot \L_f^2}{\sqrt{\rho_{\mu,X,\epsilon} N_{X,\epsilon}}} + 2 n^2 L_f^2 \Pr_{Y_{X,\epsilon}|\Cr{good},\NN_{X,\epsilon},X}\l[ \Cr{Q}^C\r],
\end{align}
for $\Cr{g2}\ge1$ and except with a probability of at most $$e^{-\Cr{g2}} + n^2  \cdot n^{-\log  \Cr{g2}} + \Pr_{Y_{X,\epsilon}|\Cr{good},\NN_{X,\epsilon},X}\l[ \Cr{Q}^C\r].$$
\end{lemma}


Before we can apply Lemmas \ref{lem:diagonal result} and \ref{lem:chaos result} to the right-hand side of \eqref{eq:decomp pre}, however, we must show that the events $\Cr{Q}$ and $\Cr{max}$ are very likely to happen. Owing to Assumption \ref{def:moments}, this is indeed the case for the right choice of $Q_{X,\epsilon}$ as shown in Appendix \ref{sec:proof of lemma bnd on Q} and summarized below.

\begin{lemma}\label{lem:bnd on Q}
Fix $\epsilon\in(0,\epsilon_\mu]$ and $X$. Suppose that $Q_{X,\epsilon}=\Cr{events}\K_\mu^{-1} \log^2(N_{X,\epsilon})$ for $\Cr{events}\ge 3$. Then, conditioned on $\Cr{good}, \NN_{X,\epsilon}, X$, it holds that
\begin{equation}
\Pr_{{Y}_{X,\epsilon}|\NN_{X,\epsilon},\Cr{good},X}\l[ \Cr{Q}^C \r] \lesssim N_{X,\epsilon}^{(1-\Cr{events}\log(N_{X,\epsilon}))}.
\label{eq:E1likely}
\end{equation}
Moreover, if $1 \le \log(n) \le \log(N_{X,\epsilon}) $ and if $N_{X,\epsilon}$ is large enough such that $$\Pr_{{Y}_{X,\epsilon}|\NN_{X,\epsilon},\Cr{good},X}\l[ \Cr{Q}^C \r] \le \frac{1}{2},$$ then conditioned on $\Cr{Q}, \Cr{good}, \NN_{X,\epsilon}, X$, the requirement in \eqref{eq:chaos event likely} is satisfied.
\end{lemma}

Revisiting \eqref{eq:decomp pre}, we put all the pieces together to conclude that if $\log(n) \ge 1$, $N \ge \log(n)$, and $\log(N_{X,\epsilon}) \ge \log(n)$, then conditioned on $\Cr{good}, \NN_{X,\epsilon}, X$,
\begin{align}
&\l\|\ddddot{\Sigma}_{X,Y_{X,\epsilon}} - \mathbb{E}_{Y_{X,\epsilon}|\Cr{good},\NN_{X,\epsilon},X}\l[
\ddddot{\Sigma}_{X,Y_{X,\epsilon}}
\r] \r\|_F \nonumber\\
& \le \l\|\ddddot{\Sigma}^d_{X,Y_{X,\epsilon}} - \mathbb{E}_{Y_{X,\epsilon}|\Cr{good},\NN_{X,\epsilon},X}\l[
\ddddot{\Sigma}^d_{X,Y_{X,\epsilon}}
\r] \r\|_F \nonumber\\
& \qquad +
\l\|\ddddot{\Sigma}^o_{X,Y_{X,\epsilon}} - \mathbb{E}_{Y_{X,\epsilon}|\Cr{good},\NN_{X,\epsilon},X}\l[
\ddddot{\Sigma}^o_{X,Y_{X,\epsilon}}
\r] \r\|_F
\qquad \mbox{(see \eqref{eq:decomp pre})}
\nonumber\\
& \lesssim   \Cr{gd} \cdot  \frac{Q_{X,\epsilon} \L_f^2 n}{\sqrt{\rho_{\mu,X,\epsilon}   N_{X,\epsilon} }} + \Cr{g2} \cdot \sqrt{\log n} \cdot \frac{n \cdot \max[Q_{X,\epsilon},Q_{X,\epsilon}^2] \cdot \L_f^2}{\sqrt{\rho_{\mu,X,\epsilon} N_{X,\epsilon}}} \nonumber\\
& \qquad 
+ 4 n^2 L_f^2 \Pr_{Y_{X,\epsilon}|\Cr{good},\NN_{X,\epsilon},X}\l[ \Cr{Q}^C\r]
\qquad \mbox{(see Lemmas \ref{lem:diagonal result} and \ref{lem:chaos result})}\nonumber\\
 &
\lesssim  \Cr{g2} \cdot  \frac{n \sqrt{\log n} }{\sqrt{\rho_{\mu,X,\epsilon} N_{X,\epsilon}}} \cdot \max[Q_{X,\epsilon},Q_{X,\epsilon}^{2}]  \L_f^2 \nonumber\\
& \qquad + 4 n^2 L_f^2 N_{X,\epsilon}^{(1-\Cr{events}\log(N_{X,\epsilon}))}
 \qquad \l(\text{set} ~ \Cr{gd} = \Cr{g2} \text{; see Lemma~\ref{lem:bnd on Q}} \r)\nonumber\\
 & \lesssim
\Cr{g2} \Cr{events}^2 \log^4(N_{X,\epsilon}) \cdot  \frac{n \sqrt{\log n} }{\sqrt{\rho_{\mu,X,\epsilon} N_{X,\epsilon}}} \cdot \max[\K_\mu^{-1},\K_\mu^{-2}]  \L_f^2 \nonumber \\
& \quad + 4 n^2 L_f^2 N_{X,\epsilon}^{(1-\Cr{events}\log(N_{X,\epsilon}))}
 \qquad \l( \mbox{choice of } Q_{X,\epsilon}\mbox{ in Lemma~\ref{lem:bnd on Q}} \r)
\label{eq:conb diag n chaos cnd on e}
\end{align}
except with a probability of at most
\begin{align}
& e^{-\Cr{gd}}+ e^{-\Cr{g2}} + n^2  \cdot n^{-\log  \Cr{g2}} + 2\Pr_{Y_{X,\epsilon}|\Cr{good},\NN_{X,\epsilon},X} \l[\Cr{Q}^C \r]
\qquad \mbox{(see Lemmas \ref{lem:diagonal result} and \ref{lem:chaos result})}
\nonumber\\
& \lesssim  e^{-\Cr{gd}}+ e^{-\Cr{g2}} +  n^{2-\log \Cr{g2}} + N_{X,\epsilon}^{(1-\Cr{events}\log(N_{X,\epsilon}))} \qquad \mbox{(see Lemma~\ref{lem:bnd on Q})} \nonumber\\
& \lesssim  e^{-\Cr{g2}} + n^{2-\log \Cr{g2}} + N_{X,\epsilon}^{(1-\Cr{events}\log(N_{X,\epsilon}))}.
\qquad \l( \mbox{choice of }\Cr{gd} \mbox{ in } \eqref{eq:conb diag n chaos cnd on e}\r)
\label{eq:pre lemma 5}
\end{align}
This completes the proof of Lemma~\ref{lem:com of Sigma prime}.

\section{Proof of Lemma~\ref{lem:diagonal result}}\label{sec:proof of lemma diagonal result}

Throughout, $X$ and the neighborhood structure   $\NN_{X,\epsilon}=\{N_{x,\epsilon}\}_{x\in X}$ (see \eqref{eq:def of neighborhoods}) are  fixed. Moreover, we assume that the event $\Cr{good}$ holds (see \eqref{eq:good event def}). In addition, for $Q_{X,\epsilon}\ge \K_\mu^{-1}$ to be  set later, we condition on the following event:
\begin{equation}\label{eq:event Q def}
\Cr{Q} := \l\{ \max_{x\in X}\max_{y\in Y_{x,\epsilon}} \l\| P_{x,y}\nabla f(x) \r\|_2^2 \le \frac{Q_{X,\epsilon}\L_f^2}{n}
 \r\}.
\end{equation}
By the definition of $\ddddot{\Sigma}_{X,Y_{X,\epsilon}}^d$ in \eqref{eq:recall Sigma prime}, we observe that
\begin{align}\label{eq:Bernie pre}
& \l\|\ddddot{\Sigma}^d_{X,Y_{X,\epsilon}} - \mathbb{E}_{Y_{X,\epsilon}|\NN_{X,\epsilon},\Cr{Q},\Cr{good},X}\l[
\ddddot{\Sigma}^d_{X,Y_{X,\epsilon}}
\r] \r\|_F \nonumber\\
& = \frac{n^2}{N} \Bigg\|
\sum_{x\in X}  \sum_{y\in Y_{x,\epsilon}}
\frac{1}{N_{x,\epsilon}^2}
\Bigg( P_{x,y} \nabla f(x) \nabla f(x)^* P_{x,y} \nonumber\\
& \qquad \qquad -\E_{y|\NN_{x,\epsilon},\Cr{Q},\Cr{good},x} \l[ P_{x,y} \nabla f(x) \nabla f(x)^* P_{x,y} \r]\Bigg)
\Bigg\|_F\qquad \mbox{(see \eqref{eq:recall Sigma prime})}
\nonumber\\
& =: \frac{n^2}{N} \l\| \sum_{x\in X}\sum_{y\in Y_{x,\epsilon}} A_{x,y}\r\|_F,
\end{align}
where $\{A_{x,y}\}_{x,y}\subset\mathbb{R}^{n\times n}$ are zero-mean independent random matrices. To bound this sum, we  appeal to Proposition~\ref{prop:Bernstein recall} by computing the $b$ and $\sigma$ parameters below. For arbitrary $x\in X$ and $y\in Y_{x,\epsilon}$,  note that
\begin{align}
& \l\| A_{x,y} \r\|_F \nonumber\\
& = \frac{1}{N_{x,\epsilon}^2} \l\| P_{x,y}  \nabla f(x) \nabla f(x)^* P_{x,y} -
\E_{y|\NN_{x,\epsilon},\Cr{Q},\Cr{good},x}\l[ P_{x,y} \nabla f(x)\nabla f(x)^* P_{x,y} \r]
\r\|_F \qquad \mbox{(see \eqref{eq:Bernie pre})}\nonumber\\
& \le \frac{1}{N_{x,\epsilon}^2} \l\|P_{x,y}\nabla f(x)\nabla f(x)^* P_{x,y} \r\|_F \nonumber\\
& \qquad \qquad + \frac{1}{N_{x,\epsilon}^2} \cdot \E_{y|\NN_{x,\epsilon},\Cr{Q},\Cr{good},x}\l\| P_{x,y}\nabla f(x) \nabla f(x)^* P_{x,y}  \r\|_F \qquad \mbox{(Jensen's inequality)}
\nonumber\\
& =
\frac{1}{N_{x,\epsilon}^2} \l\|P_{x,y}\nabla f(x) \r\|_2^2 + \frac{1}{N_{x,\epsilon}^2} \cdot \E_{y|\NN_{x,\epsilon},\Cr{Q},\Cr{good},x}\l\| P_{x,y}\nabla f(x) \r\|_2^2
\nonumber\\
& \le
\frac{1}{N_{x,\epsilon}^2} \l\|P_{x,y}\nabla f(x) \r\|_2^2 + \frac{1}{N_{x,\epsilon}^2} \cdot \E_{y|\NN_{x,\epsilon},\Cr{good},x}\l\| P_{x,y}\nabla f(x) \r\|_2^2
\qquad \mbox{(see \eqref{eq:event Q def})} \nonumber\\
& \lesssim \frac{1}{\min_{x\in X}N_{x,\epsilon}^2} \cdot \max_{x\in X} \max_{y\in Y_{x,\epsilon}}\l\| P_{x,y} \nabla f(x)\r\|_2^2 \nonumber\\
& \qquad +
\frac{1}{\min_{x\in X} N_{x,\epsilon}^2} \cdot \max_{x\in X}\frac{ \l\|\nabla f(x) \r\|_2^2}{\K_\mu n}
\qquad \mbox{(see~\cite[Lemma 5.5]{vershynin2010introduction})}
\nonumber\\
& \le
\frac{1}{\min_{x\in X} N_{x,\epsilon}^2} \cdot \max_{x\in X} \max_{y\in Y_{x,\epsilon}}\l\| P_{x,y} \nabla f(x)\r\|_2^2 +
\frac{1}{\min_{x\in X} N_{x,\epsilon}^2} \cdot \frac{\L_f^2}{\K_\mu n}
\qquad \mbox{(see  \eqref{eq:Lf})}
\nonumber\\
& =: \frac{1}{\min_{x\in X} N_{x,\epsilon}^2} \cdot
\frac{Q_{X,\epsilon}\L_f^2}{n} + \frac{1}{\min_{x \in X} N_{x,\epsilon}^2}\cdot \frac{\L_f^2}{\K_\mu n}
\qquad \mbox{(see \eqref{eq:event Q def})}
\nonumber\\
& \lesssim \frac{1}{\min_{x\in X} N_{x,\epsilon}^2} \cdot
\frac{Q_{X,\epsilon}\L_f^2}{n}
\qquad \l(\mbox{when } Q_{X,\epsilon}\ge \K_\mu^{-1} \r) \nonumber\\
&
=:b.
\end{align}
On the other hand, note that
\begin{align}
& \sum_{x\in X}\sum_{y\in Y_{x,\epsilon}} \mathbb{E}_{y|\NN_{x,\epsilon},\Cr{Q},\Cr{good},x}\l\| A_{x,y} \r\|^2_F \nonumber\\
& = \sum_{x\in X} \sum_{y\in Y_{x,\epsilon}} \frac{1}{N_{x,\epsilon}^2} \nonumber\\
& \cdot \E_{y|\NN_{x,\epsilon},\Cr{Q},\Cr{good},x} \l\| P_{x,y}\nabla f(x) \nabla f(x)^* P_{x,y} - \E_{y|\NN_{x,\epsilon},\Cr{Q},\Cr{good},x} \l[P_{x,y}\nabla f(x) \nabla f(x)^* P_{x,y} \r]
  \r\|_F^2
\nonumber\\
  & \le \sum_{x\in X} \sum_{y\in Y_{x,\epsilon}} \frac{1}{N_{x,\epsilon}^2} \cdot \E_{y|\NN_{x,\epsilon},\Cr{Q},\Cr{good},x} \l\| P_{x,y}\nabla f(x) \nabla f(x)^* P_{x,y}
  \r\|_F^2
    \nonumber\\
& = \sum_{x\in X} \sum_{y\in Y_{x,\epsilon}} \frac{1}{N_{x,\epsilon}^2} \cdot \E_{y|\NN_{x,\epsilon},\Cr{Q},\Cr{good},x} \l\|P_{x,y}\nabla f(x) \r\|_2^4 \nonumber\\
& \le \sum_{x\in X} \sum_{y\in Y_{x,\epsilon}} \frac{1}{N_{x,\epsilon}^2} \cdot \E_{y|\NN_{x,\epsilon},\Cr{good},x} \l\|P_{x,y}\nabla f(x) \r\|_2^4 \qquad \mbox{(see \eqref{eq:event Q def})} \nonumber\\
&  \lesssim \sum_{x\in X} \sum_{y\in Y_{x,\epsilon}} \frac{1}{N_{x,\epsilon}^2} \cdot \frac{1}{\K_\mu^2 n^2} \cdot \max_{x\in X}\l\| \nabla f(x)\r\|_2^4 \qquad \mbox{(see~\cite[Lemma 5.5]{vershynin2010introduction})}
\nonumber\\
& \le \frac{N}{\min_{x\in X} N_{x,\epsilon}} \cdot \frac{\L_f^4}{\K_\mu^2  n^2}. \qquad \l(\# X = N,\, \# Y_{x,\epsilon}= N_{x,\epsilon},\, \mbox{see  \eqref{eq:Lf}} \r)
\nonumber \\
& =: \sigma^2.
\end{align}
where the second line uses \eqref{eq:Bernie pre}. The third line above uses the fact that $$\E\l\|Z-\E[Z]\r\|_F^2 \le \E\l\|Z \r\|_F^2,$$ for a random  matrix  $Z$.
It follows that
\begin{equation}
\max[b,\sigma] \lesssim \sqrt{\frac{N}{\min_{x\in X} N_{x,\epsilon}}} \cdot \frac{Q_{X,\epsilon}\L_f^2}{n},
\qquad \mbox{if }Q_{X,\epsilon}\ge \K_\mu^{-1}.
\label{eq:Bernie max diag}
\end{equation}
Thus, in light of Proposition~\ref{prop:Bernstein recall}, and conditioned on $\Cr{Q},\Cr{good},\NN_{X,\epsilon},X$, it follows that
\begin{align}\label{eq:diag cnd on Q}
& \l\| \ddddot{\Sigma}^d_{X,Y_{X,\epsilon}} - \mathbb{E}_{Y_{X,\epsilon}|\Cr{Q},\NN_{X,\epsilon},\Cr{good},X}\l[
\ddddot{\Sigma}^d_{X,Y_{X,\epsilon}}
\r] \r\|_F \nonumber\\
& = \frac{n^2}{N} \l\| \sum_{x\in X}\sum_{y\in Y_{x,\epsilon}} A_{x,y}\r\|_F
\qquad \mbox{(see \eqref{eq:Bernie pre})}
\nonumber\\
& \lesssim \frac{n^2}{N}\cdot \Cr{gd} \cdot \max[b,\sigma]\nonumber\\
& \lesssim \frac{n^2}{N} \cdot  \Cr{gd} \sqrt{\frac{N}{\min_x N_{x,\epsilon}}} \cdot \frac{Q_{X,\epsilon} \L_f^2}{n}
\qquad \mbox{(see \eqref{eq:Bernie max diag})}\nonumber\\
& = \Cr{gd} \cdot  \frac{n}{\sqrt{N \cdot  \min_x N_{x,\epsilon}}} \cdot Q_{X,\epsilon} \L_f^2
\nonumber\\
& \lesssim
\Cr{gd} \cdot  \frac{n}{\sqrt{N \cdot  \min_x \frac{\mu\l(\B_{x,\epsilon}\r)}{\mu\l(\B_{X,\epsilon}\r)} \cdot N_{X,\epsilon}}} \cdot Q_{X,\epsilon} \L_f^2
\qquad \mbox{(see \eqref{eq:good event def})} \nonumber\\
& = \Cr{gd} \cdot  \frac{n}{\sqrt{\rho_{\mu,X,\epsilon}  N_{X,\epsilon}}} \cdot Q_{X,\epsilon} \L_f^2,
\qquad \mbox{(see \eqref{eq:def of rhoX thm})}
\end{align}
for $\Cr{gd}\ge1$ and except with a probability of at most $e^{-\Cr{gd}}$.
Before we can remove the conditioning on the event $\Cr{Q}$, we use the law of total expectation to write
\begin{align*}
&\mathbb{E}_{Y_{X,\epsilon}|\NN_{X,\epsilon},\Cr{good},X}\l[\ddddot{\Sigma}^d_{X,Y_{X,\epsilon}}\r] \\
&= \mathbb{E}_{Y_{X,\epsilon}|\Cr{Q},\NN_{X,\epsilon},\Cr{good},X}\l[\ddddot{\Sigma}^d_{X,Y_{X,\epsilon}}\r] \Pr_{Y_{X,\epsilon}|\Cr{good},\NN_{X,\epsilon},X}\l[ \Cr{Q} \r] \nonumber\\
& \qquad \qquad +
\mathbb{E}_{Y_{X,\epsilon}|\Cr{Q}^C, \NN_{X,\epsilon},\Cr{good},X}\l[\ddddot{\Sigma}^d_{X,Y_{X,\epsilon}}\r] \Pr_{Y_{X,\epsilon}|\Cr{good},\NN_{X,\epsilon},X}\l[ \Cr{Q}^C\r],
\end{align*}
from which it follows that
\begin{align}\label{eq:totalexp}
& \mathbb{E}_{Y_{X,\epsilon}|\NN_{X,\epsilon},\Cr{good},X}\l[\ddddot{\Sigma}^d_{X,Y_{X,\epsilon}}\r] - \mathbb{E}_{Y_{X,\epsilon}|\Cr{Q},\NN_{X,\epsilon},\Cr{good},X}\l[\ddddot{\Sigma}^d_{X,Y_{X,\epsilon}}\r] \nonumber \\
&= \Pr_{Y_{X,\epsilon}|\Cr{good},\NN_{X,\epsilon},X}\l[ \Cr{Q}^C\r] \left( \mathbb{E}_{Y_{X,\epsilon}|\Cr{Q}^C, \NN_{X,\epsilon},\Cr{good},X}\l[\ddddot{\Sigma}^d_{X,Y_{X,\epsilon}}\r] - \mathbb{E}_{Y_{X,\epsilon}|\Cr{Q},\NN_{X,\epsilon},\Cr{good},X}\l[\ddddot{\Sigma}^d_{X,Y_{X,\epsilon}}\r] \right).
\end{align}
Since for any $X$, $Y_{X,\epsilon}$, we have
\begin{align}\label{eq:totalexp1}
\l\| \ddddot{\Sigma}^d_{X,Y_{X,\epsilon}} \r\|_F
& \le \frac{n^2}{N} \sum_{x \in X} \sum_{y \in Y_{Xx,\epsilon}} \frac{1}{N_{x,\epsilon}^2} \l\| P_{x,y} \nabla f(x) \nabla f(x)^* P_{x,y} \r\|_F \qquad \mbox{(see \eqref{eq:recall Sigma prime})} \nonumber \\
& \le \frac{n^2}{N} \sum_{x \in X} \sum_{y \in Y_{Xx,\epsilon}} \frac{1}{N_{x,\epsilon}^2} \l\| P_{x,y} \nabla f(x) \r\|_2^2 \nonumber \\
& \le \frac{n^2}{N} \sum_{x \in X} \sum_{y \in Y_{Xx,\epsilon}} \frac{1}{N_{x,\epsilon}^2} \l\| \nabla f(x) \r\|_2^2 \nonumber \\
& \le \frac{n^2}{N} \sum_{x \in X} \sum_{y \in Y_{Xx,\epsilon}} \frac{1}{N_{x,\epsilon}^2} L_f^2 \qquad \mbox{(see  \eqref{eq:Lf})} \nonumber \\
& = \frac{n^2}{N} \sum_{x \in X} \frac{1}{N_{x,\epsilon}} L_f^2 \nonumber \\
& \le n^2 L_f^2,
\end{align}
we conclude that
\begin{align}\label{eq:totalexp2}
& \l\| \mathbb{E}_{Y_{X,\epsilon}|\NN_{X,\epsilon},\Cr{good},X}\l[\ddddot{\Sigma}^d_{X,Y_{X,\epsilon}}\r] - \mathbb{E}_{Y_{X,\epsilon}|\Cr{Q},\NN_{X,\epsilon},\Cr{good},X}\l[\ddddot{\Sigma}^d_{X,Y_{X,\epsilon}}\r] \r\|_F \nonumber \\
& \le \Pr_{Y_{X,\epsilon}|\Cr{good},\NN_{X,\epsilon},X} \l[ \Cr{Q}^C\r]\nonumber\\
&\,\,\,  \cdot  \l\| \mathbb{E}_{Y_{X,\epsilon}|\Cr{Q}^C, \NN_{X,\epsilon},\Cr{good},X}\l[\ddddot{\Sigma}^d_{X,Y_{X,\epsilon}}\r] - \mathbb{E}_{Y_{X,\epsilon}|\Cr{Q},\NN_{X,\epsilon},\Cr{good},X}\l[\ddddot{\Sigma}^d_{X,Y_{X,\epsilon}}\r] \r\|_F \qquad \mbox{(see  \eqref{eq:totalexp})} \nonumber \\
&  \le 2 n^2 L_f^2 \Pr_{Y_{X,\epsilon}|\Cr{good},\NN_{X,\epsilon},X}\l[ \Cr{Q}^C\r]. \qquad \mbox{(triangle inequality and \eqref{eq:totalexp1})}
\end{align}
Lastly, we remove the conditioning on the event $\Cr{Q}$ as follows:
\begin{align}
& \Pr_{Y_{X,\epsilon}|\Cr{good},\NN_{X,\epsilon},X}\Bigg[ \l\| \ddddot{\Sigma}^d_{X,Y_{X,\epsilon}} - \mathbb{E}_{Y_{X,\epsilon}|\Cr{good},\NN_{X,\epsilon},X}\l[
\ddddot {\Sigma}^d_{X,Y_{X,\epsilon}}
\r] \r\|_F
\nonumber\\
& \qquad \qquad \qquad \qquad 
 \gtrsim \Cr{gd} \cdot  \frac{ Q_{X,\epsilon} \L_f^2 n}{\sqrt{\rho_{\mu,X,\epsilon} N_{X,\epsilon}}} + 2 n^2 L_f^2 \Pr_{Y_{X,\epsilon}|\Cr{good},\NN_{X,\epsilon},X}\l[ \Cr{Q}^C\r]
\Bigg]
\nonumber\\
& \le  \Pr_{Y_{X,\epsilon}|\Cr{Q},\Cr{good},\NN_{X,\epsilon},X}\Bigg[ \l\| \ddddot{\Sigma}^d_{X,Y_{X,\epsilon}} - \mathbb{E}_{Y_{X,\epsilon}|\Cr{good},\NN_{X,\epsilon},X}\l[
\ddddot{\Sigma}^d_{X,Y_{X,\epsilon}}
\r] \r\|_F \nonumber\\
& \qquad \qquad \qquad \qquad \quad \gtrsim \Cr{gd} \cdot  \frac{ Q_{X,\epsilon} \L_f^2 n}{\sqrt{\rho_{\mu,X,\epsilon} N_{X,\epsilon}}} + 2 n^2 L_f^2 \Pr_{Y_{X,\epsilon}|\Cr{good},\NN_{X,\epsilon},X}\l[ \Cr{Q}^C\r]
\Bigg] \nonumber \\
& \qquad \qquad + \Pr_{Y_{X,\epsilon}|\Cr{good},\NN_{X,\epsilon},X}\l[ \Cr{Q}^C\r] \qquad
\mbox{(see \eqref{eq:useful ineq})}
\nonumber\\
&
\le \Pr_{Y_{X,\epsilon}|\Cr{Q},\Cr{good},\NN_{X,\epsilon},X}\Bigg[ \l\| \ddddot{\Sigma}^d_{X,Y_{X,\epsilon}} - \mathbb{E}_{Y_{X,\epsilon}|\Cr{Q},\Cr{good},\NN_{X,\epsilon},X}\l[
\ddddot{\Sigma}^d_{X,Y_{X,\epsilon}}
\r] \r\|_F \nonumber\\
& \qquad \qquad \qquad \qquad \quad \gtrsim \Cr{gd} \cdot  \frac{ Q_{X,\epsilon} \L_f^2 n}{\sqrt{\rho_{\mu,X,\epsilon} N_{X,\epsilon}}}
\Bigg] \nonumber \\
& \qquad \qquad + \Pr_{Y_{X,\epsilon}|\Cr{good},\NN_{X,\epsilon},X}\l[ \Cr{Q}^C\r] \qquad \mbox{(see \eqref{eq:totalexp2})}
 \nonumber\\
& \le e^{-\Cr{gd}}+ \Pr_{Y_{X,\epsilon}|\Cr{good},\NN_{X,\epsilon},X}\l[ \Cr{Q}^C\r]. \qquad \mbox{(see \eqref{eq:diag cnd on Q})}
\end{align}
The proof of Lemma~\ref{lem:diagonal result} is now complete.

\section{Proof of Lemma~\ref{lem:chaos result}}
\label{sec:proof of chaos result}

Throughout, $X$ and the neighborhood structure   $\NN_{X,\epsilon}=\{N_{x,\epsilon}\}_{x\in X}$ (see \eqref{eq:def of neighborhoods}) are  fixed. Moreover, we assume that the event $ \Cr{good}$ holds (see \eqref{eq:good event def}). In addition, for $Q_{X,\epsilon}\ge \K_\mu^{-1}$ to be  set later, we condition on $\Cr{Q}$ as defined in~\eqref{eq:event Q def}.

Let us index $X$ as $X=\{x_s\}_{s=1}^{N}$. For each $x_s\in X$, we index its neighbors  $Y_{x_s,\epsilon}$ as $Y_{x_s,\epsilon}=\{y_{sk}\}_{k=1}^{N_{x_s,\epsilon}}$, where $N_{x_s,\epsilon}=\# Y_{x_s,\epsilon}$ is the number of neighbors of $x_s$ (within radius of $\epsilon$).  Recalling the definition of $\ddddot{\Sigma}^o_{X,Y_{X,\epsilon}}$ from  \eqref{eq:recall Sigma prime}, we aim to find an upper bound for 
\begin{align}\label{eq:decompose err new}
&\left\|
\ddddot{\Sigma}_{X,Y_{X,\epsilon}}^o-
\mathbb{E}_{Y_{X,\epsilon}|\NN_{X,\epsilon},\Cr{Q},\Cr{good},X} \left[ \ddddot{\Sigma}_{X,Y_{X,\epsilon}}^o \right]
\right\|_F\nonumber\\
&  =\frac{n^2}{N} \Bigg\| \sum_{s=1}^{N}\sum_{k\ne l}
\frac{1}{N_{x_s,\epsilon}^2}
\Bigg( P_{x_s,y_{sk}}\nabla f(x_s) \nabla f(x_s)^* P_{x_s,y_{sl}} \nonumber\\
& \qquad \qquad \qquad \qquad \qquad 
-
\mathbb{E}_{y_s|\Cr{Q},x_s}\l[P_{x_s,y_s} \r] \nabla f(x_s) \nabla f(x_s)^* \mathbb{E}_{y_s|\Cr{Q},x_s}\l[P_{x_s,y_s} \r]
\Bigg)
\Bigg\|_F
\nonumber\\
& \le \frac{n^2}{N} \Bigg\| \sum_{s=1}^N \sum_{k\ne l} \frac{1}{N_{x_s,\epsilon}^2} \nonumber\\
& \qquad \,\, \,\,\,\cdot 
\l(P_{x_s,y_{sk}} - \E_{y_s|\Cr{Q},x_s}[P_{x_s,y_s}] \r) \nabla f(x_s)\nabla f(x_s)^* \l(P_{x_s,y_{sl}} - \E_{y_s|\Cr{Q},x_s}[P_{x_s,y_s}] \r) \Bigg\|_F  \nonumber\\
& +
 \frac{n^2}{N} \l\|
 \sum_{s=1}^N \sum_{k\ne l}\frac{1}{N_{x_s,\epsilon}^2} \l(P_{x_s,y_{sk}} - \E_{y_s|\Cr{Q},x_s}[P_{x_s,y_s}] \r) \nabla f(x_s)\nabla f(x_s)^*  \E_{y_s|\Cr{Q},x_s}[P_{x_s,y_s}]
  \r\|_F \nonumber\\
&   + \frac{n^2}{N} \l\| \sum_{s=1}^N \sum_{k\ne l} \frac{1}{N_{x_s,\epsilon}^2} \E_{y_s|\Cr{Q},x_s}[P_{x_s,y_s}]  \nabla f(x_s)\nabla f(x_s)^* \l(P_{x_s,y_{sl}} - \E_{y_s|\Cr{Q},x_s}[P_{x_s,y_s}] \r) \r\|_F,
\end{align}
after which we will remove the conditioning on $\Cr{Q}$. Above, $y_s|\Cr{Q},x_s$ is distributed according to the restriction of $\mu_{x_s,\epsilon}$ to the event $\Cr{Q}$. In the following subsections, we separately bound each of the three norms in the last line above.

\subsection{First norm \label{sec:first in decompose}}
In this section, we bound the first norm in the last line of \eqref{eq:decompose err new} by writing it as a chaos random variable. Let us first write that
\begin{align}\label{eq:decompose err}
&  \frac{n^2}{N} \Bigg\| \sum_{s=1}^{N}\sum_{k\ne l}
\frac{1}{N_{x_s,\epsilon}^2} \nonumber\\
& \qquad \cdot \left( P_{x_s,y_{sk}} - \E_{y_s|\Cr{Q},x_s}[P_{x_s,y_s}] \r) \nabla f(x_s) \nabla f(x_s)^* \left( P_{x_s,y_{sl}} - \E_{y_s|\Cr{Q},x_s}[P_{x_s,y_s}] \r) \Bigg\|_F
\nonumber\\
& =: \frac{n^2}{N} \l\|\sum_{s=1}^N \sum_{k,l=1}^{N_{x_s,\epsilon}} A_{skl} \r\|_F\nonumber\\
& = \frac{n^2}{N} \sqrt{ \sum_{i,j=1}^n\left|
\sum_{s=1}^{N}\sum_{k,l=1}^{N_{x_s}}  A_{skl} [i,j]\right|^2}.
\end{align}
Above, we also conveniently defined  the matrices $\{A_{skl}\}_{s,k,l}\subset \mathbb{R}^{n\times n}$ as
\begin{align}\label{eq:def of As}
A_{skl} & :=
\frac{1}{N_{x_s,\epsilon}^2}
\begin{cases}
\l( P_{x_s,y_{sk}}  - \E_{y_s|\Cr{Q},x_s}[P_{x_s,y_s}]\r) \nabla f(x_s) \nabla f(x_s)^*\\ 
\qquad \qquad \cdot \l( P_{x_s,y_{sl}}  - \E_{y_s|\Cr{Q},x_s}[P_{x_s,y_s}]\r) ,
& k \ne l, \\
0, & k=l,
\end{cases}
\end{align}
for every $s\in[1:N]$ and $k,l\in[1:N_{x_s,\epsilon}]$.
By their definition above, the random matrices $\{A_{skl}\}_{s,k,l}$ enjoy the following properties:
\begin{equation}\label{eq:props of As}
A_{skk} = 0,\qquad \E_{Y_{X,\epsilon}|\NN_{X,\epsilon},\Cr{Q},\Cr{good},X} \left[ A_{skl}\right]=0,\qquad   s\in[1:N], \,\,k,l\in[1:N_{x_s,\epsilon}].
\end{equation}
With fixed $s\in[1:N]$ and $i,j\in [1:n]$, we may use $\{A_{skl}[i,j]\}_{k,l}$ to form a new  matrix $A_{sij}$ as
\begin{equation}
A_{sij} := \left[ A_{skl}[i,j]\right]_{k,l} \in\R^{N_{x_s,\epsilon}\times N_{x_s,\epsilon}},\nonumber
\end{equation}
or, equivalently,
\begin{equation}\label{eq:reorder 1}
A_{sij}[k,l]:=A_{skl}[i,j],\qquad  k,l\in [1:N_{x_s,\epsilon}].
\end{equation}
Let $A_{ij}$ be the block-diagonal matrix formed from $\{A_{sij}\}_{s}\subset\R^{N_{x_s,\epsilon}\times N_{x_s,\epsilon}}$, i.e.,
\begin{equation}\label{eq:def of A_ij}
A_{ij} =
\left[
\begin{array}{cccc}
A_{1ij} & & & \\
& A_{2ij} & & \\
& & \ddots & \\
& & & A_{Nij}
\end{array}
\right] \in \R^{N_{X,\epsilon}\times N_{X,\epsilon} }.
\end{equation}
where we used the fact that $N_{X,\epsilon}=\sum_{s=1}^N N_{x_s,\epsilon}$ to calculate the dimensions of $A_{ij}$. In particular,
 \eqref{eq:props of As} implies  that
\begin{equation*}
A_{ij}[sk,sk]=0,
\end{equation*}
\begin{equation}\label{eq:props of A_i,j}
\E_{Y_{X,\epsilon}|\NN_{X,\epsilon},\Cr{Q},\Cr{good},X}\left[ A_{ij}[sk,tl]\right]=0,\qquad  s,t\in [1:N],\,\,   k\in [1:N_{x_s,\epsilon}],\,\, l\in[1:N_{x_t,\epsilon}],
\end{equation}
where, ignoring the standard convention, we indexed the entries of $A_{ij}$ so that $sk$ corresponds to the $k$th row of the $s$th block (and hence does not stand for the product of $s$ and $k$).  With this new notation, we revisit  \eqref{eq:decompose err} to write that
\begin{align}\label{eq:s20}
&
\frac{n^2}{N} \Bigg\| \sum_{s=1}^{N}\sum_{k\ne l}
\frac{1}{N_{x_s,\epsilon}^2}
\left( P_{x_s,y_{sk}} - \E_{y_s|\Cr{Q},x_s}[P_{x_s,y_s}] \r) \nabla f(x_s) \nabla f(x_s)^* \nonumber\\
& \qquad \qquad \cdot 
\left( P_{x_s,y_{sl}} - \E_{y_s|\Cr{Q},x_s}[P_{x_s,y_s}] \r) \Bigg\|_F \nonumber\\
 & =
 \frac{n^2}{N}
 \sqrt{\sum_{i,j=1}^n\left|
\sum_{s,t=1}^{N}\sum_{k=1}^{N_{x_s,\epsilon}} \sum_{l=1}^{N_{x_t,\epsilon}} A_{ij} [sk,tl]\right|^2}
\qquad \mbox{(see \eqref{eq:decompose err})}
\nonumber\\
& =:  \frac{n^2}{N} \sqrt{\sum_{i,j=1}^n a_{ij}^2}.
\end{align}
For fixed $i,j\in [1:n]$, let us next focus on the random variable $a_{ij}$.

\subsubsection{Tail Bound for $a_{ij}$}
Recall that the $p$th moment of a random variable $z$ is defined as $\E^p[z] := (\E[|z|^p])^{\frac{1}{p}}$.  Fix  $i,j\in[1:n]$. In order to bound $a_{ij}$, we
\begin{itemize}
\item First control its moments, namely
\begin{equation}
\E^p_{Y_{X,\epsilon}|\NN_{X,\epsilon},\Cr{Q},\Cr{good},X}\left[ a_{ij}\right]=\E^p_{Y_{X,\epsilon}|\NN_{X,\epsilon},\Cr{Q},\Cr{good},X}\left| \sum_{s,t=1}^{N}\sum_{k=1}^{N_{x_s,\epsilon}}\sum_{l=1}^{N_{x_t,\epsilon}} A_{ij}[sk,tl] \right|,\qquad \forall p\ge 1.
\end{equation}
\item Second we use  Markov's inequality to find a tail bound for $a_{ij}$ (given its moments).
\end{itemize}
Each step is discussed in a separate subsection below.

\subsubsection{Moments of $a_{ij}$}
In order to control the moments of $a_{ij}$, we take the following steps:
\begin{itemize}
\item \emph{symmetrization},
\item \emph{decoupling},
\item \emph{modulation} with  \emph{Rademacher sequences}, and finally
\item bounding the moments of the resulting \emph{decoupled chaos} random variable.
\end{itemize}
Each of these steps is detailed in a separate paragraph below.

\paragraph{Symmetrization}\label{sec:symm}
To control the moments of $a_{ij}$, we first use a symmetrization argument  as follows. With $s\in[1:N]$ and conditioned on $x_s$ and $\Cr{Q}$, let $Y_{x_s,\epsilon}^{\operatorname{i}}\in\R^{n\times N_{x_s,\epsilon}}$ be an independent copy of $Y_{x_s,\epsilon}$. Then note that
\begin{align}
& \E^p_{Y_{X,\epsilon}|\overline{N}_{X,\epsilon},\Cr{Q},\Cr{good},X} [a_{ij}] \nonumber\\
& =
\E^p_{Y_{X,\epsilon}|\overline{N}_{X,\epsilon},\Cr{Q},\Cr{good},X} \l[\sum_{s,t=1}^N \sum_{k=1}^{N_{x_s,\epsilon}}\sum_{l=1}^{N_{x_t,\epsilon}} A_{ij}[sk,tl] \r] \nonumber\\
& = \E^p_{Y_{X,\epsilon}|\overline{N}_{X,\epsilon},\Cr{Q},\Cr{good},X} \Bigg[ \sum_{s=1}^N \sum_{k\ne l} \frac{1}{N_{x_s,\epsilon}^2} e_i^* \l(P_{x_s,y_{sk}} -\E_{y_s|\Cr{Q},x_s}[P_{y_s|\Cr{Q},x_s}]\r) \nabla f(x_s)\nabla f(x_s)^* \nonumber\\
&\qquad\qquad \qquad \qquad \qquad   \cdot \l(P_{x_s,y_{sl}} -\E_{y_s|\Cr{Q},x_s}[P_{y_s|\Cr{Q},x_s}]\r) e_j \Bigg]\nonumber\\
& = \E^p_{Y_{X,\epsilon}|\overline{N}_{X,\epsilon},\Cr{Q},\Cr{good},X} \Bigg[ \sum_{s=1}^N \sum_{k\ne l} \frac{1}{N_{x_s,\epsilon}^2} e_i^* \l(P_{x_s,y_{sk}}-\E_{y^{\i}_{sk}|\Cr{Q},x_s}[P_{x_s, y^{\i}_{sk}}]\r) \nabla f(x_s)\nabla f(x_s)^* \nonumber\\
& \qquad \qquad \qquad \qquad \qquad \cdot \l(P_{x_s,y_{sl}} -\E_{y_{s,l}^{\i}|\Cr{Q},x_s}[P_{x_s,y_{s,l}^{\i}}]\r) e_j \Bigg] \nonumber\\
& = \E^p_{Y_{X,\epsilon}|\overline{N}_{X,\epsilon},\Cr{Q},\Cr{good},X} \Bigg[ \E_{Y^{\i}_{X,\epsilon}|\overline{N}_{X,\epsilon},\Cr{Q},\Cr{good},X} \sum_{s=1}^N \sum_{k\ne l} \frac{1}{N_{x_s,\epsilon}^2} e_i^* \l(P_{x_s,y_{sk}}-P_{x_s, y^{\i}_{sk}}\r) \nonumber\\
& \qquad \qquad \qquad \qquad \qquad \cdot \nabla f(x_s) \nabla f(x_s)^* 
 \l(P_{x_s,y_{sl}} -P_{x_s,y_{s,l}^{\i}}\r) e_j \Bigg]  \qquad \mbox{(independence)}
\nonumber\\
& \le \E^p_{Y_{X,\epsilon},Y_{X,\epsilon}^{\i}|\overline{N}_{X,\epsilon},\Cr{Q},\Cr{good},X}
\Bigg[ \sum_{s=1}^N \sum_{k\ne l} \frac{1}{N_{x_s,\epsilon}^2} e_i^* \l(P_{x_s,y_{sk}}-P_{x_s, y^{\i}_{sk}}\r) \nabla f(x_s) \nabla f(x_s)^*
\nonumber\\
& \qquad \qquad \qquad \qquad \qquad \cdot
 \l(P_{x_s,y_{sl}} -P_{x_s,y_{s,l}^{\i}} \r) e_j \Bigg]
\qquad \mbox{(Jensen's inequality)}
\nonumber\\
& = \E^p_{Y_{X,\epsilon},Y_{X,\epsilon}^{\i}|\overline{N}_{X,\epsilon},\Cr{Q},\Cr{good},X}
\l[\sum_{s,t=1}^N \sum_{k=1}^{N_{x_s,\epsilon}}\sum_{l=1}^{N_{x_t,\epsilon}} B_{ij}[sk,tl] \r],
\label{eq:s10}
\end{align}
where we defined the block-diagonal matrix $B_{ij}\in\R^{N_{X,\epsilon}\times N_{X,\epsilon}}$ such that
\begin{align}\label{eq:def of B}
& B_{ij}[sk,tl]\nonumber\\
& =\begin{cases}
 N_{x_s,\epsilon}^{-2}\cdot  e_i^*
 \l( P_{x_s,y_{sk}} - P_{x_s,y_{sk}^{\i}}\r) \nabla f(x_s) \nabla f(x_s)^*\\
\qquad \qquad \cdot  \l( P_{x_s,y_{sl}}-P_{x_s,y_{sl}^{\i}} \r)
e_j
, & s=t \mbox{ and }k\ne l,\\
 0,& s\ne t\mbox{ or }k=l,
\end{cases}
\end{align}
for every $s,t\in[1:N]$, $k\in [1:N_{x_s,\epsilon}]$, $l\in[1:N_{x_t,\epsilon}]$.
Above, $e_i\in\R^n$ is the $i$th coordinate vector.
Note that, by construction,  each $B_{ij}[sk,tl]$ is a \emph{symmetric random variable} (in the sense that its distribution is symmetric about the origin). Moreover, similar to  \eqref{eq:props of A_i,j}, it holds that
\begin{equation*}
B_{ij}[sk,sk]= 0,
\end{equation*}
\begin{equation}
\E_{Y_{X,\epsilon},Y^{\i}_{X,\epsilon}|\NN_{X,\epsilon},\Cr{Q},\Cr{good},X}\left[B_{ij}[sk,tl] \right]=0,
\qquad  s,t\in[1:N],\,\,  k\in[1:N_{x_s,\epsilon}],\,\, l\in[1:N_{x_t,\epsilon}].
\end{equation}
Our next step is to decouple the sum in the last line of \eqref{eq:s10}.

\paragraph{Decoupling}\label{sec:decouple}

Let $\Xi = \{\xi_{sk} \}_{s,k}$ (with $s\in[1:N]$ and $k\in[1:N_{x_s,\epsilon}]$) be a sequence of independent standard Bernoulli random variables: each $\xi_{sk}$ independently takes one and zero with equal probabilities. We will shortly use the following simple observation:
\begin{equation}\label{eq:s11}
\E_{\Xi}\left[ \xi_{sk} \left(1-\xi_{tl}\right)\right] = \frac{1}{4},\qquad sk\ne tl.
\end{equation}
We now revisit \eqref{eq:s10} and write  that
\begin{align}
& \E^p_{Y_{X,\epsilon}|\NN_{X,\epsilon},\Cr{Q},\Cr{good},X} \left[ a_{ij}\right]
\nonumber\\
& \le  \E^p_{Y_{X,\epsilon},Y_{X,\epsilon}^{\operatorname{i}}|\NN_{X,\epsilon},\Cr{Q},\Cr{good},X}\left[ \sum_{s,t=1}^{N} \sum_{k=1}^{N_{x_s,\epsilon}} \sum_{l=1}^{N_{x_t,\epsilon}} B_{ij}[sk,tl] \right]\qquad \mbox{(see \eqref{eq:s10})}\nonumber\\
& = 4\cdot \E^p_{Y_{X,\epsilon},Y_{X,\epsilon}^{\operatorname{i}}|\NN_{X,\epsilon},\Cr{Q},\Cr{good},X}\left[
\sum_{s,t=1}^{N} \sum_{k=1}^{N_{x_s,\epsilon}} \sum_{l=1}^{N_{x_t,\epsilon}} \E_{\Xi} \Bigg[ \xi_{sk} \left( 1-\xi_{tl}\right)\right] \cdot B_{ij}[sk,tl] \Bigg]\nonumber\\
& \qquad \qquad \qquad \qquad \qquad 
\qquad \left( B_{ij}[sk,sk]=0,   \mbox{ and  \eqref{eq:s11}}\right)\nonumber\\
& \le 4\cdot \E^p_{Y_{X,\epsilon},Y_{X,\epsilon}^{\operatorname{i}},\Xi|\NN_{X,\epsilon},\Cr{Q},\Cr{good},X}\left[ \sum_{s,t=1}^{N}\sum_{k=1}^{N_{x_s,\epsilon}}\sum_{l=1}^{N_{x_t,\epsilon}} \xi_{sk} \left( 1-\xi_{tl}\right) \cdot B_{ij}[sk,tl] \right],
\end{align}
where the last line above uses the Jensen's inequality. 
In particular, there must exist $\Xi_0=\{\xi_{0sk}\}_{s,k}$ that exceeds the expectation in the last line above, so that
\begin{align}\label{eq:s12}
&\E^p_{Y_{X,\epsilon}|\NN_{X,\epsilon},\Cr{Q},\Cr{good},X}\left[ a_{ij}\right]
\nonumber\\
& \le 4\cdot \E^p_{Y_{X,\epsilon},Y_{X,\epsilon}^{\operatorname{i}}|\NN_{X,\epsilon},\Cr{Q},\Cr{good},X}\left[\sum_{s,t=1}^{N} \sum_{k=1}^{N_{x_s,\epsilon}} \sum_{l=1}^{N_{x_t,\epsilon}} \xi_{0sk} \left( 1-\xi_{0tl}\right) \cdot B_{ij}[sk,tl] \right]\nonumber\\
& = 4\cdot \E^p_{Y_{X,\epsilon},Y_{X,\epsilon}^{\operatorname{i}}|\NN_{X,\epsilon},\Cr{Q},\Cr{good},X}\left[ \sum_{\xi_{0sk}=1,\, \xi_{0tl}=0}   B_{ij}[sk,tl] \right]\nonumber\\
& = 4\cdot \E^p_{Y_{X,\epsilon},Y_{X,\epsilon}^{\operatorname{i}}|\NN_{X,\epsilon},\Cr{Q},\Cr{good},X}\left[ \sum_{sk\in S_0 ,\, tl\notin S_0}   B_{ij}[sk,tl] \right].\nonumber\\
& \qquad \qquad \qquad 
\qquad \left( S_0:=\left\{sk\,:\, \xi_{0sk}=1\right\}\subseteq [1:N_{X,\epsilon}]\right)
\end{align}
Let $\{Y_{X,\epsilon}^{\i\i},Y_{X,\epsilon}^{\i\i\i}\}\subset\R^{n\times \# N_{X,\epsilon}}$ be an independent copy of $\{Y_{X,\epsilon},Y_{X,\epsilon}^{\operatorname{i}}\}$.
For the sake of brevity, we will use the following short hand:
\begin{equation*}
\widetilde{Y}_{X,\epsilon} := Y_{X,\epsilon} \cup Y_{X,\epsilon}^{\i} \cup Y_{X,\epsilon}^{\i\i} \cup Y_{X,\epsilon}^{\i\i\i},
\end{equation*}
\begin{equation}\label{eq:short hand}
\widetilde{Y}_{x_s,\epsilon} :=  Y_{x_s,\epsilon} \cup Y_{x_s,\epsilon}^{\i} \cup Y_{x_s,\epsilon}^{\i\i} \cup Y_{x_s,\epsilon}^{\i\i\i},\qquad \forall x_s\in X.
\end{equation}
Equipped with the construction above, we revisit \eqref{eq:s12} and write that
\begin{align}\label{eq:s21}
& \E^p_{Y_{X,\epsilon}|\NN_{X,\epsilon},\Cr{Q},\Cr{good},X}\left[ a_{ij}\right]  \nonumber\\
&
\le 4\cdot \E^p_{Y_{X,\epsilon},Y_{X,\epsilon}^{\operatorname{i}}|\NN_{X,\epsilon},\Cr{Q},\Cr{good},X}\left[ \sum_{sk\in S_0 ,\, tl\notin S_0}   B_{ij}[sk,tl] \right] \qquad \mbox{(see \eqref{eq:s12})}\nonumber\\
& = 4\cdot \E^p_{Y_{X,\epsilon},Y_{X,\epsilon}^{\operatorname{i}}|\NN_{X,\epsilon},\Cr{Q},\Cr{good},X}
\Bigg[ \sum_{sk\in S_0 ,\, sl\notin S_0}
N_{x_s,\epsilon}^{-2}\cdot  e_i^* \l(P_{x_s,y_{sk}} -P_{x_s,y_{sk}^{\i}} \r)\nonumber\\
& \qquad \qquad \qquad \qquad   \cdot
  \nabla f(x_s) \nabla f(x_s)^* \l(P_{x_s,y_{sl}} -P_{x_s,y_{sl}^{\i}} \r) e_j \Bigg]
 \qquad 
\mbox{(see \eqref{eq:def of B})}\nonumber\\
& = 4\cdot \E^p_{\widetilde{Y}_{X,\epsilon}|\NN_{X,\epsilon},\Cr{Q},\Cr{good},X} \Bigg[ \sum_{sk\in S_0 ,\, sl\notin S_0} \nonumber\\
& \qquad  
\underset{C_{ij}[sk,sl]}
{\underbrace{
N_{x_s,\epsilon}^{-2}\cdot  e_i^* \l(P_{x_s,y_{sk}} -P_{x_s,y_{sk}^{\i}} \r) \nabla f(x_s) \nabla f(x_s)^* \l(P_{x_s,y_{sl}^{\i\i}} -P_{x_s,y_{sl}^{\i\i\i}} \r) e_j
}} \Bigg] \nonumber\\
& \qquad 
\mbox{(independence)}\nonumber\\
& = 4\cdot \E^p_{\widetilde{Y}_{X,\epsilon}|\NN_{X,\epsilon},\Cr{Q},\Cr{good},X}
\Bigg[ \sum_{sk\in S_0 ,\, sl\notin S_0}   C_{ij}[sk,sl]
\nonumber\\
& \qquad \qquad + \sum_{sk\notin S_0}  \E_{{Y}_{X,S_0^C},{Y}_{X,S_0^C}^{\i}|{Y}_{X}^{\i\i},{Y}_{X}^{\i\i\i},\NN_{X,\epsilon},\Cr{Q},\Cr{good},X}\left[ C_{ij}[sk,sl]  \right] 
\nonumber\\
& \qquad\qquad  +    \sum_{sk\in S_0,\, sl\in S_0} \E_{{Y}^{\i\i}_{X,S_0},{Y}_{X,S_0}^{\i\i\i}|{Y}_{X},{Y}_{X}^{\i},\NN_{X,\epsilon},\Cr{Q},\Cr{good},X}\left[ C_{ij}[sk,sl]  \right]\Bigg]\nonumber\\
& \le 4\cdot \E^p_{\widetilde{Y}_{X,\epsilon}|\NN_{X,\epsilon},\Cr{Q},\Cr{good},X}\left[ \sum_{t,s=1}^{N}\sum_{k=1}^{N_{x_s,\epsilon}}  \sum_{l=1}^{N_{x_t,\epsilon}}  C_{ij}[sk,tl] \right],
\end{align}
where we added two zero expectation terms in the last equality above. The last line above uses independence and Jensen's inequality. 
Above, we also defined the block-diagonal matrix $B_{ij}\in\R^{N_{X,\epsilon}\times N_{X,\epsilon}}$ such that
\begin{align}\label{eq:def of C}
& C_{ij}[sk,tl]\nonumber\\
& =\begin{cases}
 N_{x_s,\epsilon}^{-2}\cdot  e_i^*
 \l( P_{x_s,y_{sk}} - P_{x_s,y_{sk}^{\i}}\r) \nabla f(x_s) \nabla f(x_s)^* \\
 \qquad \qquad \cdot \l( P_{x_s,y_{sl}^{\i\i}}-P_{x_s,y_{sl}^{\i\i\i}} \r)
e_j
, & s=t \mbox{ and }k\ne l,\\
 0,& s\ne t\mbox{ or }k=l,
\end{cases}
\end{align}
for every $s,t\in[1:N]$, $k\in [1:N_{x_s,\epsilon}]$, $l\in[1:N_{x_t,\epsilon}]$.
For every $s\in[N]$, we can also define a family of matrices $\{C_{skl}\}_{k,l\in [N_{x_s,\epsilon}]}\subset\R^{n\times n}$ such that
\begin{equation}\label{eq:re-order C}
 C_{skl}[i,j] = C_{sij}[k,l],\qquad
\forall k,l\in[N_{x_s,\epsilon}].
\end{equation}
Note that
\begin{equation}
C_{skl} :=
\frac{1}{N_{x_s,\epsilon}^2}
\begin{cases}
\l( P_{x_s,y_{sk}}  - P_{x_s,y_{sk}^{\i}}\r) \nabla f(x_s) \nabla f(x_s)^* \l( P_{x_s,y_{sl}^{\i\i}}  - P_{x_s,y_{sl}^{\i\i\i}}\r) ,
& k \ne l, \\
0, & k=l.
\end{cases}
\end{equation}
The next step is to modulate the sum in the last line of \eqref{eq:s21} with a Rademacher sequence.

\paragraph{Modulation with Rademacher Sequences}\label{sec:modulation}

Fix $i,j\in[1:n]$, and recall  the definitions of $C_{ij}\in\R^{ N_{X,\epsilon}\times 	N_{X,\epsilon}}$ from \eqref{eq:def of C}.
Let $H=\{\eta_{sk}\}_{s,k}$ (with $s\in[1:N]$ and $k\in[1:N_{x_s,\epsilon}]$) be a Rademacher sequence, that is $\{\eta_{sk}\}_{s,k}$ are independent Bernoulli random variables taking $\pm1$ with equal chances. Also let $H^{\i}=\{\eta_{sk}^{\i}\}_{s,k}$ be an independent copy of $H$. Then, we argue that
\begin{align}
& \E^p_{Y_{X,\epsilon}|\NN_{X,\epsilon},\Cr{Q},\Cr{good},X}\left[ a_{ij}\right] \nonumber\\
&
\le  4\cdot \E^p_{\widetilde{Y}_{X,\epsilon}|\NN_{X,\epsilon},\Cr{Q},\Cr{good},X}\left[\sum_{s,t=1}^{N} \sum_{k=1}^{N_{x_s,\epsilon}}  \sum_{l=1}^{N_{x_t,\epsilon}}  C_{ij}[sk,tl] \right]\qquad \mbox{(see \eqref{eq:s21})}\nonumber\\
& = 4\cdot \E^p_{Y_{X,\epsilon}^{\i\i},Y^{\operatorname{i}\i\i}_{X,\epsilon}|\NN_{X,\epsilon},\Cr{Q},\Cr{good},X}\nonumber\\
& \qquad 
\left[
\E^p_{Y_{X,\epsilon},Y^{\i}_{X,\epsilon}|Y^{\i\i}_{X,\epsilon},Y^{\i\i\i}_{X,\epsilon},\Cr{Q},\Cr{good},X}
\left[ \sum_{s,k}  \left(
\sum_{t,l} C_{ij}[sk,tl]\right)
\right] \right]
\qquad \mbox{(see \eqref{eq:short hand})}
\nonumber\\
 & = 4\cdot \E^p_{Y_{X,\epsilon}^{\i\i},Y^{\operatorname{i}\i\i}_{X,\epsilon}|\NN_{X,\epsilon},\Cr{Q},\Cr{good},X}\nonumber\\
& \qquad  \left[
\E^p_{Y_{X,\epsilon},Y^{\i}_{X,\epsilon},H|Y^{\i\i}_{X,\epsilon},Y^{\i\i\i}_{X,\epsilon},\Cr{Q},\Cr{good},X}
\left[
 \sum_{s,k}  \eta_{sk}\cdot \left(
\sum_{t,l} C_{ij}[sk,tl]\right)
 \right] \right]
 \nonumber \\
& \qquad
 \mbox{(independence and symmetry)}\nonumber\\
 & =4\cdot \E^p_{Y_{X,\epsilon},Y^{\operatorname{i}}_{X,\epsilon},H|\NN_{X,\epsilon},\Cr{Q},\Cr{good},X}\nonumber\\
& \qquad  \left[
\E^p_{Y_{X,\epsilon}^{\i\i},Y^{\i\i\i}_{X,\epsilon}|Y_{X,\epsilon},Y_{X,\epsilon}^{\i},\Cr{Q},\Cr{good},X}
\left[ \sum_{t,l}   \left(
\sum_{s,k} \eta_{sk}  C_{ij}[sk,tl]\right)
 \right] \right]\nonumber\\
& = 4\cdot \E^p_{Y_{X,\epsilon},Y^{\operatorname{i}}_{X,\epsilon},H|\NN_{X,\epsilon},\Cr{Q},\Cr{good},X}\nonumber\\
& \qquad \left[
\E^p_{Y_{X,\epsilon}^{\i\i},Y^{\i\i\i}_{X,\epsilon},H^{\i}|Y_{X,\epsilon},Y^{\i}_{X,\epsilon},\Cr{Q},\Cr{good},X}
\left[ \sum_{t,l}  \eta^{\i}_{tl} \cdot \left(
\sum_{s,k} \eta_{sk}  C_{ij}[sk,tl]\right)
 \right] \right]
 \nonumber \\
& \qquad
\mbox{(independence and symmetry)}\nonumber\\
& = 4\cdot \E^p_{\widetilde{Y}_{X,\epsilon},H,H^{\i}|\NN_{X,\epsilon},\Cr{Q},\Cr{good},X}\left[ \sum_{s,t=1}^{N}\sum_{k=1}^{N_{x_s,\epsilon}}  \sum_{l=1}^{N_{x_l,\epsilon}} \eta_{sk}\eta^{\i}_{tl}\cdot  C_{ij}[sk,tl] \right]
\nonumber\\
& =:  4\cdot \E^p_{\widetilde{Y}_{X,\epsilon},H,H^{\i}|\NN_{X,\epsilon},\Cr{Q},\Cr{good},X}\left[c_{ij}\right],
 \label{eq:s22}
\end{align}
where we set
\begin{equation}
c_{ij} := \l| \sum_{s,t=1}^{N}\sum_{k=1}^{N_{x_s,\epsilon}}  \sum_{l=1}^{N_{x_l,\epsilon}} \eta_{sk}\eta^{\i}_{tl}\cdot  C_{ij}[sk,tl]
\r|.
\label{eq:def of d}
\end{equation}
Conditioned on everything but $H$ and $H^{\i}$, $c_{ij}$ is a \emph{decoupled chaos}: decoupled because $H=\{\eta_{sk}\}_{s,k}$ and $H^{\i}=\{\eta^{\i}_{sk}\}_{s,k}$ are independent (Rademacher) sequences. The behavior of the moments of a chaos random variable is well-understood.

\paragraph{Moments of a Decoupled Chaos}

The fist moment of $c_{ij}$, namely its expectation, can be estimated as follows. First observe that
\begin{align}\label{eq:Ed pre}
& \E_{H,H^{\i}|\widetilde{Y}_{X,\epsilon},\Cr{Q},\Cr{good},X}[c_{ij}] \nonumber\\
& \le  \sqrt{\E_{H,H^{\i}|\widetilde{Y}_{X,\epsilon},\Cr{Q},\Cr{good},X}[c_{ij}^2]} \qquad \mbox{(Jensen's inequality)}
\nonumber\\
& = \l\| C_{ij} \r\|_F \qquad \l(H \mbox{ and } H^{\i} \mbox{ are independent Rademacher sequences} \r)\nonumber\\
& \le \sqrt{N}\cdot \max_{s\in[1:N]} \left\|C_{sij} \right\|_F.\qquad \left(C_{ij} \mbox{ is block-diagonal}  \right)
\end{align}
Let us therefore focus on $\|C_{sij}\|_F$ for fixed $s\in[1:N]$:
\begin{align}
\|C_{sij}\|_F^2
&
= \sum_{k,l=1}^{N_{x_s,\epsilon}} \left|
C_{sij}[k,l]
\right|^2 \nonumber\\
&
= \sum_{k,l=1}^{N_{x_s,\epsilon}} \left|
C_{skl}[i,j]
\right|^2 \qquad \mbox{(see \eqref{eq:re-order C})}
\nonumber\\
& \le N_{x_s,\epsilon}^2\cdot \max_{k,l\in[1:N_{x_s,\epsilon}]}
\left|
C_{skl}[i,j]
\right|^2\nonumber\\
& \le N_{x_s,\epsilon}^2 \cdot \max_{k,l\in[1:N_{x_s,\epsilon}]} \l\| C_{skl} \r\|_{\infty}^2,
\label{eq:Csij F norm}
\end{align}
where $\|A\|_{\infty}$ is the largest entry of $A$ in magnitude. With $e_i\in\mathbb{R}^{n}$ denoting the $i$th canonical vector, we continue by noting that
\begin{align}
& \l\| C_{skl}\r\|_{\infty}\nonumber\\
& = \max_{i,j\in[1:n]} \left| C_{skl}[i,j]\right|\nonumber\\
& =\max_{i,j\in[1:n]} \left| e_i^* C_{skl}e_j\right|\nonumber\\
& =
N_{x_s,\epsilon}^{-2}
\nonumber\\
&  
\cdot
\max_{i,j\in[1:n]}
\left|
e_i^*
\left(
P_{x_s,y_{sk}} - P_{x_s,y_{sk}^{\i}}\r) \nabla f(x_s) \nabla f(x_s)^* \l( P_{x_s,y^{\i\i}_{sl}} - P_{x_s,y^{\i\i\i}_{sl}}\r)
e_j
\right|
\quad \mbox{(see \eqref{eq:def of C})}
\nonumber\\
& \le
4  N_{x_s,\epsilon}^{-2} \cdot
\max_{s\in[1:N]}\max_{i\in[1:n]}
\max_{y_s\in \widetilde{Y}_{x_s  ,\epsilon}}
\left|e_i^*
P_{x_s,y_s}\nabla f(x_s)\right|^2
\qquad \mbox{(see \eqref{eq:short hand})}
\nonumber\\
& \le 4  N_{x_s,\epsilon}^{-2}\cdot
\max_{s\in[1:N]}\max_{i\in[1:n]}
\max_{y_s\in \widetilde{Y}_{x_s  ,\epsilon}}
\left\|
P_{x_s,y_s}e_i\r\|_2^2 \cdot \l\|P_{x_s,y_s}\nabla f(x_s)\right\|^2_2
\quad \mbox{(Cauchy-Schwarz's)}
\nonumber\\
& \le 4  N_{x_s,\epsilon}^{-2}\cdot
 \frac{Q_{X,\epsilon}}{n}\cdot \frac{Q_{X,\epsilon} \L_f^2}{n},
\qquad \l(\mbox{conditioned on the event } \Cr{max}\r)
\label{eq:Cskl max norm}
\end{align}
where we defined the  event $\Cr{max}$  as
\begin{align}
\Cr{max} & =
 \l\{ \max_{ s\in[1:N]}\max_{i\in [1:n]}
\max_{y_s\in \widetilde{Y}_{x_s  ,\epsilon}}
\l\| P_{x_s,y_s} e_i  \r\|_2^2 \le \frac{Q_{X,\epsilon}}{n}\r\} \nonumber\\
& \qquad 
\bigcap
 \l\{ \max_{ s\in[1:N]}
\max_{y_s\in \widetilde{Y}_{x_s  ,\epsilon}}
\l\| P_{x_s,y_s} \nabla f(x_s)  \r\|_2^2 \le \frac{Q_{X,\epsilon}\L_f^2}{n}\r\},
\end{align}
for $Q_{X,\epsilon}>0$ to be set later. For $p\ge1$ to be assigned later,  we also assume that $\Cr{max}$ is very likely to happen:
\begin{equation}\label{eq:likely bnd}
\Pr_{\widetilde{Y}_{X,\epsilon}|\NN_{X,\epsilon},\Cr{Q},\Cr{good},X}\l[ \Cr{max}^C \r] \lesssim
 \l( \frac{p}{N_{X,\min,\epsilon}\rho_{\mu,X,\epsilon} N_{X,\epsilon}} \r)^{\frac{p}{2}}.
 \qquad
\mbox{(see \eqref{eq:def of rhoX thm})}
\end{equation}
We now complete our calculation of the first moment of $c_{ij}$:
\begin{align}
& \mathbb{E}_{H,H^{\i}|\Cr{max},\widetilde{Y}_{X,\epsilon},\Cr{Q},\Cr{good},X}[c_{ij}] \nonumber\\
& \le \l\| C_{ij} \r\|_F \nonumber\\
& \le \sqrt{N} \cdot \max_{s\in [1:N]} \l\| C_{sij} \r\|_F \qquad \mbox{(see \eqref{eq:Ed pre})}\nonumber\\
& \le \sqrt{N} \cdot  \max_{s\in[1:N]} \max_{k,l\in[1:N_{x_s,\epsilon}]} N_{x_s,\epsilon} \cdot
\l\| C_{skl}\r\|_{\infty}
\qquad \mbox{(see \eqref{eq:Csij F norm})}
\nonumber\\
& \le  \sqrt{N} \cdot \max_{s\in[1:N]}  N_{x_s,\epsilon} \cdot  \frac{4Q_{X,\epsilon}^2\L_f^2}{n^2N_{x_s,\epsilon}^2}  \qquad \mbox{(see  \eqref{eq:Cskl max norm})}\nonumber\\
& \le \frac{4\sqrt{N}Q_{X,\epsilon}^2\L_f^2}{n^2 \cdot \min_{x\in X} N_{x,\epsilon}}.
\label{eq:bnd on 1st moment of chaos}
\end{align}
To control the higher order moments of $c_{ij}$, we invoke the following result \cite[Corollary 2]{adamczak2005logarithmic}.
\begin{proposition}\label{prop:chaos}\emph{\textbf{(Moments of a decoupled chaos)}} For a square matrix $C$, a Rademacher sequence $H=\{\eta_k\}_k$, and an independent copy $H^{\i}=\{\eta^{\i}_l\}_l$, consider the  decoupled (second-order) chaos
$$
c=\l| \sum_{k,l} \eta_k \eta^{\i}_l \cdot C[k,l]\r|.
$$
Then, it holds that
\begin{equation}
\E^p[c-\E[c]] \lesssim  p\cdot b +\sqrt{p}\cdot \sigma  ,\qquad \forall p\ge 1,
\end{equation}
where
\begin{equation} \label{eq:def of b prop}
b := \|C\|,
\end{equation}
\begin{equation}
\sigma :=  \sqrt{\E_H \left[\left\|C \eta\right\|^2_2\right]} = \|C\|_F,\label{eq:def of sigma proposition}
\end{equation}
and $\eta$ is the vector formed from the Rademacher sequence $H$.
\end{proposition}

We now appeal to Proposition~\ref{prop:chaos} in order to  bound the moments of the chaos random variable $c_{ij}$ in \eqref{eq:def of d} (conditioned on $X,\widetilde{Y}_{X,\epsilon}$ and the event $\Cr{Q}\cap\Cr{good}\cap\Cr{max}$). To that end, note that
\begin{align}\label{eq:b est pre pre}
b & = \|C_{ij}\| \qquad \mbox{(see \eqref{eq:def of b prop})}
\nonumber\\
& = \max_{s\in[1:N]} \|C_{sij}\|.
\qquad \left(C_{ij} \mbox{  is block-diagonal}\right)
\end{align}
Let us then focus on  $\|C_{sij}\|$ for fixed $s\in[1:N]$. Observe that
\begin{align}
\left\|C_{sij}\right\|
& \le N_{x_s,\epsilon}  \cdot \l\| C_{sij} \r\|_{\infty}
\qquad \l(\|A\| \le a \cdot \|A\|_{\infty},\,\,\forall A\in \R^{a\times a}  \r)
\nonumber\\
& \le N_{x_s,\epsilon}\cdot \max_{k,l\in[1:N_{x_s,\epsilon}]} \l\| C_{skl} \r\|_{\infty} \qquad \mbox{(see \eqref{eq:re-order C})}\nonumber\\
& \le N_{x_s,\epsilon} \cdot \frac{4Q_{X,\epsilon}^2\L_f^2}{n^2N_{x_s,\epsilon}^2} \qquad
\mbox{(see \eqref{eq:Cskl max norm})}
\nonumber\\
& \le \frac{4Q_{X,\epsilon}^2\L_f^2}{n^2\cdot \min_{x\in X} N_{x,\epsilon}}.\qquad \l(\mbox{see \eqref{eq:good event def}}\r)
\label{eq:est b pre}
\end{align}
In light of  \eqref{eq:b est pre pre}, it follows that
\begin{align}\label{eq:est b}
b
& \le \max_{s\in[1:N]} \l\| C_{sij}\r\|
\qquad \mbox{(see \eqref{eq:b est pre pre})} \nonumber\\
& \le \frac{4Q_{X,\epsilon}^2\L_f^2}{n^2 \cdot \min_{x\in X}N_{x,\epsilon}}. \qquad \mbox{(see \eqref{eq:est b pre})}
\end{align}
We argue likewise to find $\sigma$:
\begin{align}
\sigma & = \left\| C_{ij} \right\|_F
\qquad \mbox{(see \eqref{eq:def of sigma proposition})}
\nonumber\\
& \le \frac{4\sqrt{N}Q_{X,\epsilon}^2\L_f^2}{ n^2\cdot \min_{x\in X}N_{x,\epsilon}}.
\qquad \mbox{(see \eqref{eq:bnd on 1st moment of chaos})}
\label{eq:est sigma}
\end{align}
With $b$ and $\sigma$ at hand, we now invoke Proposition~\ref{prop:chaos} to write that
\begin{align}
&  \E^p_{H,H^{\i}|\Cr{max},\widetilde{Y}_{X,\epsilon},\Cr{Q},\Cr{good},X} \left[ c_{ij}\r]\nonumber\\
& = \E^p_{H,H^{\i}|\Cr{max},\widetilde{Y}_{X,\epsilon},\Cr{Q},\Cr{good},X} \left[
\sum_{s,t=1}^{N}\sum_{k=1}^{N_{x_s,\epsilon}} \sum_{l=1}^{N_{x_t,\epsilon}} \eta_{sk} \eta^{\i}_{tl} \cdot C_{ij}[sk,tl]
\right]
\qquad \mbox{(see \eqref{eq:def of d})}
\nonumber\\
& \le \E^p_{H,H^{\i}|\Cr{max},\widetilde{Y}_{X,\epsilon},\Cr{Q},\Cr{good},X}
\left[ c_{ij} - \mathbb{E}_{H,H^{\i}|\Cr{max},\widetilde{Y}_{X,\epsilon},\Cr{Q},\Cr{good},X}\l[c_{ij}\r] \r] + \E_{H,H^{\i}|\Cr{max},\widetilde{Y}_{X,\epsilon},\Cr{Q},\Cr{good},X} \left[ c_{ij}\r]
\nonumber\\
& \lesssim \l(p\cdot b +\sqrt{p}\cdot
\sigma\r)
+ \frac{\sqrt{N}Q_{X,\epsilon}^2\L_f^2}{n^2 \cdot \min_{x}N_{x,\epsilon}}
\qquad \mbox{(see Proposition~\ref{prop:chaos}  and \eqref{eq:bnd on 1st moment of chaos})}
\nonumber\\
& \lesssim
\l( p \cdot \frac{Q_{X,\epsilon}^2\L_f^2}{ n^2 \cdot \min_{x}N_{x,\epsilon}} +  \sqrt{p} \cdot \frac{\sqrt{N}Q_{X,f,\epsilon}^2\L_f^2}{  n^2\cdot \min_x N_{x,\epsilon}}
\r)
+ \frac{\sqrt{N}Q_{X,\epsilon}^2\L_f^2}{n^2\cdot \min_x N_{x,\epsilon}}
\qquad
\mbox{(see (\ref{eq:est b},\ref{eq:est sigma}))}\nonumber\\
&
\lesssim    \sqrt{p} \cdot \frac{\sqrt{N}Q_{X,\epsilon}^2\L_f^2}{ n^2\cdot \min_x N_{x,\epsilon}}
\qquad
\l(\mbox{if } 1 \le p\le N \r)
\nonumber\\
& \lesssim
\sqrt{p} \cdot \frac{N Q_{X,\epsilon}^2\L_f^2}{ n^2 \sqrt{N_{X,\min,\epsilon} \cdot \rho_{\mu,X,\epsilon}N_{X,\epsilon}}}.
\qquad
\l(
\mbox{see \eqref{eq:good event def} and \eqref{eq:def of rhoX thm}}
\r)
\label{eq:where p is constrained}
\end{align}
Conditioned on $\NN_{X,\epsilon}$, the bound above  is independent of $\widetilde{Y}_{X,\epsilon}$, which allows us to remove the conditioning and find that
\begin{equation}\label{eq:est moment pre}
\E^p_{\widetilde{Y}_{X,\epsilon},H,H^{\i}|\Cr{max}, \NN_{X,\epsilon},\Cr{Q},\Cr{good},X} \left[
c_{ij}
\right]
 \lesssim \sqrt{p} \cdot \frac{N Q_{X,\epsilon}^2 \L_f^2}{n^2 \sqrt{N_{X,\min,\epsilon}\cdot  \rho_{\mu,X,\epsilon} N_{X,\epsilon}}}.
\end{equation}
As a useful aside, we also record a uniform bound on $c_{ij}$ for every $i,j\in[1:n]$:
\begin{align}
\l| c_{ij}\r| &
= \l| \sum_{s,t=1}^{N}\sum_{k=1}^{N_{x_s,\epsilon}} \sum_{l=1}^{N_{x_t,\epsilon}} \eta_{sk} \eta^{\i}_{tl} \cdot C_{ij}[sk,tl] \r|
\qquad \mbox{(see \eqref{eq:def of d})}
\nonumber\\
& \le
\sum_{s,t=1}^{N}
\left|
\sum_{k=1}^{N_{x_s,\epsilon}} \sum_{l=1}^{N_{x_t,\epsilon}} \eta_{sk} \eta^{\i}_{tl} \cdot C_{ij}[sk,tl]
\right| \qquad \mbox{(triangle inequality)}\nonumber\\
& =
\sum_{s=1}^N \left|
\sum_{k=1}^{N_{x_s,\epsilon}} \sum_{l=1}^{N_{x_s,\epsilon}} \eta_{sk} \eta^{\i}_{sl} \cdot C_{sij}[k,l]
\right|
\nonumber\\
& \le \sum_{s=1}^N  N_{x_s,\epsilon} \cdot \l\|C_{sij} \r\|
\qquad \l( H,H^{\i} \mbox{ are Rademacher sequences} \r)\nonumber\\
& \le \sum_{s=1}^N N_{x_s,\epsilon}^2  \cdot  \frac{4Q_{X,\epsilon}^2\L_f^2}{n^2N_{x_s,\epsilon}^2}
\qquad \mbox{(see \eqref{eq:est b pre})}
\nonumber\\
& = \frac{4NQ_{X,\epsilon}^2\L_f^2}{n^2},
\label{eq:general bnd on d}
\end{align}
where the third line uses the fact that $C_{ij}$  is block-diagonal with blocks $C_{sij}\in\mathbb{R}^{N_{x_s,\epsilon}\times N_{x_s,\epsilon}}$ and also uses \eqref{eq:re-order C}. 
Putting everything back together, we finally argue that
\begin{align}
& \E^p_{Y_{X,\epsilon}|\NN_{X,\epsilon},\Cr{Q},\Cr{good},X}\left[ a_{ij}\right] \nonumber\\
& \le 4\cdot \E^p_{\widetilde{Y}_{X,\epsilon},H,H^{\i}|\overline{N}_{X,\epsilon},  \Cr{Q},\Cr{good},X} \l[ c_{ij}\r] \qquad
\mbox{(see \eqref{eq:s22})}\nonumber\\
&
\le  4 \cdot \E^p_{\widetilde{Y}_{X,\epsilon},H,H^{\i}|\Cr{max},\overline{N}_{X,\epsilon},\Cr{Q},\Cr{good},X}\left[ c_{ij}\right]  + 4 \cdot \sup \l|c_{ij}\r|\cdot
\l(\Pr_{\widetilde{Y}_{X,\epsilon}|\overline{N}_{X,\epsilon},\Cr{Q},\Cr{good},X}\l[ \Cr{max}^C \r]\r)^{\frac{1}{p}}
\nonumber \\
& \qquad
\mbox{(see \eqref{eq:useful ineq})}
\nonumber\\
& \lesssim \sqrt{p} \cdot \frac{N Q_{X,\epsilon}^2\L_f^2}{n^2 \sqrt{N_{X,\min,\epsilon} \rho_{\mu,X,\epsilon} N_{X,\epsilon}}} +
\frac{N Q_{X,\epsilon}^2\L_f^2}{n^2}
\cdot
\sqrt{\frac{{p}}{{ N_{X,\min,\epsilon}\rho_{\mu,X,\epsilon} N_{X,\epsilon}}}}
\nonumber \\
& \qquad
 \mbox{(see \eqref{eq:est moment pre}, \eqref{eq:general bnd on d}, and \eqref{eq:likely bnd})}\nonumber\\
& \lesssim \sqrt{p} \cdot \frac{N Q^2_{X,\epsilon}\L_f^2}{n^2 \sqrt{N_{X,\min,\epsilon}  \rho_{\mu,X,\epsilon}N_{X,\epsilon}}},
\label{eq:est moment}
\end{align}
when $1\le p\le N$ (see \eqref{eq:where p is constrained}).  At last, \eqref{eq:est moment} describes the  moments of the random variable $a_{ij}$ for fixed $i,j$ (and conditioned on $\NN_{X,\epsilon},\Cr{Q},\Cr{good},X$).

\subsubsection{Applying Markov's Inequality}

Given the estimates of the moments of $a_{ij}$ in \eqref{eq:est moment}, we can simply apply Markov's inequality to translate this information into a tail bound for $a_{ij}$. Indeed, for arbitrary $1\le p\le N$ and $\Cl[gam]{g1}>0$, it holds that
\begin{align}
\Pr_{Y_{X,\epsilon}|\NN_{X,\epsilon},\Cr{Q},\Cr{good},X} \left[  \left| a_{ij}\right|> \Cr{g1} \right]
& = \Pr_{Y_{X,\epsilon}|\NN_{X,\epsilon},\Cr{Q},\Cr{good},X} \left[  \left| a_{ij}\right|^p> \Cr{g1}^p\right]\nonumber\\
& \le \left(\frac{\E^p_{Y_{X,\epsilon}|\NN_{X,\epsilon},\Cr{Q},\Cr{good},X} \left[ a_{ij} \right]}{\Cr{g1}}\right)^p \qquad \mbox{(Markov's inequality)}\nonumber\\
& \le \left(   \frac{\Cl[cte]{exp}\sqrt{p}N Q_{X,\epsilon}^2 \L_f^2}{\Cr{g1} n^2 \sqrt{N_{X,\min,\epsilon} \rho_{\mu,X,\epsilon}  N_{X,\epsilon}}} \right)^p, \qquad \mbox{(see \eqref{eq:est moment})}
\end{align}
for an absolute constant $\Cr{exp}$. In particular,  the choice of
$$
\Cr{g1}=\Cr{exp} \Cr{g2} \cdot \sqrt{\log n}\cdot  \frac{N Q_{X,\epsilon}^2 \L_f^2}{n^2 \sqrt{N_{X,\min,\epsilon}\rho_{\mu,X,\epsilon}N_{X,\epsilon}}},\qquad
p =  \max\l[\log n,1\r] \le N,
\qquad
\Cr{g2}\ge 1,
$$
yields
\begin{align}\label{eq:tb of a_x}
& \Pr_{Y_{X,\epsilon}|\NN_{X,\epsilon},\Cr{Q},\Cr{good},X} \left[  \left| a_{ij}\right|\gtrsim \Cr{g2} \cdot \sqrt{\log n} \cdot \frac{N Q_{X,\epsilon}^2\L_f^2}{n^2\sqrt{N_{X,\min,\epsilon} \rho_{\mu,X,\epsilon} N_{x,\epsilon}}}  \right] \nonumber\\
& \le \Cr{g2}^{-\log n} = n^{-\log \Cr{g2}}.
\end{align}
With the tail bound of $a_{ij}$ finally available above (for fixed $i,j\in[1:n]$ and conditioned on $\overline{N}_{X,\epsilon},\Cr{Q},\Cr{good},X$), we next quantify how ${\ddddot{\Sigma}}^o_{X,Y_{X,\epsilon}}$  concentrates about its expectation.

\subsubsection{Applying the Union Bound}

In light of \eqref{eq:tb of a_x} and by applying the union bound to $\{a_{ij}\}_{i,j}$, we arrive at the following statement.
\begin{align}\label{eq:pre complete}
 &   \Pr_{Y_{X,\epsilon}|\NN_{X,\epsilon},\Cr{Q},\Cr{good},X} \left[  \max_{i,j\in[1:n]} \left| a_{ij}\right|\lesssim \Cr{g2} \cdot \sqrt{\log n}\cdot \frac{  N Q_{X,\epsilon}^2\L_f^2}{ n^2 \sqrt{N_{X,\min,\epsilon} \rho_{\mu,X,\epsilon}N_{x,\epsilon}}}\right]
 \nonumber\\
& \ge 1 -  n^2  \cdot n^{-\log  \Cr{g2}}.
\qquad \mbox{(union bound and  \eqref{eq:tb of a_x})}
\end{align}

\subsection{Second and third norms}
\label{sec:sec n third in decompose}

In this section, we bound the second and third norms in the last line of \eqref{eq:decompose err new} using the Bernstein inequality. Let us bound the second norm as
\begin{align}
& \frac{n^2}{N} \l\|
 \sum_{s=1}^N \sum_{k=1}^{N_{x_s,\epsilon}}\frac{1}{N_{x_s,\epsilon}^2} \l(P_{x_s,y_{sk}} - \E_{y_s|\Cr{Q},x_s}[P_{x_s,y_s}] \r) \nabla f(x_s)\nabla f(x_s)^*  \sum_{l\ne k}\E_{y_s|\Cr{Q},x_s}[P_{x_s,y_s}]
  \r\|_F  \nonumber\\
 &\le  \frac{n^2}{N} \l\|
 \sum_{s=1}^N \sum_{k=1}^{N_{x_s,\epsilon}}\frac{1}{N_{x_s,\epsilon}} \l(P_{x_s,y_{sk}} - \E_{y_s|\Cr{Q},x_s}[P_{x_s,y_s}] \r) \nabla f(x_s)\nabla f(x_s)^*  \E_{y_s|\Cr{Q},x_s}[P_{x_s,y_s}]
  \r\|_F \nonumber\\
  & =: \frac{n^2}{N} \l\| \sum_{s=1}^N \sum_{k=1}^{N_{x_s,\epsilon}} A_{x_s,y_{sk}} \r\|_F,
\end{align}
where $\{A_{x_s,y_{sk}}\}_{sk} \subset\R^{n\times n}$ is a sequence of zero-mean and independent random matrices. To apply the Bernstein inequality (Proposition~\ref{prop:Bernstein recall}) conditioned on the event $\Cr{Q}$, we write that
\begin{align}
\l\| A_{x_s,y_{sk}} \r\|_F & = \frac{1}{N_{x_s,\epsilon}} \l\| \l(P_{x_s,y_{sk}} - \E_{y_s|\Cr{Q},x_s} [P_{x_s,y_s}]\r) \nabla f(x_s) \nabla f(x_s)^*  \E_{y_s|\Cr{Q},x_s}[P_{x_s,y_s}] \r\|_F \nonumber\\
& \le \frac{1}{N_{x_s,\epsilon}} \l\|P_{x_s,y_{sk}}  \nabla f(x_s) \nabla f(x_s)^* \E_{y_s|\Cr{Q},x_s}[P_{x_s,y_s}]  \r\|_F
\nonumber \\
& \qquad
+ \frac{1}{N_{x_s,\epsilon}} \l\| \E_{y_s|\Cr{Q},x_s}[P_{x_s,y_{s}}]  \nabla f(x_s) \nabla f(x_s)^* \E_{y_s|\Cr{Q},x_s}[P_{x_s,y_s}]  \r\|_F \nonumber\\
& \le  \frac{1}{N_{x_s,\epsilon}} \l\|P_{x_s,y_{sk}}  \nabla f(x_s)\r\|_2 \l\|  \E_{y_s|\Cr{Q},x_s}[P_{x_s,y_s}]  \nabla f(x_s) \r\|_2\nonumber\\
 & \qquad  + \frac{1}{N_{x_s,\epsilon}} \l\| \E_{y_s|\Cr{Q},x_s}[P_{x_s,y_{s}}]  \nabla f(x_s) \r\|_2^2  \nonumber\\
&\le  \frac{1}{N_{x_s,\epsilon}} \l\|P_{x_s,y_{sk}}  \nabla f(x_s)\r\|_2 \cdot \sqrt{\E_{y_s|\Cr{Q},x_s} \l\|  P_{x_s,y_s}  \nabla f(x_s) \r\|^2_2} \nonumber\\
& \qquad 
+ \frac{1}{N_{x_s,\epsilon}} \E_{y_s|\Cr{Q},x_s} \l\|  P_{x_s,y_s} \nabla f(x_s) \r\|_2^2
\qquad \mbox{(Jensen's inequality)}
\nonumber\\
& \le \frac{1}{N_{x_s,\epsilon}} \sqrt{\frac{Q_{X,\epsilon}\L_f^2}{n}} \sqrt{\frac{\L_f^2}{\K_\mu n}}+{\frac{\L_f^2}{\K_\mu N_{x_s,\epsilon}n}}
\,\, \mbox{(see (\ref{eq:event Q def},\ref{eq:Lf}),~\cite[Lemma 5.5]{vershynin2010introduction})} \nonumber\\
& \le {\frac{2Q_{X,\epsilon} \L_f^2}{\min_{s\in [N]}N_{x_s,\epsilon}n}} =: b.
\qquad \l( \mbox{if } Q_{X,\epsilon} \ge \K_\mu^{-1}\r)
\end{align}
On the other hand,
\begin{align}
& \sum_{s=1}^{N} \sum_{k=1}^{N_{x_s,\epsilon}} \E_{Y_{X,\epsilon}|\overline{N}_{X,\epsilon},\Cr{Q},\Cr{good},X} \l\| A_{x_s,y_{sk}} \r\|_F^2  \nonumber\\
&\le  \sum_{s=1}^{N} \frac{1}{N_{x_s,\epsilon}^2} \sum_{k=1}^{N_{x_s,\epsilon}} \E_{y_s|\Cr{Q},x_s}  \l\| P_{x_s,y_{s}} \nabla f(x_s) \nabla f(x_s)^* \E_{y_s|\Cr{Q},x_s} [P_{x_s,y_s}] \r\|_F^2  \nonumber\\
& \le  \sum_{s=1}^{N} \frac{1}{N_{x_s,\epsilon}^2} \sum_{k=1}^{N_{x_s,\epsilon}} \E_{y_s|\Cr{Q},x_s}  \l\| P_{x_s,y_{sk}} \nabla f(x_s)  \r\|_2^2 \l\|  \E_{y_s|\Cr{Q},x_s} [P_{x_s,y_s}] \nabla f(x_s)\r\|_2^2  \nonumber\\
& \le \sum_{s=1}^{N} \frac{1}{N_{x_s,\epsilon}^2} \sum_{k=1}^{N_{x_s,\epsilon}} \E_{y_s|\Cr{Q},x_s}  \l\| P_{x_s,y_{sk}} \nabla f(x_s)  \r\|_2^4
\qquad \mbox{(Jensen's inequality)} \nonumber\\
& \le \sum_{s=1}^N \frac{\L_f^4 }{\K_\mu^2 N_{x_s,\epsilon}n^2} \qquad
\mbox{(see \eqref{eq:event Q def}, \eqref{eq:Lf}, and~\cite[Lemma 5.5]{vershynin2010introduction})} \nonumber\\
& \le \frac{N Q_{X,\epsilon}^2 \L_f^4}{\min_{s\in[N]}N_{x_s,\epsilon} n^2} =:\sigma^2.
\qquad \l( \mbox{if } Q_{X,\epsilon} \ge \K_\mu^{-1}\r)
\end{align}
The second line above uses the fact that $\E\l\|Z-\E[Z]\r\|_F^2 \le \E\l\|Z \r\|_F^2$ for a random  matrix  $Z$. It follows that
\begin{align}
\max[b,\sigma] \le  2\sqrt{\frac{N}{\min_{s\in [N]}N_{x_s,\epsilon}}} \frac{Q_{X,\epsilon}\L_f^2}{n}.
\end{align}
An application of the Bernstein inequality now yields that conditioned on $\Cr{Q},\Cr{good}$, $\NN_{X,\epsilon},X$,
\begin{align}
& \frac{n^2}{N} \l\|
 \sum_{s=1}^N \sum_{k=1}^{N_{x_s,\epsilon}}\frac{1}{N_{x_s,\epsilon}^2} \l(P_{x_s,y_{sk}} - \E_{y_s|\Cr{Q},x_s}[P_{x_s,y_s}] \r) \nabla f(x_s)\nabla f(x_s)^*  \sum_{l\ne k}\E_{y_s|\Cr{Q},x_s}[P_{x_s,y_s}]
  \r\|_F \nonumber\\
& \le \frac{n^2}{N} \l\|
 \sum_{s=1}^N \sum_{k=1}^{N_{x_s,\epsilon}} A_{x_s,y_{sk}}\r\|_F
\nonumber\\
& \lesssim \frac{n^2}{N}\cdot \gamma \max[b,\sigma] \nonumber\\
& \lesssim \frac{\gamma n^2}{N} \sqrt{\frac{N}{\min_{s\in [N]}N_{x_s,\epsilon}}} \frac{Q_{X,\epsilon}\L_f^2}{n} \nonumber\\
& = \frac{\gamma n Q_{X,\epsilon}\L_f^2}{\sqrt{N \min_{s\in [N]} N_{x_s,\epsilon}}} \nonumber\\
& = \frac{\gamma n Q_{X,\epsilon}\L_f^2}{\sqrt{\rho_{\mu,X,\epsilon} N_{X,\epsilon}}}
\label{eq:secondtermfinal}
\end{align}
for $\gamma\ge 1$ and except with a probability of at most $e^{-\gamma}$. An identical bound holds for the third norm in the last line of \eqref{eq:decompose err new}.
\subsection{Bound on \eqref{eq:decompose err new}}
We now combine the bounds for the terms in \eqref{eq:decompose err new} obtained in Sections \ref{sec:first in decompose} and \ref{sec:sec n third in decompose}. Applying~\eqref{eq:s20}, we have that conditioned on $\Cr{Q},\Cr{good},\NN_{X,\epsilon},X$,
\begin{align}\label{eq:decompose err finalE1}
&\left\|
\ddddot{\Sigma}_{X,Y_{X,\epsilon}}^o-
\mathbb{E}_{Y_{X,\epsilon}|\NN_{X,\epsilon},\Cr{Q},\Cr{good},X} \left[ \ddddot{\Sigma}_{X,Y_{X,\epsilon}}^o \right]
\right\|_F\nonumber\\
& \le \frac{n^2}{N} \sqrt{\sum_{i,j=1}^n a_{ij}^2} \nonumber\\
& \,\, +
 \frac{n^2}{N} \l\|
 \sum_{s=1}^N \sum_{k\ne l}\frac{1}{N_{x_s,\epsilon}^2} \l(P_{x_s,y_{sk}} - \E_{y_s|\Cr{Q},x_s}[P_{x_s,y_s}] \r) \nabla f(x_s)\nabla f(x_s)^*  \E_{y_s|\Cr{Q},x_s}[P_{x_s,y_s}]
  \r\|_F \nonumber\\
&  \,\, + \frac{n^2}{N} \l\| \sum_{s=1}^N \sum_{k\ne l} \frac{1}{N_{x_s,\epsilon}^2} \E_{y_s|\Cr{Q},x_s}[P_{x_s,y_s}]  \nabla f(x_s)\nabla f(x_s)^* \l(P_{x_s,y_{sl}} - \E_{y_s|\Cr{Q},x_s}[P_{x_s,y_s}] \r) \r\|_F \nonumber \\
& \le \frac{n^3}{N}\cdot
\max_{i,j\in[1:n]} \left| a_{ij}\right| + 2 \frac{\gamma n Q_{X,\epsilon}\L_f^2}{\sqrt{\rho_{\mu,X,\epsilon} N_{X,\epsilon}}} \quad \mbox{(see \eqref{eq:secondtermfinal})} \nonumber \\
& \lesssim \frac{n^3}{N}\cdot \Cr{g2} \cdot \sqrt{\log n}\cdot \frac{  N Q_{X,\epsilon}^2\L_f^2}{ n^2 \sqrt{N_{X,\min,\epsilon} \rho_{\mu,X,\epsilon}N_{x,\epsilon}}} + 2 \frac{\gamma n Q_{X,\epsilon}\L_f^2}{\sqrt{\rho_{\mu,X,\epsilon} N_{X,\epsilon}}}
\quad \mbox{(see \eqref{eq:pre complete})} \nonumber \\
& = \Cr{g2} \cdot \sqrt{\log n}\cdot \frac{n Q_{X,\epsilon}^2\L_f^2}{\sqrt{N_{X,\min,\epsilon} \rho_{\mu,X,\epsilon}N_{x,\epsilon}}} + 2 \frac{\gamma n Q_{X,\epsilon}\L_f^2}{\sqrt{\rho_{\mu,X,\epsilon} N_{X,\epsilon}}} \nonumber \\
& \lesssim \max[\Cr{g2},\gamma] \cdot \sqrt{\log n} \cdot \frac{n \cdot \max[Q_{X,\epsilon},Q_{X,\epsilon}^2] \cdot \L_f^2}{\sqrt{\rho_{\mu,X,\epsilon} N_{X,\epsilon}}}
\end{align}
for $\gamma, \Cr{g2} \ge 1$ and except with a probability of at most $e^{-\gamma} + n^2  \cdot n^{-\log  \Cr{g2}}$. This holds under \eqref{eq:likely bnd} (with $p=\max[\log n,1] \le N$).

Finally, we proceed to remove the conditioning on $\Cr{Q}$. Similar to \eqref{eq:totalexp}, we have
\begin{align}\label{eq:totalexpK}
& \mathbb{E}_{Y_{X,\epsilon}|\NN_{X,\epsilon},\Cr{good},X}\l[\ddddot{\Sigma}^o_{X,Y_{X,\epsilon}}\r] - \mathbb{E}_{Y_{X,\epsilon}|\Cr{Q},\NN_{X,\epsilon},\Cr{good},X}\l[\ddddot{\Sigma}^o_{X,Y_{X,\epsilon}}\r] \nonumber \\
&= \Pr_{Y_{X,\epsilon}|\NN_{X,\epsilon},\Cr{good},X}\l[ \Cr{Q}^C\r] \nonumber\\
& \qquad \cdot  \left( \mathbb{E}_{Y_{X,\epsilon}|\Cr{Q}^C, \NN_{X,\epsilon},\Cr{good},X}\l[\ddddot{\Sigma}^o_{X,Y_{X,\epsilon}}\r] - \mathbb{E}_{Y_{X,\epsilon}|\Cr{Q},\NN_{X,\epsilon},\Cr{good},X}\l[\ddddot{\Sigma}^o_{X,Y_{X,\epsilon}}\r] \right).
\end{align}
Since for any $X$, $Y_{X,\epsilon}$, we have
\begin{align}\label{eq:totalexp1K}
\l\| \ddddot{\Sigma}^o_{X,Y_{X,\epsilon}} \r\|_F
& \le \frac{n^2}{N} \sum_{s=1}^{N} \sum_{k\ne l} \frac{1}{N_{x_s,\epsilon}^2} \l\| P_{x_s,y_{sk}}\nabla f(x_s) \nabla f(x_s)^* P_{x_s,y_{sl}} \r\|_F \qquad \mbox{(see \eqref{eq:recall Sigma prime})} \nonumber \\
& \le \frac{n^2}{N} \sum_{s=1}^{N} \sum_{k\ne l} \frac{1}{N_{x_s,\epsilon}^2} \l\| P_{x_s,y_{sk}}\nabla f(x_s) \r\|_2 \l\|\nabla f(x_s)^* P_{x_s,y_{sl}} \r\|_2 \nonumber \\
& \le \frac{n^2}{N} \sum_{s=1}^N \sum_{k\ne l} \frac{1}{N_{x,\epsilon}^2} \l\| \nabla f(x_s) \r\|_2^2 \nonumber \\
& \le \frac{n^2}{N} \sum_{s=1}^N \sum_{k\ne l} \frac{1}{N_{x,\epsilon}^2}  L_f^2 \qquad \mbox{(see  \eqref{eq:Lf})} \nonumber \\
& \le \frac{n^2}{N} \sum_{s=1}^N L_f^2 \nonumber \\
& = n^2 L_f^2,
\end{align}
we conclude that
\begin{align}\label{eq:totalexp2K}
& \l\| \mathbb{E}_{Y_{X,\epsilon}|\NN_{X,\epsilon},\Cr{good},X}\l[\ddddot{\Sigma}^o_{X,Y_{X,\epsilon}}\r] - \mathbb{E}_{Y_{X,\epsilon}|\Cr{Q},\NN_{X,\epsilon},\Cr{good},X}\l[\ddddot{\Sigma}^o_{X,Y_{X,\epsilon}}\r] \r\|_F \nonumber \\
& \le \Pr_{Y_{X,\epsilon}|\Cr{good},\NN_{X,\epsilon},X}\l[ \Cr{Q}^C\r] \nonumber\\
& \qquad \qquad \cdot \l\| \mathbb{E}_{Y_{X,\epsilon}|\Cr{Q}^C, \NN_{X,\epsilon},\Cr{good},X}\l[\ddddot{\Sigma}^o_{X,Y_{X,\epsilon}}\r] - \mathbb{E}_{Y_{X,\epsilon}|\Cr{Q},\NN_{X,\epsilon},\Cr{good},X}\l[\ddddot{\Sigma}^o_{X,Y_{X,\epsilon}}\r] \r\|_F  \nonumber \\
&  \le 2 n^2 L_f^2 \Pr_{Y_{X,\epsilon}|\Cr{good},\NN_{X,\epsilon},X}\l[ \Cr{Q}^C\r], \qquad \mbox{(triangle inequality and \eqref{eq:totalexp1K})}
\end{align}
where the second line above uses \eqref{eq:totalexpK}).

Lastly, we remove the conditioning on the event $\Cr{Q}$ as follows:
\begin{align}
& \Pr_{Y_{X,\epsilon}|\Cr{good},\NN_{X,\epsilon},X}\l[ \vphantom{\frac{n \cdot \max[Q_{X,\epsilon},Q_{X,\epsilon}^2] \cdot \L_f^2}{\sqrt{\rho_{\mu,X,\epsilon} N_{X,\epsilon}}}} \l\| \ddddot{\Sigma}^o_{X,Y_{X,\epsilon}} - \mathbb{E}_{Y_{X,\epsilon}|\Cr{good},\NN_{X,\epsilon},X}\l[
\ddddot {\Sigma}^o_{X,Y_{X,\epsilon}}
\r] \r\|_F \right. \nonumber \\
& \quad\quad\quad \left. \gtrsim \max[\Cr{g2},\gamma] \cdot \sqrt{\log n} \cdot \frac{n \cdot \max[Q_{X,\epsilon},Q_{X,\epsilon}^2] \cdot \L_f^2}{\sqrt{\rho_{\mu,X,\epsilon} N_{X,\epsilon}}} + 2 n^2 L_f^2 \Pr_{Y_{X,\epsilon}|\Cr{good},\NN_{X,\epsilon},X}\l[ \Cr{Q}^C\r]
\r]
\nonumber\\
& \le  \Pr_{Y_{X,\epsilon}|\Cr{Q},\Cr{good},\NN_{X,\epsilon},X}\l[ \vphantom{\frac{n \cdot \max[Q_{X,\epsilon},Q_{X,\epsilon}^2] \cdot \L_f^2}{\sqrt{\rho_{\mu,X,\epsilon} N_{X,\epsilon}}}} \l\| \ddddot{\Sigma}^o_{X,Y_{X,\epsilon}} - \mathbb{E}_{Y_{X,\epsilon}|\Cr{good},\NN_{X,\epsilon},X}\l[
\ddddot{\Sigma}^o_{X,Y_{X,\epsilon}}
\r] \r\|_F \right. \nonumber \\
& \quad\quad\quad \left. \gtrsim \max[\Cr{g2},\gamma] \cdot \sqrt{\log n} \cdot \frac{n \cdot \max[Q_{X,\epsilon},Q_{X,\epsilon}^2] \cdot \L_f^2}{\sqrt{\rho_{\mu,X,\epsilon} N_{X,\epsilon}}} + 2 n^2 L_f^2 \Pr_{Y_{X,\epsilon}|\Cr{good},\NN_{X,\epsilon},X}\l[ \Cr{Q}^C\r]
\r] \nonumber \\
& \quad + \Pr_{Y_{X,\epsilon}|\Cr{good},\NN_{X,\epsilon},X}\l[ \Cr{Q}^C\r] \qquad
\mbox{(see \eqref{eq:useful ineq})}
\nonumber\\
&
\le \Pr_{Y_{X,\epsilon}|\Cr{Q},\Cr{good},\NN_{X,\epsilon},X}\Bigg[ \l\| \ddddot{\Sigma}^o_{X,Y_{X,\epsilon}} - \mathbb{E}_{Y_{X,\epsilon}|\Cr{Q},\Cr{good},\NN_{X,\epsilon},X}\l[
\ddddot{\Sigma}^o_{X,Y_{X,\epsilon}}
\r] \r\|_F \nonumber\\
& \qquad \qquad \qquad \qquad \qquad \gtrsim \max[\Cr{g2},\gamma] \cdot \sqrt{\log n} \cdot \frac{n \cdot \max[Q_{X,\epsilon},Q_{X,\epsilon}^2] \cdot \L_f^2}{\sqrt{\rho_{\mu,X,\epsilon} N_{X,\epsilon}}}
\Bigg] \nonumber \\
& \quad + \Pr_{Y_{X,\epsilon}|\Cr{good},\NN_{X,\epsilon},X}\l[ \Cr{Q}^C\r] \qquad \mbox{(see \eqref{eq:totalexp2K})}
 \nonumber\\
& \le e^{-\gamma} + n^2  \cdot n^{-\log  \Cr{g2}} + \Pr_{Y_{X,\epsilon}|\Cr{good},\NN_{X,\epsilon},X}\l[ \Cr{Q}^C\r]. \qquad \mbox{(see \eqref{eq:decompose err finalE1})}
\end{align}
This holds for $\gamma, \Cr{g2} \ge 1$ and under \eqref{eq:likely bnd} (with $p=\max[\log n,1] \le N$). Setting $\gamma = \Cr{g2}$ completes the proof of Lemma~\ref{lem:chaos result}.

\section{Proof of Lemma~\ref{lem:bnd on Q}}
\label{sec:proof of lemma bnd on Q}

First, to prove~\eqref{eq:E1likely}, suppose $X$ and the neighborhood structure $\NN_{X,\epsilon}=\{N_{x,\epsilon} \}_{x\in X}$ are fixed. Then, for every $x\in X$,
 the columns of the matrix $Y_{x,\epsilon}\in \R^{n\times N_{x,\epsilon}}$ are random vectors drawn from the conditional probability measure  $\mu_{x,\epsilon}$ (see \eqref{eq:cond mu}). For fixed $x\in X$  and with $y\sim \mu_{x,\epsilon}$, recall from Assumption \ref{def:moments} that
\begin{equation}
\Pr{}_{y|x}\l[
\l\| P_{x,y} v \r\|_2^2
> \frac{\Cr{beta}}{n}
\r] \lesssim e^{-\K_\mu \Cr{beta}},
\label{eq:pxyv}
\end{equation}
for arbitrary (but fixed)  $v\in\R^n $ with $\|v\|_2=1$ and   $\Cr{beta}\ge 0$. The inequality~\eqref{eq:E1likely} readily follows with an application of the union bound: For all possible choices of $x,y$, it holds that
\begin{equation}
\l\| P_{x,y} \nabla f(x)\r\|_2^2
\le \frac{\Cr{beta} \l\| \nabla f(x)\r\|_2^2}{n} \le \frac{\Cr{beta} \L_f^2}{n},
\qquad \mbox{(see \eqref{eq:Lf})}
\end{equation}
except with a probability $\lesssim N_{X,\epsilon} e^{-\K_\mu\Cr{beta}}$. With the choice of $$\Cr{beta}=Q_{X,\epsilon}=\Cr{events} \K_\mu^{-1} \log^2(N_{X,\epsilon})$$ for $\Cr{events}\ge 3$, we establish~\eqref{eq:E1likely}.

Our next goal is to prove that \eqref{eq:chaos event likely} is satisfied. Note that the probability in \eqref{eq:chaos event likely} is conditioned on $\Cr{Q}$. We can remove this conditioning using the law of total probability:
\begin{align*}
\Pr_{\widetilde{Y}_{X,\epsilon}|\NN_{X,\epsilon},\Cr{good},X}\l[ \Cr{max}^C \r] &=
\Pr_{\widetilde{Y}_{X,\epsilon}|\NN_{X,\epsilon},\Cr{Q},\Cr{good},X}\l[ \Cr{max}^C \r]
\Pr_{\widetilde{Y}_{X,\epsilon}|\NN_{X,\epsilon},\Cr{good},X}\l[ \Cr{Q} \r] \nonumber\\
& \qquad +
\Pr_{\widetilde{Y}_{X,\epsilon}|\NN_{X,\epsilon},\Cr{Q}^C,\Cr{good},X}\l[ \Cr{max}^C \r]
\Pr_{\widetilde{Y}_{X,\epsilon}|\NN_{X,\epsilon},\Cr{good},X}\l[ \Cr{Q}^C \r] \nonumber \\
& \ge \Pr_{\widetilde{Y}_{X,\epsilon}|\NN_{X,\epsilon},\Cr{Q},\Cr{good},X}\l[ \Cr{max}^C \r]
\Pr_{\widetilde{Y}_{X,\epsilon}|\NN_{X,\epsilon},\Cr{good},X}\l[ \Cr{Q} \r].
\end{align*}
Rearranging terms, we have that
\begin{align}
\Pr_{\widetilde{Y}_{X,\epsilon}|\NN_{X,\epsilon},\Cr{Q},\Cr{good},X}\l[ \Cr{max}^C \r] &\le \frac{\Pr_{\widetilde{Y}_{X,\epsilon}|\NN_{X,\epsilon},\Cr{good},X}\l[ \Cr{max}^C \r]}{\Pr_{\widetilde{Y}_{X,\epsilon}|\NN_{X,\epsilon},\Cr{good},X}\l[ \Cr{Q} \r]} \nonumber \\
& = \frac{\Pr_{\widetilde{Y}_{X,\epsilon}|\NN_{X,\epsilon},\Cr{good},X}\l[ \Cr{max}^C \r]}{1 - \Pr_{\widetilde{Y}_{X,\epsilon}|\NN_{X,\epsilon},\Cr{good},X}\l[ \Cr{Q}^C \r]} \nonumber \\
& \lesssim \Pr_{\widetilde{Y}_{X,\epsilon}|\NN_{X,\epsilon},\Cr{good},X}\l[ \Cr{max}^C \r],
\label{eq:E3uncond}
\end{align}
where the last line follows under the assumption that $N_{X,\epsilon}$ large enough that, under~\eqref{eq:E1likely}, $\Pr_{\widetilde{Y}_{X,\epsilon}|\NN_{X,\epsilon},\Cr{good},X}\l[ \Cr{Q}^C \r]$ is bounded above by a constant smaller than $1$. To bound the right hand side in~\eqref{eq:E3uncond}, suppose $X$ and the neighborhood structure $\NN_{X,\epsilon}=\{N_{x,\epsilon} \}_{x\in X}$ are fixed. Then, for every $x\in X$, the columns of  the matrix $\widetilde{Y}_{x,\epsilon}\in \mathbb{R}^{n\times (4 N_{x,\epsilon})}$ are random vectors drawn from the conditional probability measure $\mu_{x,\epsilon}$ (see \eqref{eq:cond mu}). For fixed $x\in X$  and with $y\sim \mu_{x,\epsilon}$, recall from Assumption \ref{def:moments} that~\eqref{eq:pxyv} holds for arbitrary (but fixed) $v\in\R^n $ with $\|v\|_2=1$ and $\Cr{beta}\ge 0$. For all possible choices of $x,y,i$, it follows that
\begin{equation}
\l\| P_{x,y}e_i\r\|_2^2 \le \frac{\Cr{beta}}{n},
\qquad
\l\| P_{x,y} \nabla f(x)\r\|_2^2
\le \frac{\Cr{beta} \l\| \nabla f(x)\r\|_2^2}{n} \le \frac{\Cr{beta} \L_f^2}{n},
\qquad \mbox{(see \eqref{eq:Lf})}
\end{equation}
except with a probability $\lesssim n N_{X,\epsilon} e^{-\K_\mu\Cr{beta}}$. With the choice of $\Cr{beta}=Q_{X,\epsilon}=\Cr{events} \K_\mu^{-1} \log^2(N_{X,\epsilon})$ for $\Cr{events}\ge 3$, we find that
\begin{align}
\Pr_{\widetilde{Y}_{X,\epsilon}|\NN_{X,\epsilon},\Cr{good},X}\l[ \Cr{max}^C \r]
& \lesssim n N_{X,\epsilon}^{(1-\Cr{events}\log(N_{X,\epsilon}))} \nonumber \\
& \lesssim n^{(2-\Cr{events}\log(N_{X,\epsilon}))} \quad \mbox{($N_{X,\epsilon} \ge n$)} \nonumber \\
& \lesssim n^{(2-\Cr{events})\log(N_{X,\epsilon})} \quad (\log(N_{X,\epsilon}) \ge 1) \nonumber \\
& \lesssim n^{-\log(N_{X,\epsilon})} \quad \mbox{($\Cr{events} \ge 3$)} \nonumber \\
& = N_{X,\epsilon}^{-\log(n)} \nonumber \\
& = \left( \frac{1}{N_{X,\epsilon}^2} \right)^{\frac{1}{2} \log(n)} \nonumber \\
& \le \l( \frac{\log n}{N_{X,\min,\epsilon}\rho_{\mu,X,\epsilon} N_{X,\epsilon}} \r)^{\frac{\log n}{2}},
\label{eq:E3uncond2}
\end{align}
where the last line follows since $N_{X,\epsilon} \ge N_{X,\min,\epsilon}$, $\log(n) \ge 1$, and $\rho_{\mu,X,\epsilon} \le 1$. Combining~\eqref{eq:E3uncond} and~\eqref{eq:E3uncond2} proves that \eqref{eq:chaos event likely} is satisfied and thus completes the proof of Lemma~\ref{lem:bnd on Q}.

\medskip
Received xxxx 20xx; revised xxxx 20xx.
\medskip

\end{document}